\documentclass[aip,jcp,amsmath,amssymb,reprint]{revtex4-1}
\usepackage{graphicx}
\usepackage{dcolumn}
\usepackage{bm}
\usepackage{mathrsfs}
\usepackage{amsmath, amssymb, amsthm, amsfonts}
\usepackage[dvipsnames]{xcolor}
\usepackage{ulem}
\usepackage{caption}
\usepackage[textfont=rm]{subcaption}

\begin{document}
\title{Structure-Dynamics Correlation and Its Link to Fragility and Dynamic Heterogeneity}
\author{Mohit Sharma}
\address{\textit{Polymer Science and Engineering Division, CSIR-National Chemical Laboratory, Pune-411008, India}}
\affiliation{\textit{Academy of Scientific and Innovative Research (AcSIR), Ghaziabad 201002, India}}
\author{Srikanth Sastry}
\affiliation{\textit{Theoretical Sciences Unit and School of Advanced Materials, Jawaharlal Nehru Centre for Advanced Scientific Research, Jakkur Campus, Bengaluru 560064, Karnataka, India}}

\author{Sarika Maitra Bhattacharyya}
\email{mb.sarika@ncl.res.in}
\address{\textit{Polymer Science and Engineering Division, CSIR-National Chemical Laboratory, Pune-411008, India}}
\affiliation{\textit{Academy of Scientific and Innovative Research (AcSIR), Ghaziabad 201002, India}}

\begin{abstract}

Understanding the connection between structure, dynamics, and fragility, the rate at which the relaxation time grows with decreasing temperature, is central to unravelling the glass transition. Fragility is often associated with dynamic heterogeneity, implying that if structure influences dynamics, more fragile systems should exhibit stronger structure–dynamics correlations. In this study, we test the generality of this assumption using: Lennard-Jones (LJ) and Weeks–Chandler–Andersen (WCA) systems, where fragility is tuned via density, and a modified LJ $(q,p)$ system, where fragility is varied by changing the potential softness. We define a structural order parameter based on a mean-field caging potential and analyse energy barriers at both macroscopic and microscopic levels. While the macroscopic free energy barrier slope correlates with fragility, the microscopic free energy barrier does not show a consistent trend. Instead, it exhibits a strong correlation with a structure–dynamics correlation measure obtained from isoconfigurational ensemble simulations. Interestingly, the two systems showing the highest structure–dynamics correlation, LJ at $\rho = 1.1$ and the $(8,5)$ model, are respectively the least and most fragile within their classes. These systems exhibit broad mobility distributions, large non-Gaussian parameters, yet low four-point susceptibilities, suggesting a decoupling between spatial correlation length and mobility contrast. Both systems lie in the enthalpy dominated regime and are close to the spinodal, pointing to mechanical instability as a source of heterogeneity. Our results reveal that structure–dynamics correlation is more closely linked to the contrast in individual particle mobility than to the spatial extent of dynamic correlations that typically scale with fragility.

\end{abstract}
\maketitle
\section{Introduction}

The behaviour of supercooled liquids near the glass transition is characterised by a dramatic slowing down in dynamics without any obvious structural transition \cite{Ediger2000_Physchem, Debenedetti2001_Nature, Dyre2006_modern, Lubchenko2007_Physchem, Tong2018_prx}. The rapidity with which the relaxation time increases with decreasing temperature is quantified in terms of the fragility parameter \cite{Angell1995_science}. Fragile liquids show a super-Arrhenius increase in relaxation time, while strong liquids follow an Arrhenius behaviour \cite{Martinez2001_nature}. 

One of the hallmarks of supercooled liquids is dynamic heterogeneity, the presence of regions with significantly different mobilities. Traditionally, this heterogeneity has been linked to fragility: more fragile systems tend to exhibit stronger dynamic heterogeneity \cite{Kob1997_prl, Donati1998_pre, Sillescu1999_noncryst, Ediger2000_Physchem, Martinez2001_nature, Sastry2001_nature, Biroli2006_prl}. This link is often assessed through quantities like the four-point dynamic susceptibility, $\chi_{4}$, which captures spatial correlations in particle dynamics \cite{Lacevic2003_jcp, Toninelli2005_pre}.

Although, as discussed in the literature, the glass transition and the rapid growth of the dynamics happens without substantial structural change \cite{Debenedetti2001_Nature, Lubchenko2007_Physchem}, studies have shown that structure does indeed contribute to the dynamics, and small changes in structure can lead to large changes in other quantities like entropy which eventually strongly effects the dynamics \cite{Banerjee2014_prl, Nandi2015_jcp, Banerjee2016_jcp}. There have also been studies trying to understand the structural origin of dynamical heterogeneity \cite{Cooper2008_nature, Schoenholz2015_nature, Richard2020_prm, Nandi2021_prl}. Among these studies, the machine learning study expressing the structural order by a softness parameter \cite{Schoenholz2015_nature} and the microscopic study involving some of us expressing the structural order in terms of the depth of a mean field caging potential \cite{Nandi2021_prl,Sharma2022_pre} sought to directly connect local structure to the local dynamics, in addition to investigating the global structure-dynamics correlation. For a fragile system, super-Arrhenius growth in the relaxation time implies a large change in the activation energy. It was suggested that the slope in the activation energy obtained from the macroscopic dynamics, when plotted as a function of the inverse depth of the caging potential, is high for more fragile liquids \cite{Nandi2021_prl}. In a later study involving the machine learned softness parameter, it was shown that the slope of the activation energy obtained from microscopic dynamics, when plotted against the ML softness, increases with fragility \cite{Tah2022_jcp}. A higher slope indicates that even small variations in the order parameter can lead to significant changes in the free energy barrier and dynamics, reflecting a stronger coupling between structure and dynamics. Thus, these studies suggested that for a more fragile system, the structure-dynamics correlation is stronger, both macroscopically and microscopically. 

Our goal in this study is to understand how generic this statement is. To do that, we take three different sets of systems and vary fragility by changing different properties. We take a system interacting via the attractive Lennard Jones (LJ) potential and its repulsive counterpart, the Weeks-Chandler-Andersen(WCA) potential, and vary their fragility by changing the density \cite{Berthier2009_jcp, Banerjee2016_jcp}. We also vary the stepness of an attractive LJ-like potential and vary the fragility, which we refer to as the modified LJ $(q,p)$ system consisting of $(12,11)$, $(12,6)$ and $(8,5)$ models (where the pairs of integers refer to the exponent of the repulsive and attractive interactions)\cite{Bordat2004_prl, Sengupta2011_jcp}. This way, we have both attractive and repulsive potentials and also systems where the fragility is varied without changing the density.

We find that there is no consistent correlation between fragility and the slope of the scaled activation free energy barrier. However, the latter does correlate well with a complementary measure of structure-dynamics correlation, as obtained from isoconfigurational ensemble simulations. Notably, two systems, LJ at $\rho=1.1$ and $(8,5)$, exhibit the highest structure-dynamics correlation, despite having very different fragility values. What is striking is that both systems also display a broad distribution of particle mobilities, a distinct double-peak in the logarithmic displacement distribution, and large values of the non-Gaussian parameter but small values of four-point dynamic susceptibility. We also find that these two systems are in the enthalpy-dominated regime and very close to the spinodal line \cite{Sastry2000_prl_press,Sengupta2011_jcp} where it is known that dynamic heterogeneity may arise due to mechanical instability and growing fluctuations \cite{Sastry2000_prl_press}.

These findings lead us to emphasise a critical distinction: dynamic heterogeneity is a broader concept than what is captured by $\chi_{4}$, which quantifies spatial correlations in particle motion. Dynamic heterogeneity also includes the distribution and variability of individual particle mobilities, which can arise due to different reasons and can exist even in the absence of significant spatial correlations. Thus, whereas $\chi_{4}$ highlights cooperative dynamics, other measures, such as the van Hove correlation function, the non-Gaussian parameter and mobility distributions from isoconfigurational runs, reflect heterogeneity in a more general sense.

This insight is significant because it challenges the commonly held assumption that fragility, dynamic heterogeneity, and structure-dynamics correlation should be universally connected. Instead, our results point to a more nuanced picture, where distinct types of heterogeneity may dominate under different conditions, and where strong structure-dynamics correlation can arise from mechanisms not directly tied to fragility.

The rest of the paper is organised as follows. Section~\ref{section_simulation} describes the simulation details, while Section~\ref{section_methodology} outlines the methodology. Section~\ref{section_results} presents the observations and results. Finally, Section~\ref{conclusion_section} provides the conclusions. In addition, the paper includes five appendices that provide supplementary information.

\section{Simulation Details}
\label{section_simulation}
\subsection{Simulation Details for the LJ and WCA Systems}
The first two classes of systems we study are the 3-dimensional Kob-Andersen model for glass-forming liquid, which is a binary mixture (80:20) of Lennard-Jones (LJ) particles \cite{Kob1994_prl} and its repulsive counterpart Weeks-Chandler-Andersen potential (WCA)\cite{Weeks1971_jcp}. The interaction between the particles i and j, where i,j = A,B (the type of the particles), is given by
\begin{equation}
U_{ij}(r) =
\begin{cases}
\begin{aligned}
&U_{ij}^{(LJ)}(r;\sigma_{ij},\epsilon_{ij}) \\
&\quad - U_{ij}^{(LJ)}(r^{(c)}_{ij};\sigma_{ij},\epsilon_{ij}),
\end{aligned} & r \leq r^{(c)}_{ij} \\
0, & r > r^{(c)}_{ij},
\end{cases}
\label{pot}
\end{equation}

where $U_{ij}^{(LJ)}(r)$ = $4\epsilon_{ij}[(\sigma_{ij}/r)^{12}-(\sigma_{ij}/r)^6]$, $r$ is the distance between particles i and j and $\sigma_{ij} $ is the effective diameter of the particle and $r_{ij}^{(c)}=2.5\sigma_{ij}$ for LJ and for the WCA system $r^{(c)}_{ij}$ is the position of minima of $U_{ij}^{(LJ)}(r)$.

The length, temperature, and time are given in units of $\sigma_{AA}$, $\epsilon_{AA}/k_B$, $(m\sigma_{AA}^2 /\epsilon_{AA})^{1/2}$, respectively.
We use $\sigma_{AA}=1.0 $, $\sigma_{AB}=0.8 $, $\sigma_{BB}=0.88 $, $\epsilon_{AA}=1.0 $, $\epsilon_{AB}=1.5 $, $\epsilon_{BB}=0.5 $, $ m_{A}=m_{B}=1 $ and Boltzmann constant $k_{B} = 1$.

We have performed MD simulation(using the LAMMPS package\cite{Majure2008_lammps}), we have used periodic boundary conditions and Nos\'{e}-Hoover thermostat with integration timestep 0.005$\tau$. The time constants for Nos\'{e}-Hoover thermostat are taken to be 100 timesteps. We worked with 4 different total number densities, i.e. $\rho = 1.1,1.2,1.4,1.6$ for both systems, where V is the system volume, and N=4000 is the total number of particles.

\subsection{Simulation Details for the $(q,p)$ Models}
We also investigate a series of binary mixtures in which particles interact {\it via.} a modified Lennard-Jones-type potential\cite{Sengupta2011_jcp} defined as:
\begin{equation}
U_{ij}(r) =
\begin{cases}
\begin{aligned}
\frac{\epsilon_{ij}}{q-p} \left[p \left( \dfrac{r^{\min}_{ij}}{r} \right)^q - q \left( \dfrac{r^{\min}_{ij}}{r} \right)^p \right] \\+ c_{1,ij} r^2 + c_{2,ij},\end{aligned} & r < r_c^{ij} \\
0, & r \geq r_c^{ij}
\end{cases}
\label{eq:ksystem_potential}
\end{equation}

Here, \( r^{\min}_{ij} = 2^{1/6} \sigma_{ij} \) is the position of the potential minimum, and the correction terms \( c_{1,ij} \) and \( c_{2,ij} \) are chosen such that the potential and its derivative vanish smoothly at the cutoff radius \( r_c^{ij} = 2.5\sigma_{ij} \):
\[
U_{ij}(r_c^{ij}) = 0, \quad \left. \frac{dU_{ij}}{dr} \right|_{r = r_c^{ij}} = 0.
\]

We consider three models denoted as $(12,11)$, $(12,6)$, and $(8,5)$, corresponding to exponent pairs $(q,p) = (12,11)$, $(12,6)$, and $(8,5)$, respectively. The parameters \( \sigma_{ij} \), \( \epsilon_{ij} \), and the units of length, temperature, and time are the same as defined for the LJ and WCA systems.

Molecular dynamics simulations were performed using the LAMMPS package~\cite{Majure2008_lammps}. A cubic box with periodic boundary conditions was used with a Nos\'{e}-Hoover thermostat. The integration timestep was set to $0.005\tau$. The thermostat time constant was 100 timesteps. The total number of particles was $N = 4000$, and the number density was fixed at \( \rho = N/V = 1.2 \).

\section{Methodology}
\label{section_methodology}
\subsection{Structural order parameter}
\label{section_sop}
To quantify the static structure in glassy systems, we analyse the Structural Order Parameter (SOP) at both macroscopic and microscopic levels. The SOP is the inverse of the depth of the mean-field caging potential experienced by a particle due to its surrounding neighbours and is derived from the Ramakrishnan-Yussouff free energy functional~\cite{Ramakrishnan1979_prb}. We define the macroscopic SOP as $(\beta\Phi)^{-1}$ and the microscopic SOP as $(\beta\phi)^{-1}$ throughout this study. According to our earlier studies the inverse depth of the mean field caging potential is proportional to the curvature of the potential and thus related to its softness. \cite{Nandi2021_prl,Sharma2022_pre}
The effective caging potential at zero displacement is computed as
\begin{equation}
\begin{split}
\beta \Phi &= -\rho \int \textbf{dr} \sum_{uv} C_{uv}(r)\, x_u x_v\, g_{uv}(r) ,
\end{split}
\label{eq:macro_sop}
\end{equation}
where \( \rho \) is the number density, and \( x_u, x_v \) are the mole fractions of species \( u \) and \( v \), respectively.

The radial distribution function \( g_{uv}(r) \) is defined as\cite{HansenMcDonald}
\begin{equation}
g_{uv}(r) = \frac{1}{4\pi r^2 \rho_v N_u} \left\langle \sum_{i=1}^{N_u} \sum_{j=1}^{N_v} \delta(r - |\mathbf{r}_i^u - \mathbf{r}_j^v|) \right\rangle,
\label{eq:bulk_gr}
\end{equation}
where \( \rho_v \) is the number density of species \( v \), \( N_u \) and \( N_v \) are the number of particles of species \( u \) and \( v \), respectively, and the angular brackets denote an ensemble average over configurations.

\( C_{uv}(r) \) is the direct correlation function, which for the present study is approximated by using the Hypernetted Chain (HNC) closure relation\cite{HansenMcDonald}:
\begin{equation}
C_{uv}(r) = -\beta U_{uv}(r) + \left( g_{uv}(r) - 1 \right) - \ln g_{uv}(r),
\label{eq:hnc}
\end{equation}
where \( U_{uv}(r) \) is the pairwise interaction potential between particles of species \( u \) and \( v \), and \( \beta = 1/(k_B T) \) is the inverse thermal energy.

At the microscopic level, the effective caging potential experienced by particle \( i \) of type $u$ is given by
\begin{equation}
\beta \phi^i_{u} = -\rho \int \textbf{dr} \sum_{v} x_v C_{uv}^i(r)\, g_{uv}^i(r),
\label{eq:micro_sop}
\end{equation}
where \( \rho \) is the number density, \( C_{uv}^i(r) \) is the local direct correlation function, and \( g_{uv}^i(r) \) is the local radial distribution function for particle \( i \).

To compute \( g_{uv}^i(r) \), we employ a Gaussian smoothing scheme\cite{Piaggi2017_prl}:
\begin{equation}
g_{uv}^i(r) = \frac{1}{4\pi \rho r^2} \sum_j \frac{1}{\sqrt{2\pi \delta^2}} \exp\left(-\frac{(r - r_{ij})^2}{2\delta^2}\right),
\label{eq:micro_gr}
\end{equation}
where \( r_{ij} \) is the distance between particle \( i \) and particle \( j \), and \( \delta = 0.09\sigma_{AA} \) is the Gaussian width parameter. This smoothing technique provides a continuous and differentiable estimate of the local radial distribution function, mitigating the noise associated with discrete binning methods.

\subsection{Entropy and local energy}
\label{section_entropy}
The two-body excess entropy at the particle level for a binary system, corresponding to a particle $i$ of type $u$, is computed as \cite{Goel2008_jcp}:
\begin{equation}
\begin{split}
S^{i}_{u2} = 
-k_{B} \frac{\rho}{2} \sum_{v = 1}^{2} x_{v} \int_{0}^{\infty} \left\{ g^{i}_{uv}(r) \ln g^{i}_{uv}(r) - g^{i}_{uv}(r) + 1 \right\} \textbf{dr},
\end{split}
\label{s2_particle}
\end{equation}
\noindent
where the local radial distribution function $g^{i}_{uv}(r)$ is calculated as described in Section~\ref{section_sop}.

Similarly, the local potential energy $e^{i}_{uloc}$ for a particle $i$ of type $u$ is given by:
\begin{equation}
e^{i}_{uloc} = \frac{\rho}{2} \sum_{v = 1}^{2} x_{v} \int_{0}^{\infty} U_{uv}(r)\, g^{i}_{uv}(r) \textbf{dr},
\label{E_2}
\end{equation}
\noindent
where the interaction potential $U_{uv}(r)$ has been defined in the simulation details section(Section~\ref{section_simulation}).

\subsection{Relaxation time and Four-Point Susceptibility}
\label{section_relaxation}
The structural relaxation time, $\tau_{\alpha}$, is determined from the condition $q(t = \tau_{\alpha}) = 1/e$, where $q(t)$ denotes the time-dependent overlap function \cite{Sengupta2011_jcp}. It is defined as
\begin{equation}
\begin{aligned}
q(t) = \frac{1}{N} \sum_{i=1}^{N} \omega(|\mathbf{r}_i(t) - \mathbf{r}_i(0)|),
\end{aligned}
\label{overlap}
\end{equation}
\noindent
where the indicator function $\omega(x)$ takes the value 1 if $0 \leq x \leq a$ and 0 otherwise. The choice of the cut-off parameter $a$ is crucial, as it determines the sensitivity of the overlap function to particle displacements. In this study, we set $a = 0.3$ to account for low-amplitude vibrational motion such that particles undergoing only minor displacements are still considered to remain in the same position. This value is selected to be comparable to the mean squared displacement (MSD) observed in the plateau regime, which separates the ballistic and diffusive regimes in the dynamics.

As discussed earlier, in supercooled liquids, the relaxation time grows rapidly with temperature. This growth can be described by the Vogel-Fulcher-Tammann (VFT) relation,\cite{GarciaColin1989_vft}
\begin{equation}
 \tau_{\alpha}(T) = \tau_{0}\exp\Big(\frac{T_{VFT}}{K_{VFT}(T-T_{VFT}))}\Big), 
 \label{vft_T}
\end{equation}
where $\tau_{0}$ is a prefactor, $T_{VFT}$ is the temperature where the dynamics should diverge, and $K_{VFT}$ is the kinetic fragility which quantifies how rapidly the relaxation time grows with temperature. A high value of $K_{VFT}$ corresponds to a more fragile system, while a smaller value of $K_{VFT}$ indicates a stronger liquid.

To quantify spatial correlations in mobility, we compute the four-point susceptibility $\chi_4(t)$, defined as the variance of the time-dependent overlap function\cite{Flenner2014_prl} $q(t)$:
\begin{equation}
\chi_4(t) = N \left[ \langle q^2(t) \rangle - \langle q(t) \rangle^2 \right],
\end{equation}
where $N$ is the total number of particles. The peak of $\chi_4(t)$ identifies the time scale of maximum dynamic heterogeneity.
\subsection{van Hove correlation Function and Non-Gaussian Parameter} 
\label{section_alpha2}
The self-part of the van Hove correlation function\cite{vanHove1954_review, HansenMcDonald}, which gives the probability distribution of single-particle displacements, is written as:
\begin{equation}
G_s(r, t) = \left\langle \frac{1}{N} \sum_{i=1}^{N} \delta\left(r - |\mathbf{r}_i(t) - \mathbf{r}_i(0)|\right) \right\rangle.
\end{equation}
For systems in the high temperature regime, when the particle motion is homogeneous and diffusive, the van Hove correlation function is Gaussian. However, at the low temperature regime, due to heterogeneity in particle motion, the van Hove correlation function becomes non-Gaussian and it develops a tail which is best visualised in the probability distribution of logarithmic displacements\cite{Flenner2005_pre,Barrat2010_soft}, $P(\log_{10} \delta r)$, which is related to the displacement distribution $P(\delta r)=4\pi\delta r^{2} G_{s}(r,t)$ and can be written as:
\begin{eqnarray}
P(\log_{10} \delta r) =ln(10) \delta r P(\delta r) = ln(10)4\pi \delta r^3 G_s(\delta r, t).
\end{eqnarray}
This representation emphasises both mobile and immobile particle populations and is particularly useful for identifying the broad, often bimodal, displacement distributions observed at the $\alpha$-relaxation time in dynamically heterogeneous systems.

To measure the deviation of the particle displacement distribution from Gaussian, we use the non-Gaussian parameter\cite{Rahman1964_review}, $\alpha_2(t)$, defined as:
\begin{equation}
\alpha_2(t) = \frac{3\langle \delta r^4(t) \rangle}{5\langle \delta r^2(t) \rangle^2} - 1,
\end{equation}
where $\langle \delta r^n(t) \rangle$ denotes the $n$-th moment of the particle displacement distribution at time $t$. A higher value of $\alpha_2(t)$ indicates an increased dynamic heterogeneity and non-Gaussian behaviour in the displacement distribution.

\section{RESULTS}
\label{section_results}
As discussed in the Introduction, in this study we work with three different sets of systems, i) the attractive LJ potential at different densities ii) the repulsive WCA potential, at different densities \cite{Banerjee2016_jcp,Berthier2009_jcp}, iii) modified LJ $(q,p)$ sytem consisting of $(12,11)$,$(12,6)$ and $(8,5)$, systems- LJ-like potential where the steepness of the potential is varied \cite{Sengupta2011_jcp, Bordat2004_prl}, and vary fragility by changing different properties like the density and the steepness of the potential. While describing the results in all figures, we use a consistent colour and symbol scheme to represent increasing fragility for each set: black circles for the least fragile systems, followed by red squares, green up-triangles, and blue down-triangles for progressively more fragile systems.

\subsection {The temperature and SOP dependence of the free energy barrier and its correlation with fragility}
\label{section_deltaE}
\begin{figure}
\centering
\vspace{0.2cm}
\begin{subfigure}{0.23\textwidth}
\includegraphics[width=0.98\linewidth]{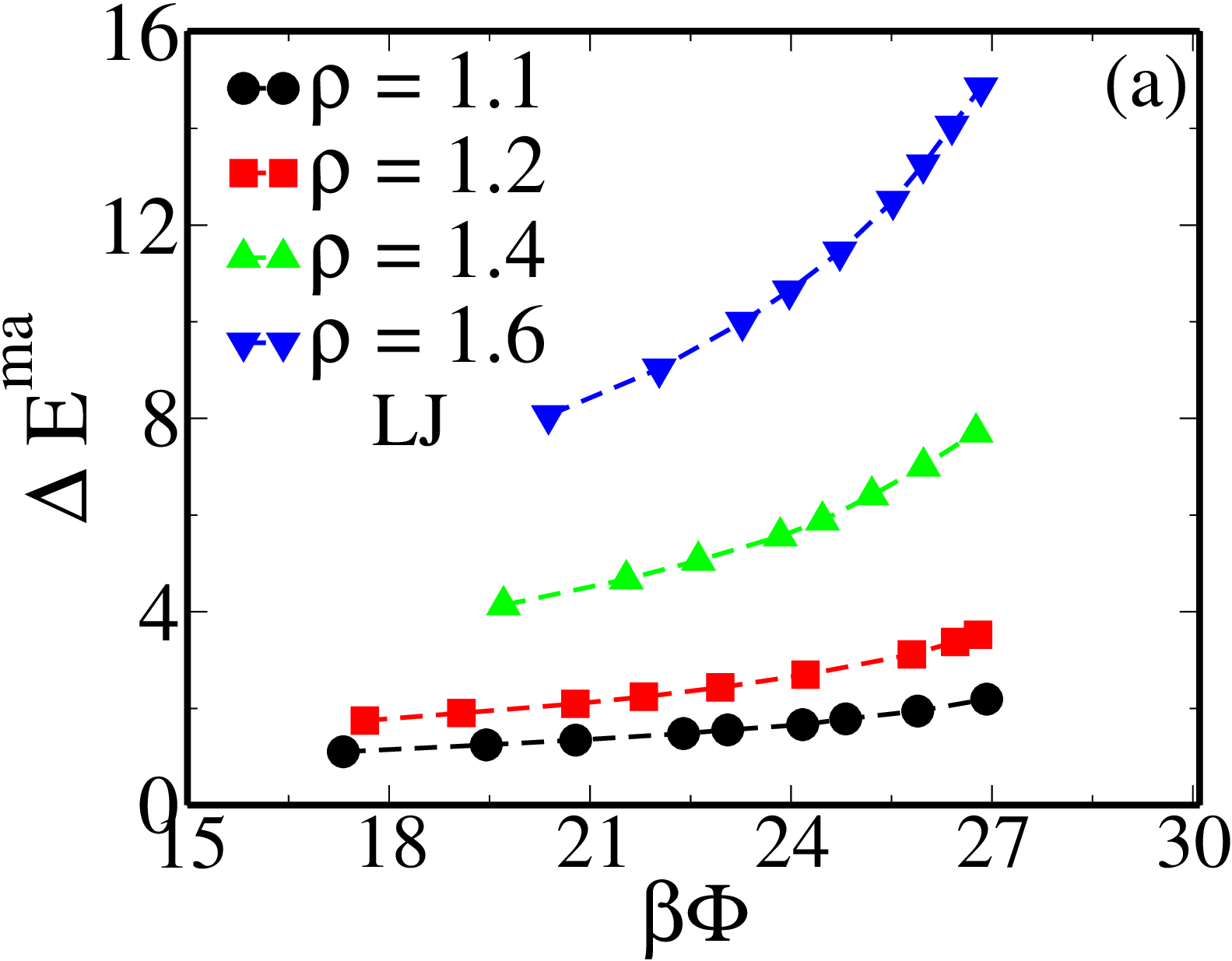}
\end{subfigure}
\hspace{0.2cm}
\begin{subfigure}{0.23\textwidth}
\includegraphics[width=0.98\linewidth]{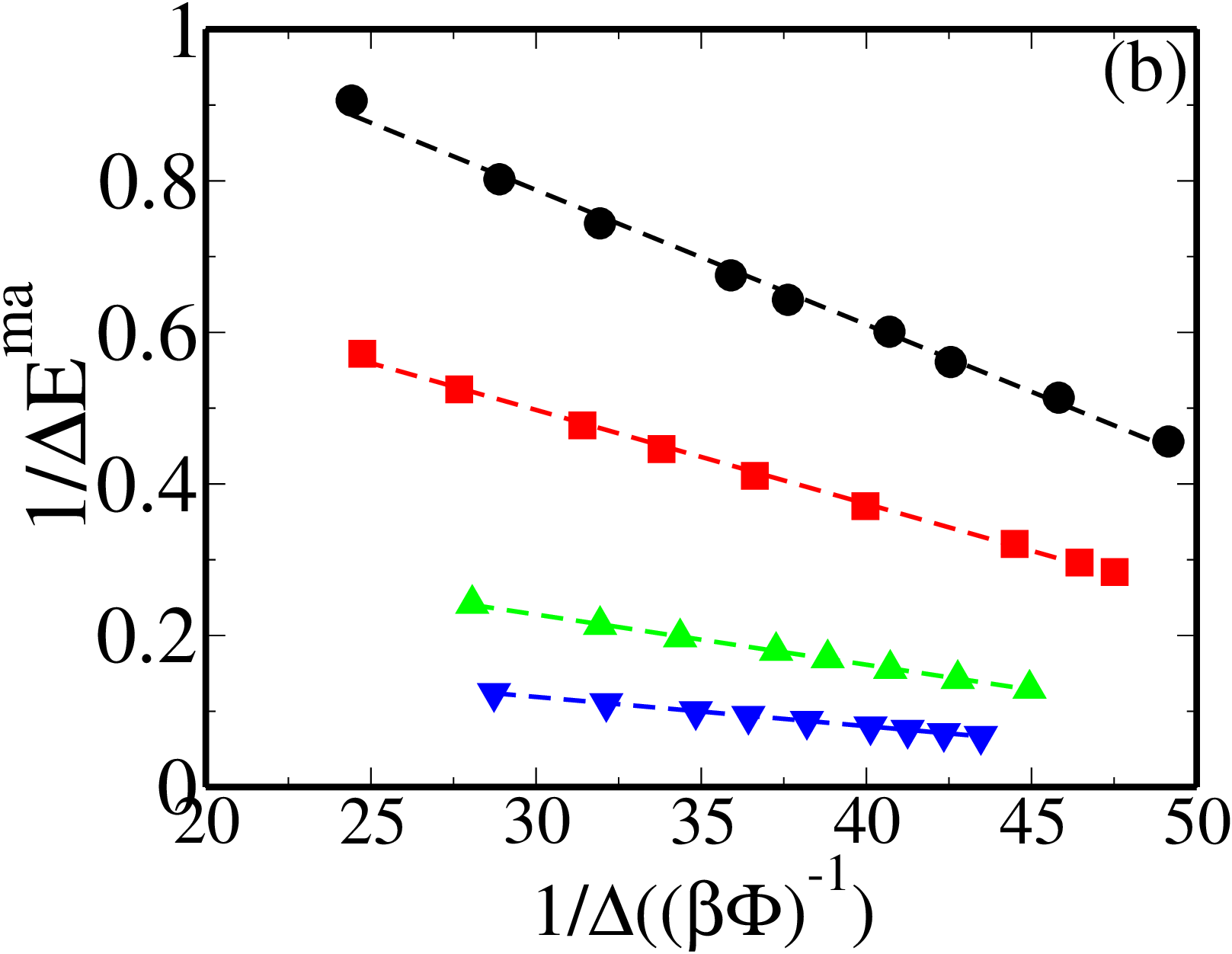}
\end{subfigure}
\vspace{0.2cm}
\begin{subfigure}{0.23\textwidth}
\includegraphics[width=0.98\linewidth]{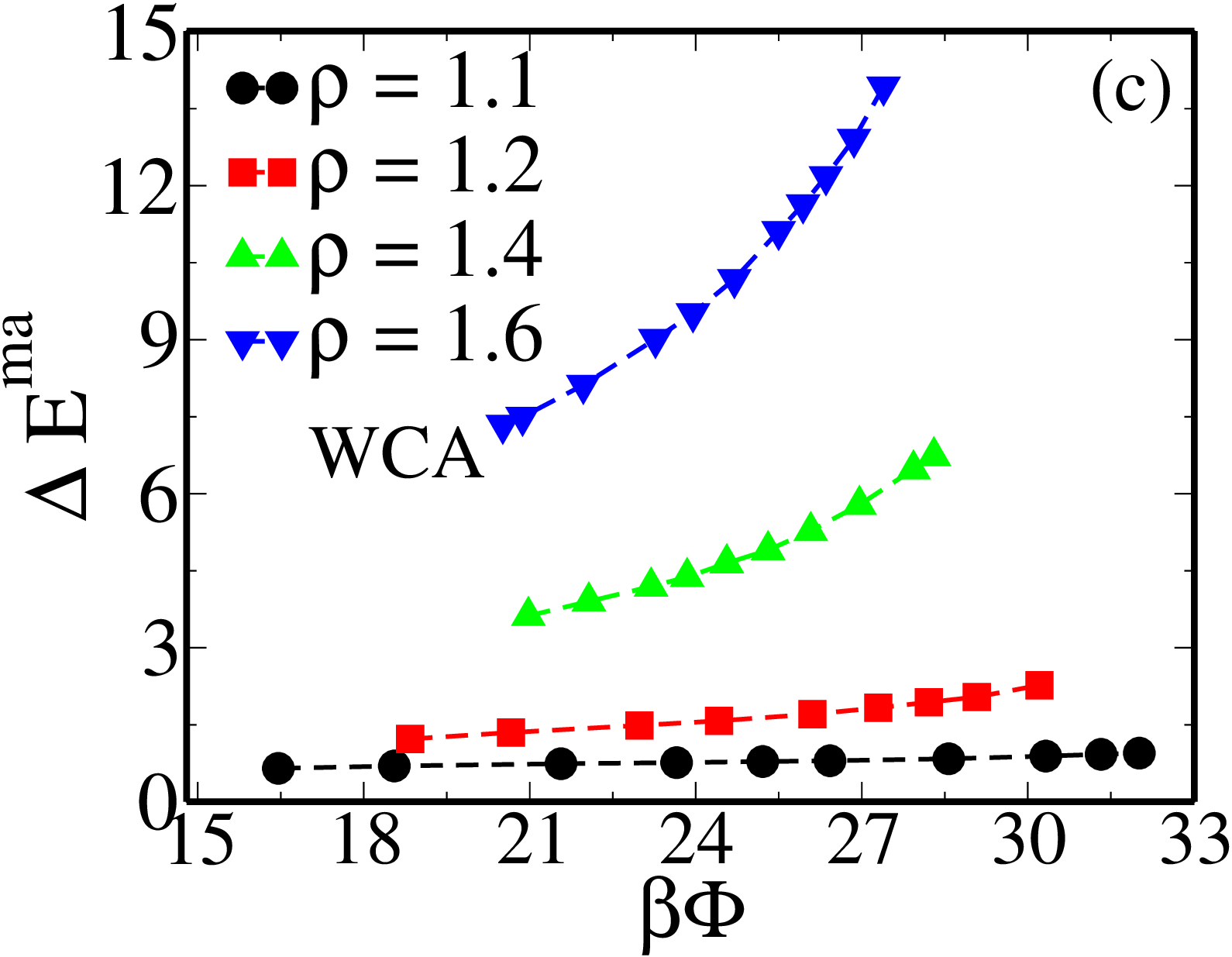}
\end{subfigure}
\hspace{0.2cm}
\begin{subfigure}{0.22\textwidth}
\includegraphics[width=0.98\linewidth]{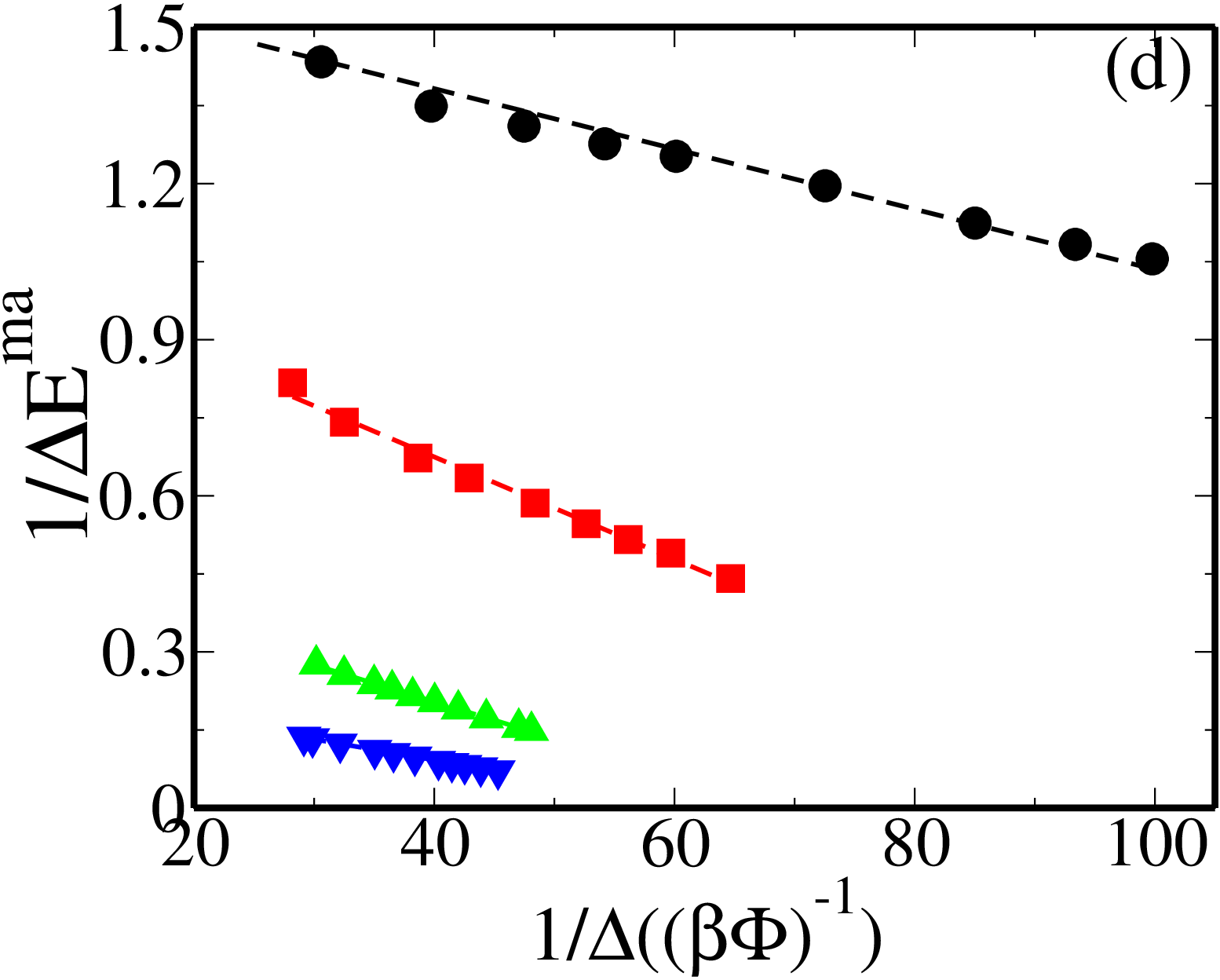}
\end{subfigure}
\vspace{0.2cm}
\begin{subfigure}{0.22\textwidth}
\includegraphics[width=0.98\linewidth]{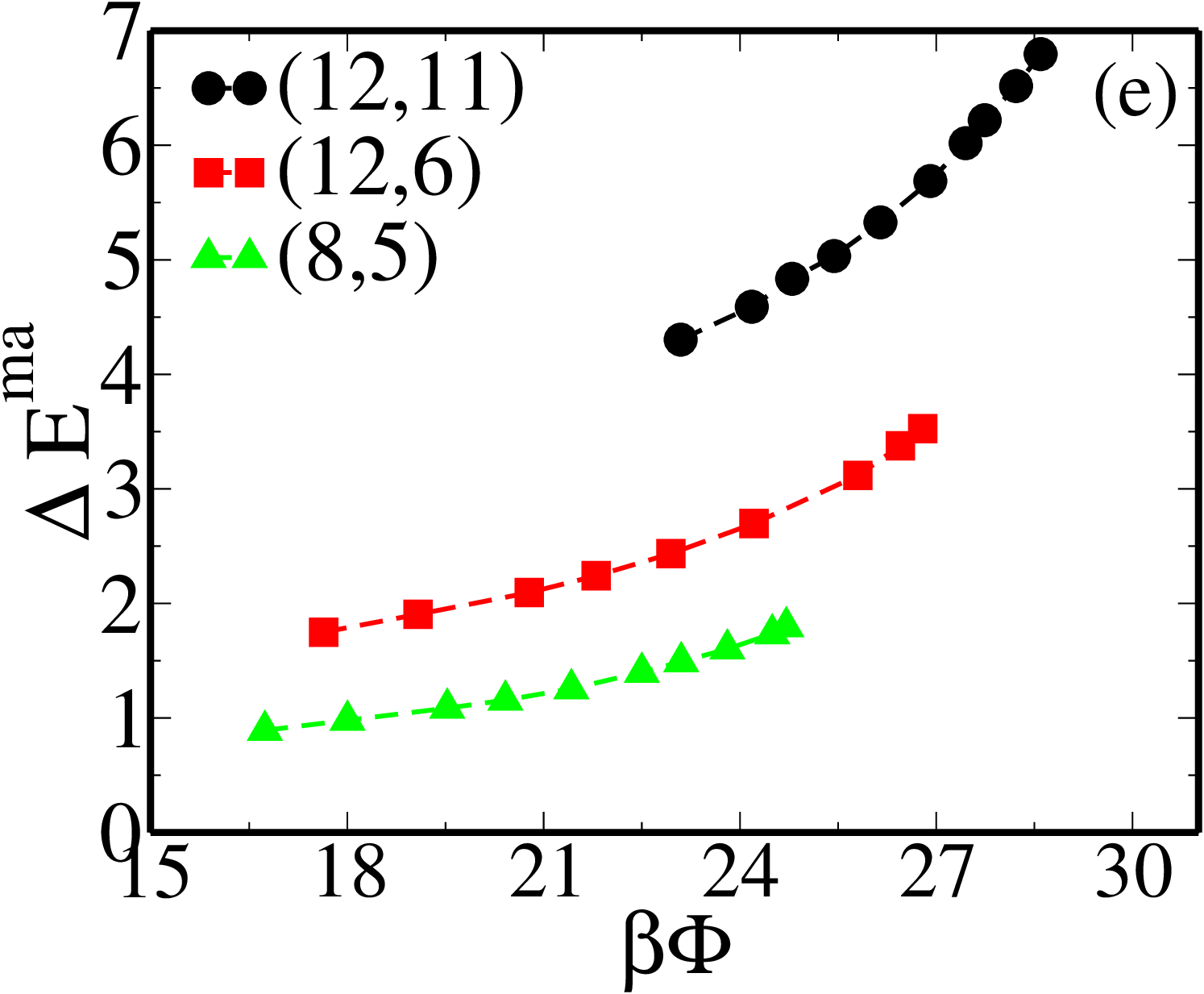}
\end{subfigure}
\hspace{0.2cm}
\begin{subfigure}{0.23\textwidth}
\includegraphics[width=0.98\linewidth]{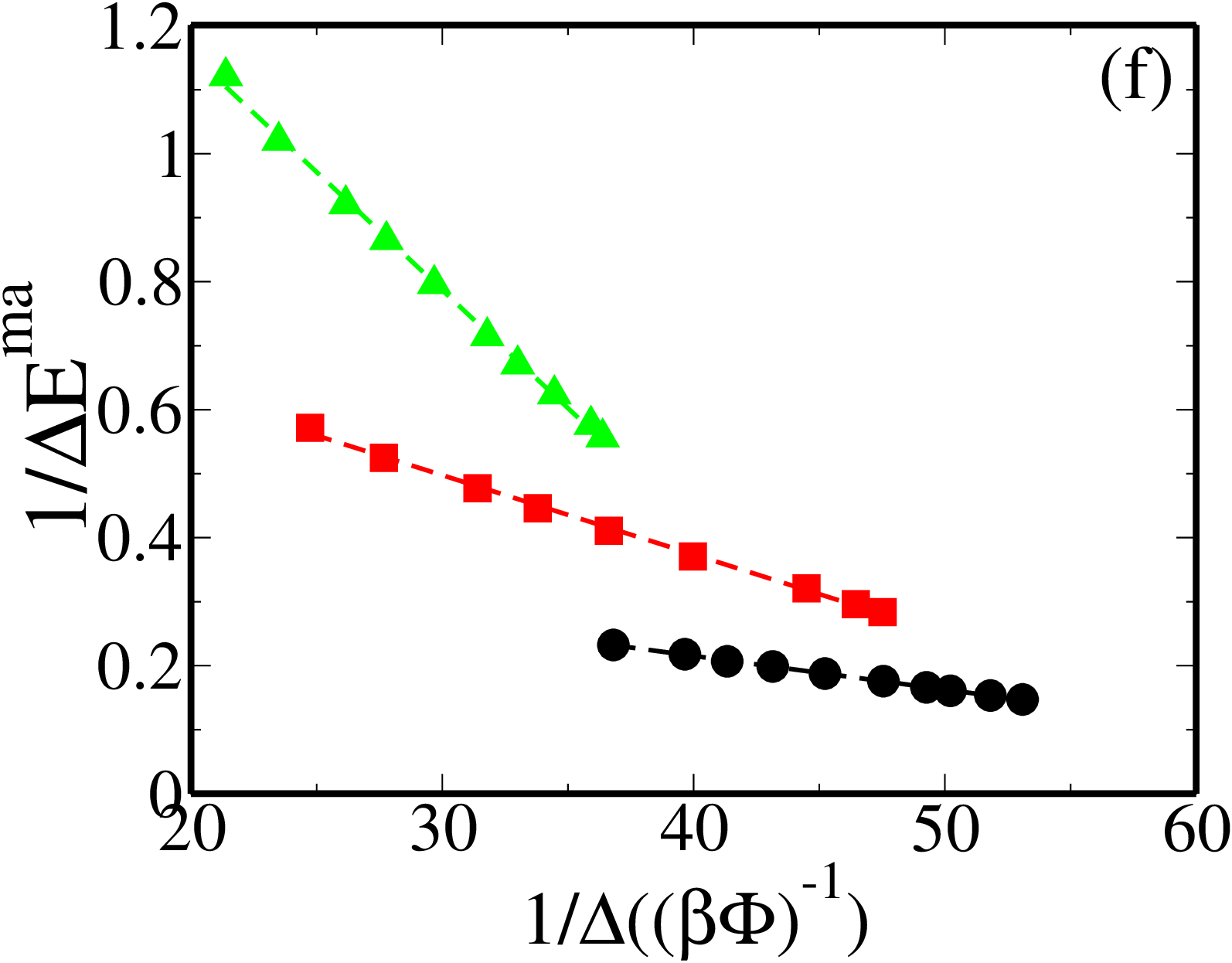}
\end{subfigure}
\vspace{0.2cm}
\caption{\textit{\textbf{Left Panels:} Macroscopic free energy barrier, $\Delta E^{\text{ma}}$, plotted against the inverse macroscopic structural order parameter, $\beta\Phi$, for three sets of systems: (a)LJ and (c)WCA, each examined at four different densities, and (e) modified LJ $(q,p)$ system.
\textbf{Right Panels:} The inverse activation energy $1/\Delta E^{ma}$ plotted against $1/\Delta((\beta\phi)^{-1})$ for the same systems. The slope of the linear fits yields the $B'/A'$ values used in the fragility analysis(see Fig.\ref{Fig_fragility}). The colour coding in the right panels matches that of the left panels. 
}}
\label{Fig_macro}
\end{figure}

We can describe the macroscopic free energy barrier, $\Delta E^{ma}=T*ln(\tau_{\alpha}/\tau_{0})$, where $\tau_{\alpha}$ is the $\alpha$-relaxation time of the systems as obtained from the overlap function (Section \ref{section_relaxation}). $\tau_{0}$ is a prefactor. In Fig.\ref{Fig_macro}(left panels), we plot the macroscopic free energy barrier as a function of inverse macroscopic SOP, $\beta\Phi$, for the three different sets of systems. We can also obtain the free energy barrier from the microscopic motion of particles. This is a standard procedure \cite {Schoenholz2015_nature,Sharma2022_pre}, and the details are given in Appendix II. We can write the probability of a particle with a certain value of the inverse SOP to be a fast moving particle as, $P_{R}(\beta\phi)=P_{0}(\beta\phi)exp[-\Delta E(\beta\phi)/T]$. Here $\Delta E(\beta\phi)$ is the free energy barrier for particles having the value of the inverse microscopic order parameter $\beta\phi$ (given in Eq.\ref{eq:macro_sop}). The microscopic free energy barriers thus obtained are plotted against the inverse microscopic per particle level SOP, $\beta\phi$(left panels of Fig.\ref{Fig_micro}). We observe that the macroscopic free energy barrier grows more strongly than linear, approximately quadratic, while the microscopic free energy barrier increases linearly. This weaker growth of the microscopic barrier arises from its inability to capture cooperative molecular motion, consistent with earlier findings \cite{Tah2022_jcp}. As we will show later, the stronger than linear growth of the macroscopic free energy barrier as a function SOP can be derived from the VFT expression, which clearly shows that the prediction of the linear growth is an approximation. Nevertheless, both macroscopic and microscopic free energy barriers exhibit similar qualitative trends. In both LJ and WCA systems, we find that the free energy barrier grows more steeply with increasing fragility, in agreement with previous studies \cite{Nandi2021_prl,Tah2022_jcp}. However, the behaviour appears to be completely reversed when we consider the set of modified LJ, $(q,p)$ systems where the fragility is varied by changing the steepness of the potential. In this case, the slope of both the macroscopic and microscopic free energy barriers is lowest for the $(8,5)$ system, the most fragile, and highest for the $(12,11)$ system, the least fragile. This finding challenges the commonly held assumption \cite{Nandi2021_prl,Tah2022_jcp} that the most fragile system should exhibit the steepest slope in the free energy barrier versus the SOP, raising the question of whether this relationship is truly generic.

 Note that this analysis depends on the order parameter we choose, and the fragility in terms of the order parameter can be different from that in terms of temperature. First, we present the temperature dependence of the SOP. Similar to earlier studies \cite{Nandi2021_prl,Sharma2022_pre}, we find that for all the systems the SOP, at macroscopic level, $(\beta\Phi)^{-1}$ and microscopic level, $<(\beta\phi)^{-1}>$ are linearly proportional to the temperature (right panels of Fig.\ref{Fig_micro}). Although the value of the order parameter at the macroscopic and microscopic levels is different, which is expected (as elaborated in Appendix I), their temperature dependences are similar. Both at the macroscopic and the microscopic levels, the SOP is linear with $T$ and can be fitted to the equations $(\beta\Phi)^{-1} = (\beta\Phi_{0})^{-1} + \gamma T$, and $<(\beta\phi)^{-1}> = <(\beta\phi_{0})^{-1}> + \gamma^{'}T$ where $\gamma$ and $\gamma^{'}$ are the corresponding slopes (given in Table \ref{tablegamma}). $(\beta\Phi_{0})^{-1}$ and $<(\beta\phi_{0})^{-1}>$ are the values of the macroscopic and average microscopic SOP, at T=0, respectively. We find that the values of the slopes $\gamma$ and $\gamma^{'}$ are not identical but similar, which can be expected as both come from the temperature evolution of the radial distribution function.
 
We have redefined our order parameter at the macroscopic and microscopic level as,
 
\begin{eqnarray}
\Delta((\beta\Phi)^{-1}) &=&(\beta\Phi)^{-1} - (\beta\Phi_{0})^{-1}=\gamma T,\nonumber\\
\Delta(<(\beta\phi)^{-1}>) &=&<(\beta\phi)^{-1}> - <(\beta\phi_{0})^{-1}>=\gamma^{'} T,\nonumber\\
\label{gamma}
\end{eqnarray}

We next investigate the fragility of the systems both as a function of temperature, given by Eq.\ref{vft_T}, and also as a function of $\Delta((\beta\Phi)^{-1})$, which can be represented by the following VFT expression \cite{GarciaColin1989_vft}, 
{\small
\begin{equation}
 \tau_{\alpha}(\Delta((\beta\Phi)^{-1})) = \tau_{0}\exp\Big(\frac{\Delta((\beta\Phi)^{-1}_{VFT})}{K_{VFT}(\Delta((\beta\Phi)^{-1})-\Delta((\beta\Phi)^{-1}_{VFT}))}\Big) 
 \label{vft_sop}
\end{equation}
\noindent
Here $\Delta((\beta\Phi)^{-1}_{VFT})$ is the value of the order parameter at the VFT temperature. Note that Eq.\ref{vft_sop} can be written when we replace the temperature in Eq.\ref{vft_T} by the SOP as given in Eq.\ref{gamma}. In Fig.\ref{Fig_masterplot}, we present a master plot of the VFT fit for all the systems as a function of the SOP. The inset shows the corresponding plot as a function of temperature. The good data collapse in both representations confirms the validity of the equations (Eq.\ref{vft_T} and Eq.\ref{vft_sop}) and the consistency of the extracted VFT parameters, as listed in Table I. We find that the fragility in terms of the temperature and the SOP is the same. 
\begin{figure}
\centering
\vspace{0.2cm}
\begin{subfigure}{0.22\textwidth}
\includegraphics[width=0.98\linewidth]{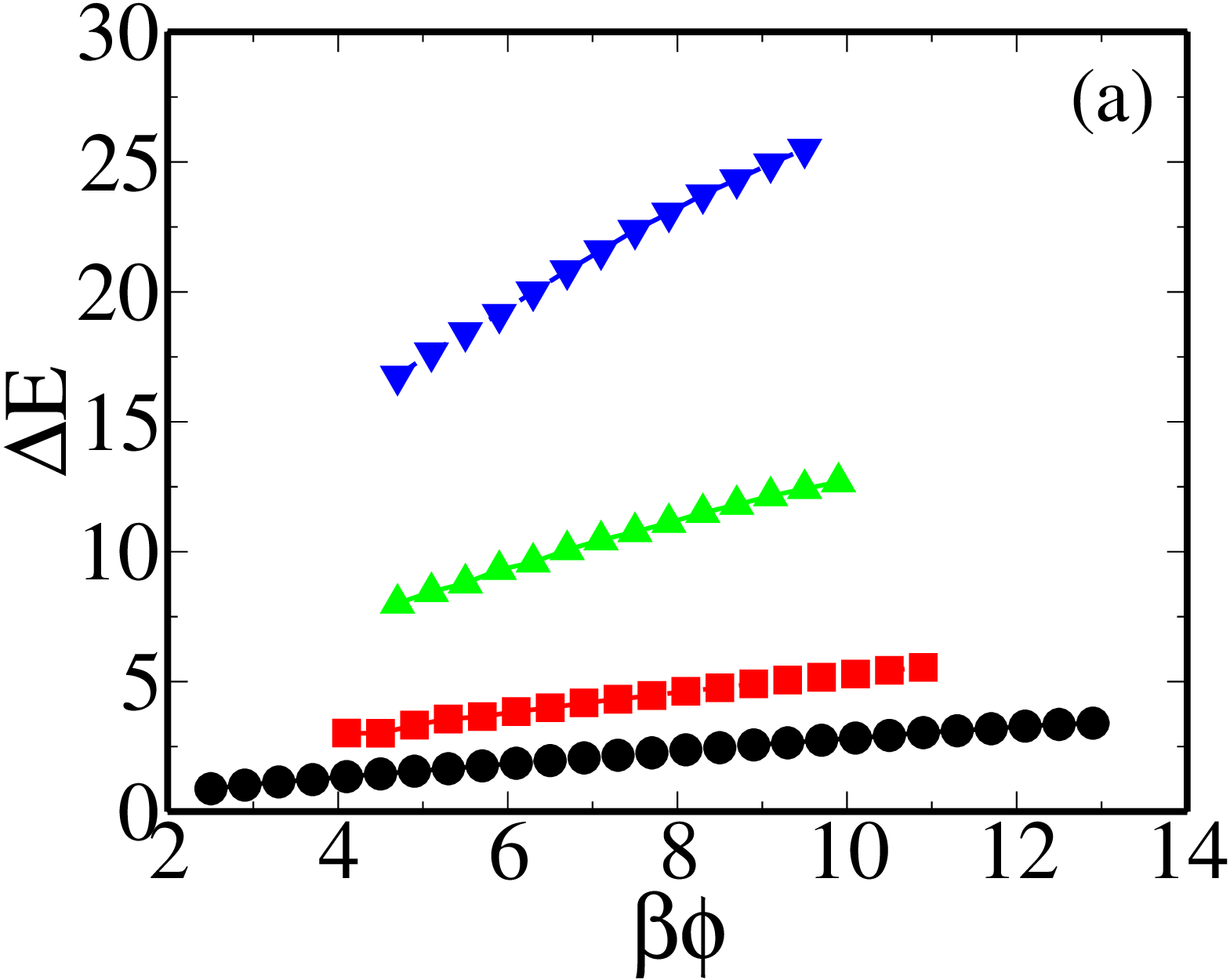}
\end{subfigure}
\hspace{0.2cm}
\begin{subfigure}{0.23\textwidth}
\includegraphics[width=0.98\linewidth]{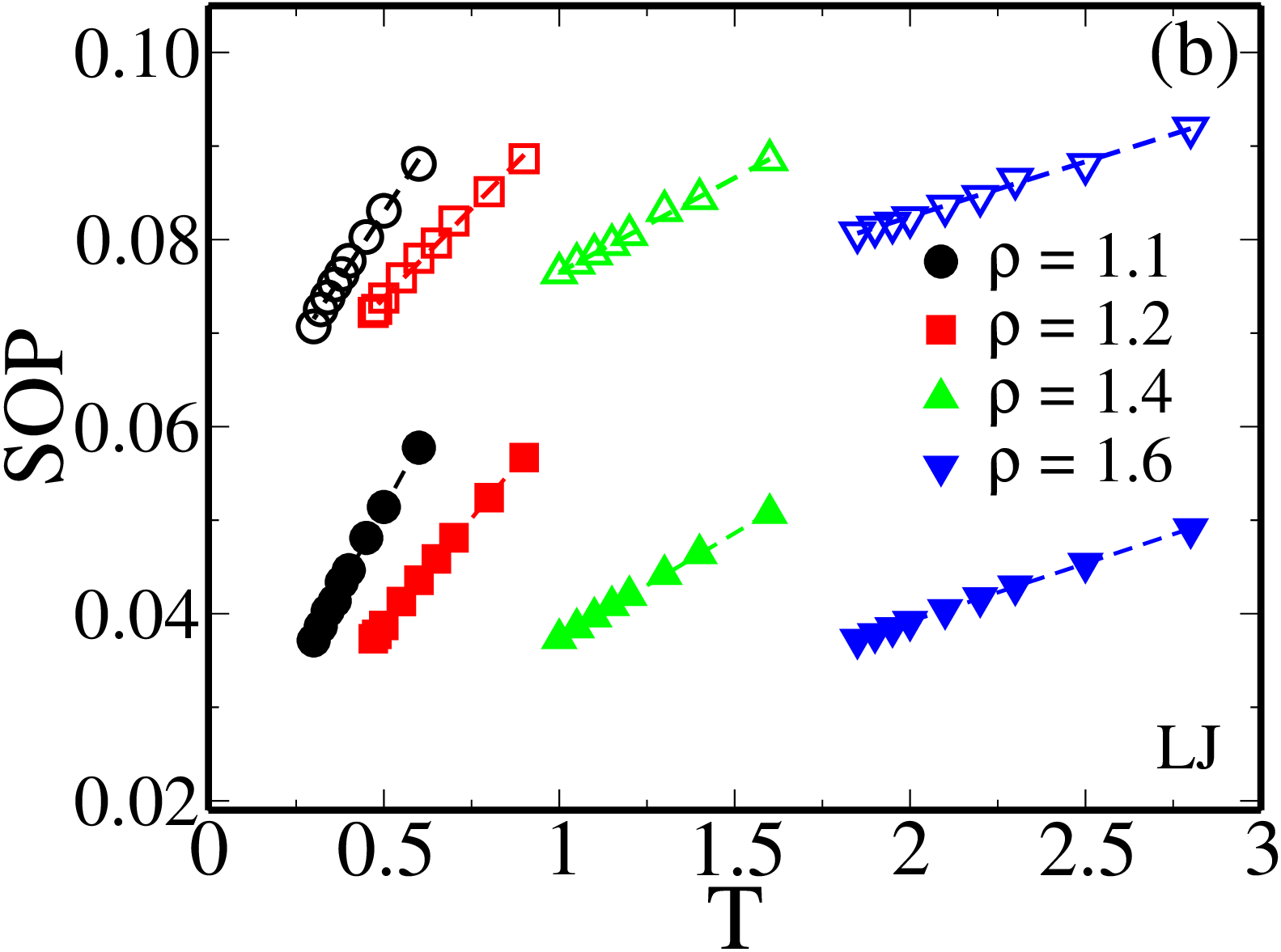}
\end{subfigure}
\vspace{0.2cm}
\begin{subfigure}{0.22\textwidth}
\includegraphics[width=0.98\linewidth]{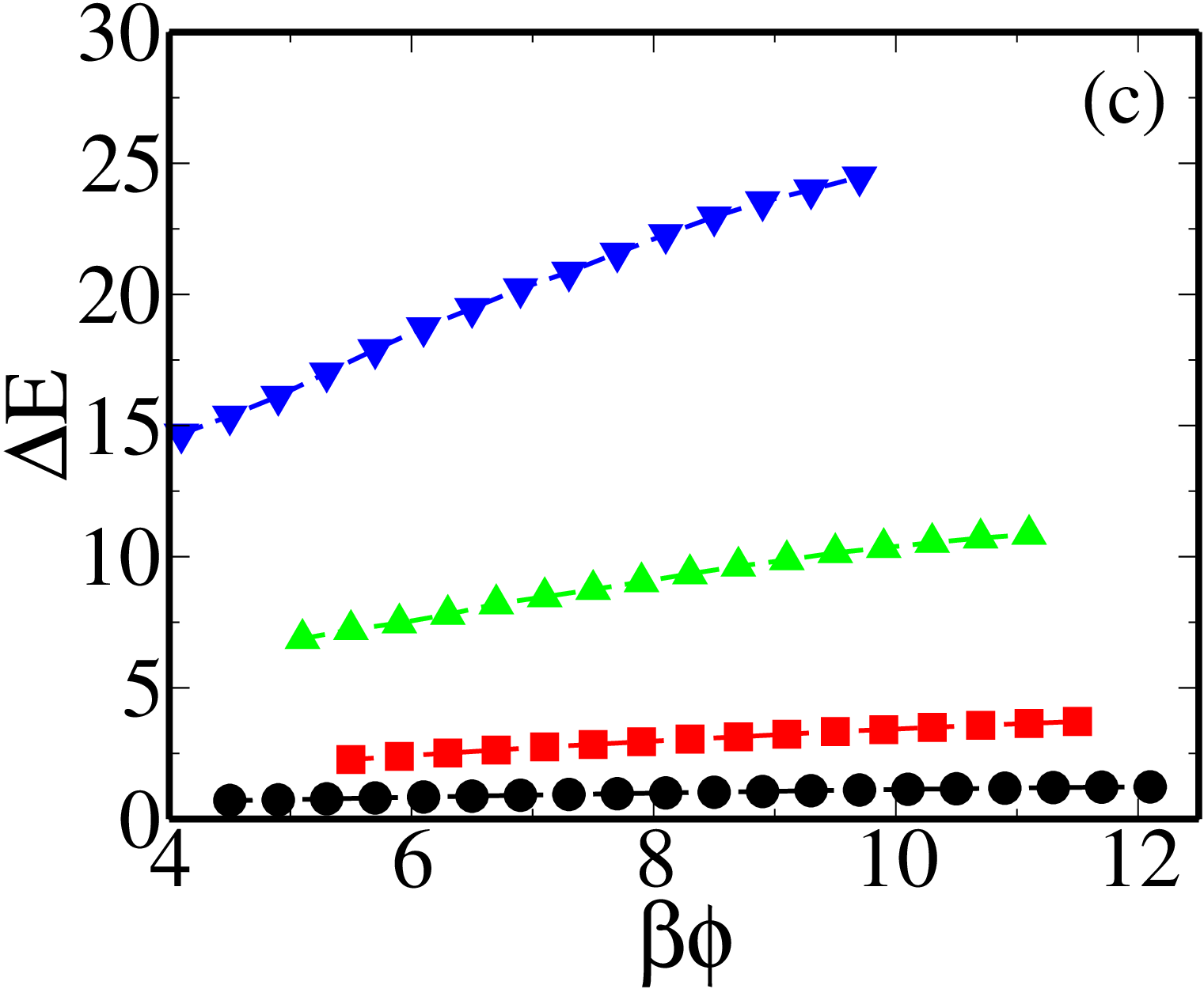}
\end{subfigure}
\hspace{0.2cm}
\begin{subfigure}{0.23\textwidth}
\includegraphics[width=0.98\linewidth]{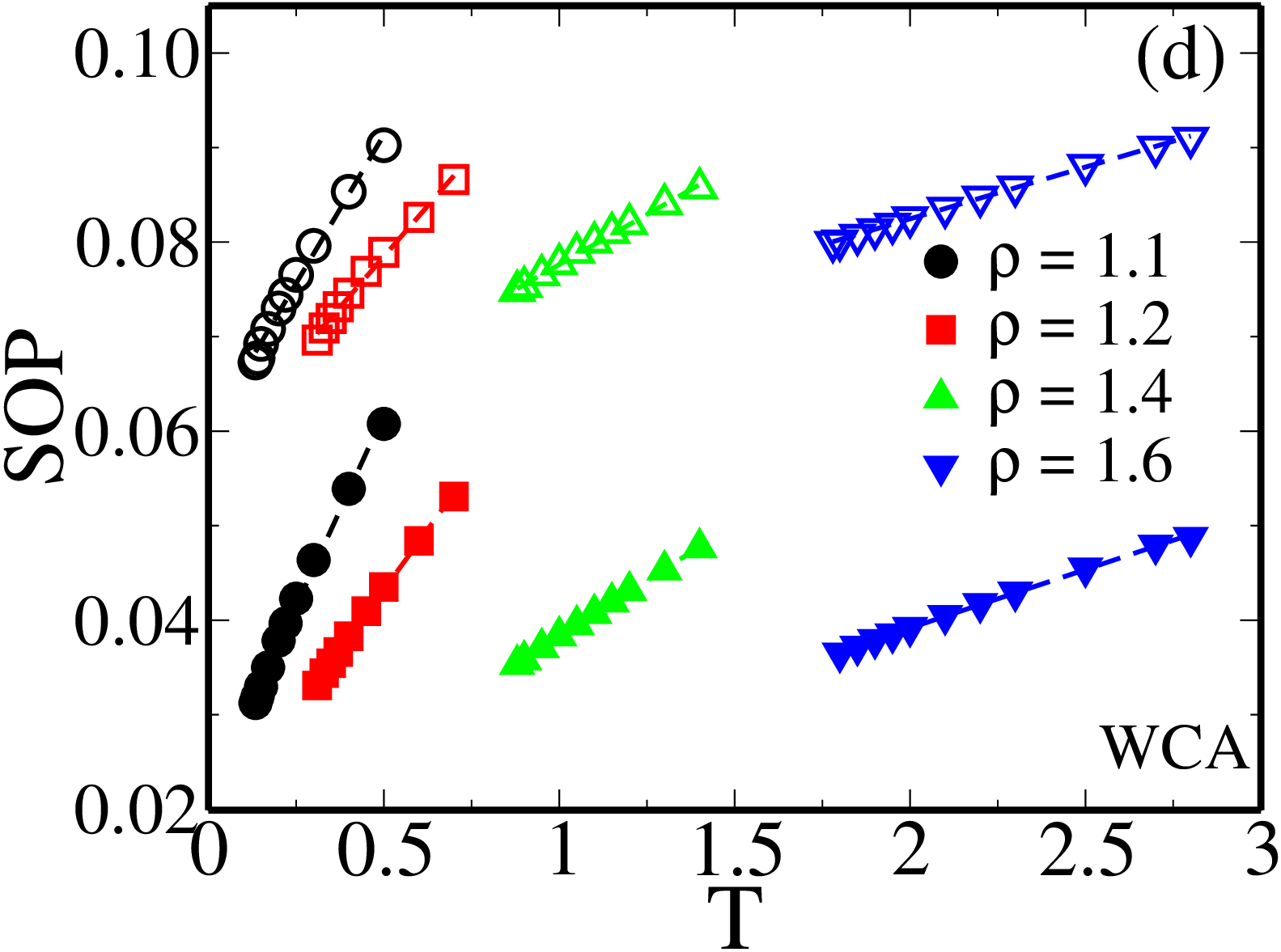}
\end{subfigure}
\vspace{0.2cm}
\begin{subfigure}{0.22\textwidth}
\includegraphics[width=0.98\linewidth]{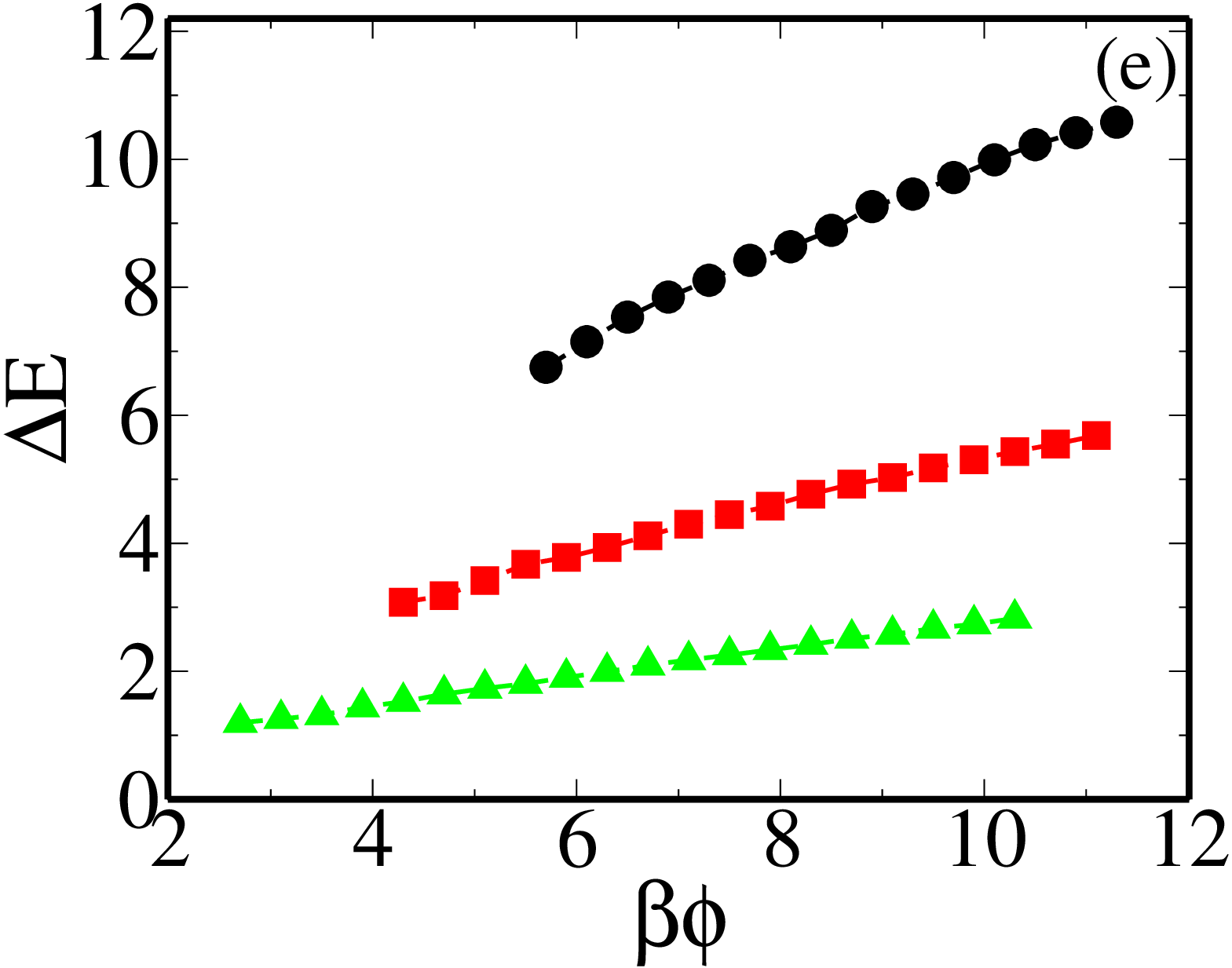}
\end{subfigure}
\hspace{0.2cm}
\begin{subfigure}{0.23\textwidth}
\includegraphics[width=0.98\linewidth]{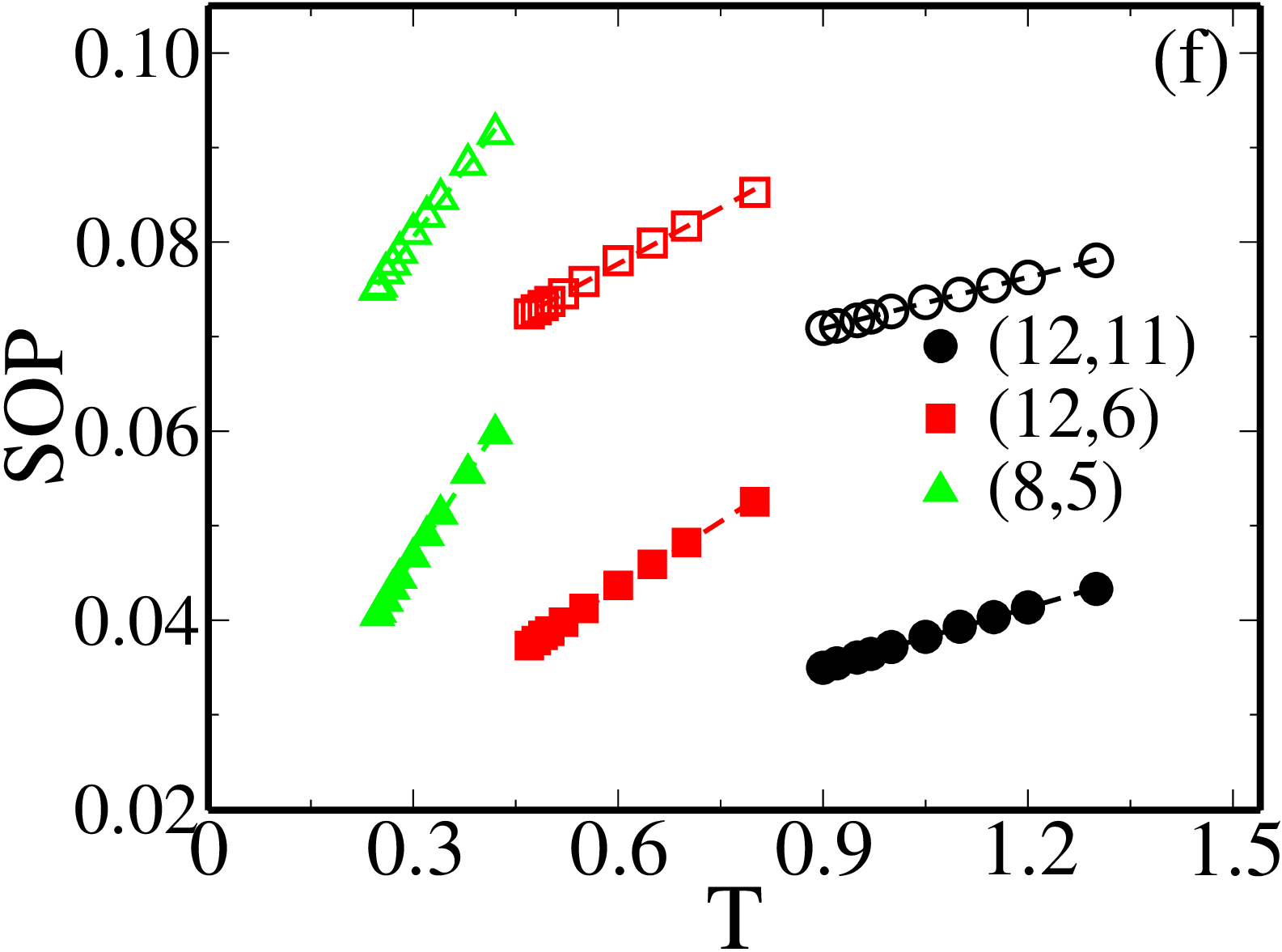}
\end{subfigure}
\vspace{0.2cm}
\caption{\textit{\textbf{Left Panels:}Microscopic free energy barrier, $\Delta E$, versus the inverse microscopic structural order parameter, $\beta\phi$, for three sets of systems: (b)LJ and (d)WCA, each examined at four different densities, and (f) modified LJ $(q,p)$ system. \textbf{Right Panels:}Temperature dependence of both the macroscopic structural order parameter, $(\beta\Phi)^{-1}$, (closed symbols) and the average microscopic structural order parameter $\langle(\beta\phi)^{-1}\rangle$,(open symbols) for the same systems. Dotted lines represent linear fits used to extract slope parameters for further analysis. The colour coding in the left panels matches that of the right panels. 
}}
\label{Fig_micro}
\end{figure}

\begin{figure}
\centering
\vspace{0.2cm}
\begin{subfigure}{0.48\textwidth}
\includegraphics[width=1.0\linewidth]{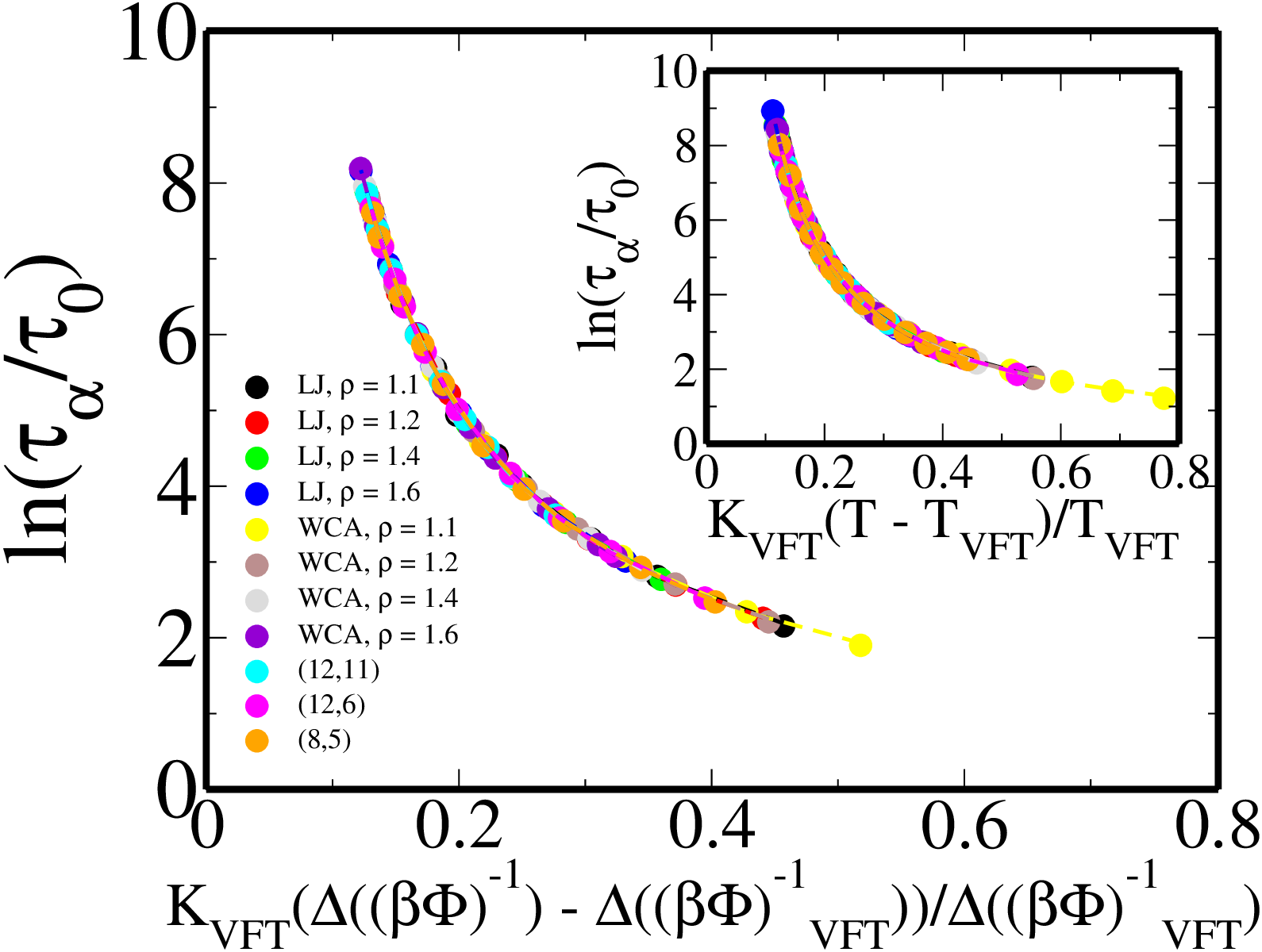}
\end{subfigure}
\vspace{0.2cm}
\caption{\textit{Master plot of the VFT equation w.r.t SOP, (Eq.\ref{vft_sop}) for all the systems—LJ and WCA at four different densities, and modified LJ $(q,p)$ system. The inset displays the corresponding master of the VFT equation with respect to temperature T (Eq.\ref{vft_T}). The collapse of the data confirms the validity of the equations and the consistency of the extracted VFT parameters, as listed in Table I.
}}
\label{Fig_masterplot}
\end{figure}

If we now assume that $\Delta E^{ma}=T*ln(\tau_{\alpha}/\tau_{0})$ then using Eq.\ref{vft_sop} we can write an expression for $\Delta E^{ma}$ in terms of the SOP where we replace the temperature from Eq.\ref{gamma}. We can then write,

\begin{equation}
1/\Delta E^{ma} = \frac{K_{VFT}}{T_{VFT}}-\gamma.K_{VFT}.\frac{1}{\Delta((\beta\Phi)^{-1})} =A'-B'.\frac{1}{\Delta((\beta\Phi)^{-1})}
\label{ideltaE_phi}
\end{equation}
\noindent
and
\begin{equation}
 \Delta E^{ma} \approx \frac{T_{VFT}}{K_{VFT}} + 
 \gamma.\frac{T_{VFT}^{2}}{K_{VFT}}.\frac{1}{\Delta((\beta\Phi)^{-1})}=A+B.\frac{1}{\Delta((\beta\Phi)^{-1})}
 \label{deltaE_phi}
\end{equation}
\noindent

From Eq.\ref{ideltaE_phi} we get that $A' = \frac{K_{VFT}}{T_{VFT}}$ and $B'= \gamma.K_{VFT}$, which shows that the absolute slope $B'$ of the inverse of the free energy barrier is proportional to fragility.
From Eq.\ref{deltaE_phi} we get A = $\frac{T_{VFT}}{K_{VFT}}$ and B = $\gamma.\frac{T_{VFT}^2}{K_{VFT}}$ and the slope just like that for temperature (see Appendix III) is again not proportional to but inversely proportional to the fragility. For both the LJ and WCA systems, fragility increases with density, accompanied by a corresponding rise in $T_{VFT}$ as the systems at higher densities operate in higher temperature regimes. This trend explains why with increasing fragility, we observe in Fig.\ref{Fig_macro}(left panels) and Fig.\ref{Fig_micro} (left panels) a sharper increase in the free energy barrier, which is approximated by a linear expression given by Eq.\ref{deltaE_phi} with slope $\gamma.\frac{T_{VFT}^2}{K_{VFT}}$. For the modified LJ $(q,p)$ system, the fragility is higher for the $(8,5)$ system, which has the lowest value of $T_{VFT}$, and this is the reason the growth of the free energy barrier for this system is low. Thus, in cases where the growth of the free energy barrier follows fragility, it is because for more fragile systems the $T_{VFT}$ is higher. 

Note that $A$($A'$) describes the part of the activation free energy(inverse energy) which is independent of the temperature and also the structural order and $B$($B'$) describes the sensitivity of the dynamics to change in the structural order. To do a meaningful comparison of the free energy barrier and also the dynamics of different systems, we need to scale the free energy barrier by the temperature independent value. Thus we find that instead of $B$($B'$) it is the scaled quantity $B/A$($B'/A'$) that describes the sensitivity of the dynamics to change in structural order. Interestingly, we find that $B/A=B'/A'=\gamma T_{VFT}$. The values of $\gamma$ obtained from the temperature dependence of the macroscopic SOP (right panels in Fig.\ref{Fig_micro}) are given in Table \ref{tablegamma}. We observe that $\gamma T_{VFT}$ varies linearly with the fragility parameter $K_{VFT}$ across all three sets of systems, as demonstrated in Fig.\ref{Fig_fragility}. Note that at the macroscopic level $\Delta E^{ma}$ as a function of inverse SOP does not actually show a linear dependence (left panels of Fig.\ref{Fig_macro}) and Eq.\ref{deltaE_phi} is an approximate expression. However, as shown in the right panels of Fig.\ref{Fig_macro}, $1/\Delta E^{ma}$ as a function of SOP does show a linear behaviour, validating Eq.\ref{ideltaE_phi}. Thus, we obtain the values of $A'$ and $B'$ from the right panels in Fig.\ref{Fig_macro} (given in Table.\ref{AB} in Appendix III). In the inset of Fig.\ref{Fig_fragility} we also plot $B'/A'$ against $\gamma T_{VFT}$.
We find that all the points fall on the y=x line, thus confirming our analysis. The physical reason for $\gamma T_{VFT}$ being a good fragility metric needs further investigation. Earlier studies \cite{Zaccone_PNAS,Zaccone_PRB} have shown that the fragility of the modified LJ systems is linearly proportional to the product of the thermal expansion coefficient and the glass transition temperature. It will be interesting to understand the correlation between the thermal expansion coefficient and the $\gamma$ parameter. 

\begin{figure}
\centering
\vspace{0.2cm}
\begin{subfigure}{0.48\textwidth}
\includegraphics[width=1.0\linewidth]{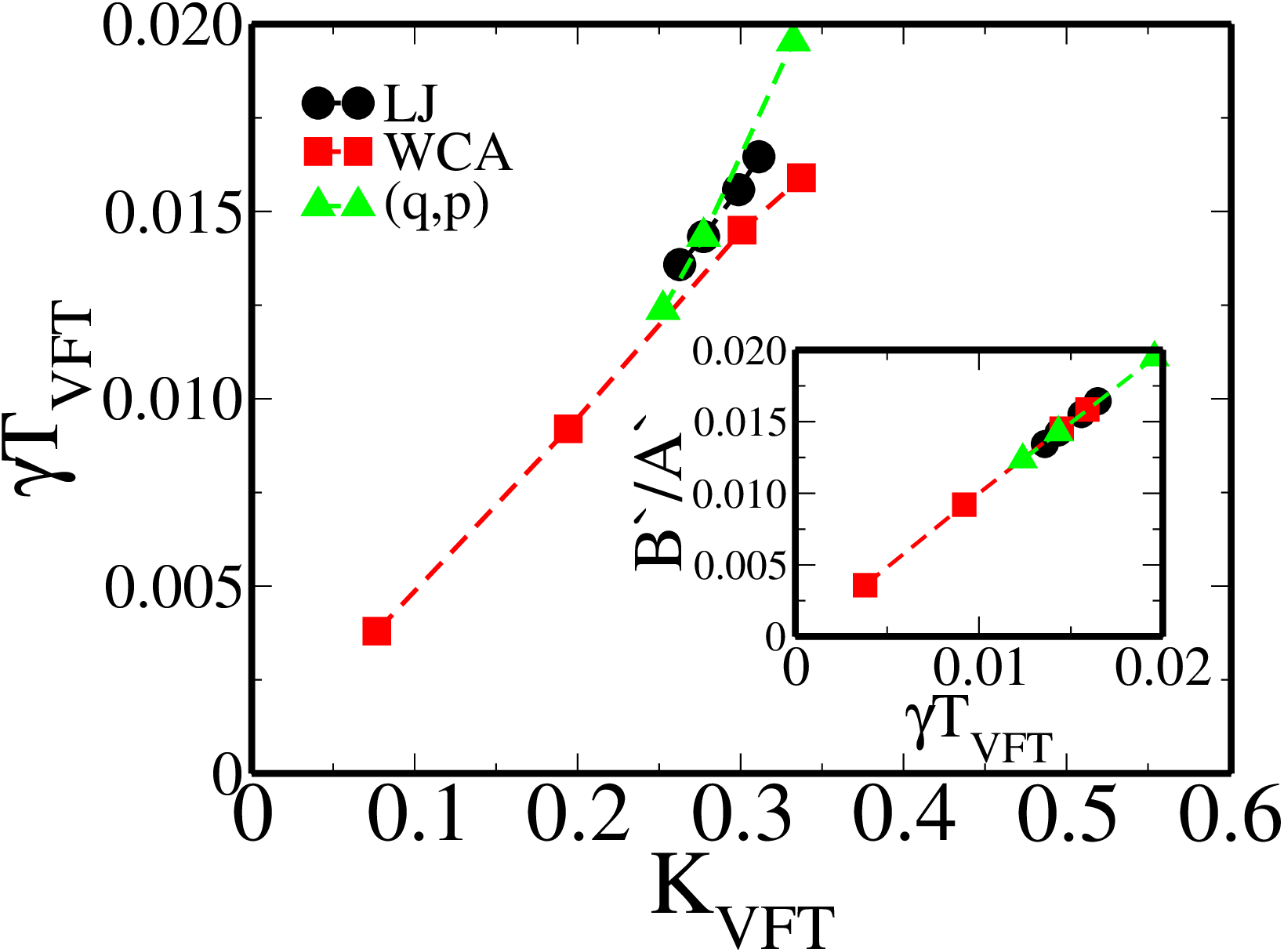}
\end{subfigure}
\vspace{0.2cm}
\caption{\textit{The main panel displays $\gamma T_{\mathrm{VFT}}$ plotted against $K_{\mathrm{VFT}}$ for three sets of systems: LJ, WCA, and modified LJ $(q,p)$ system. Each set exhibits a clear linear relationship, indicating a consistent correlation between the parameters.
\textbf{Inset:} The scaled slope of the macroscopic free energy barrier plot, $B'/A'$, derived from the $1/\Delta E^{\mathrm{ma}}$ versus $1/\Delta((\beta\Phi)^{-1})$ plot in the right panels of Fig.\ref{Fig_macro}, is plotted against $\gamma T_{\mathrm{VFT}}$. All data points lie on the $y = x$ line, demonstrating that the two quantities are numerically identical, validating our analysis. 
}
}
\label{Fig_fragility}
\end{figure}

\begin{table}[h]
\centering
\caption{\textit{Fitted values of the VFT temperature, $T_{\mathrm{VFT}}$, and fragility, $K_{\mathrm{VFT}}$, for various systems, obtained using Eq.\ref{vft_T}. The parameters $\gamma$ and $\gamma^{'}$ represent the temperature dependence of our SOP (Eq.\ref{gamma}) at macroscopic and microscopic levels, respectively (right panels of Fig.\ref{Fig_micro}). Note that $(12,6)$ system is same as LJ at $\rho=1.2$}}
\vspace{0.2cm}
\label{tab:vftfragility}
\begin{tabular}{|c|c|c|c|c|}
\hline
\textbf{System} & $\boldsymbol{T_{\mathrm{VFT}}}$ & $\boldsymbol{K_{\mathrm{VFT}}}$ & $\boldsymbol{\gamma}$ & $\boldsymbol{\gamma'}$\\
\hline
LJ, $\rho = 1.1$  & 0.197 & 0.262 & 0.068 & 0.056\\
\hline
LJ, $\rho = 1.2$  & 0.317 & 0.277 & 0.045 & 0.038\\
\hline
LJ, $\rho = 1.4$  & 0.697 & 0.298 & 0.022 & 0.020\\
\hline
LJ, $\rho = 1.6$  & 1.322 & 0.311 & 0.012 & 0.011\\
\hline
WCA, $\rho = 1.1$ & 0.047 & 0.076 & 0.081 & 0.062\\
\hline
WCA, $\rho = 1.2$ & 0.181 & 0.194 & 0.051 & 0.042\\
\hline
WCA, $\rho = 1.4$ & 0.613 & 0.300 & 0.023 & 0.021\\
\hline
WCA, $\rho = 1.6$ & 1.289 & 0.337 & 0.012 & 0.010\\
\hline
$(12,11)$         & 0.590 & 0.252 & 0.020 & 0.018\\
\hline
$(12,6)$          & 0.317 & 0.277 & 0.045 & 0.038\\
\hline
$(8,5)$           & 0.174 & 0.332 & 0.111 & 0.095\\
\hline
\end{tabular}
\label{tablegamma}
\end{table}

\begin{figure}
\centering
\vspace{0.2cm}
\begin{subfigure}{0.225\textwidth}
\includegraphics[width=0.98\linewidth]{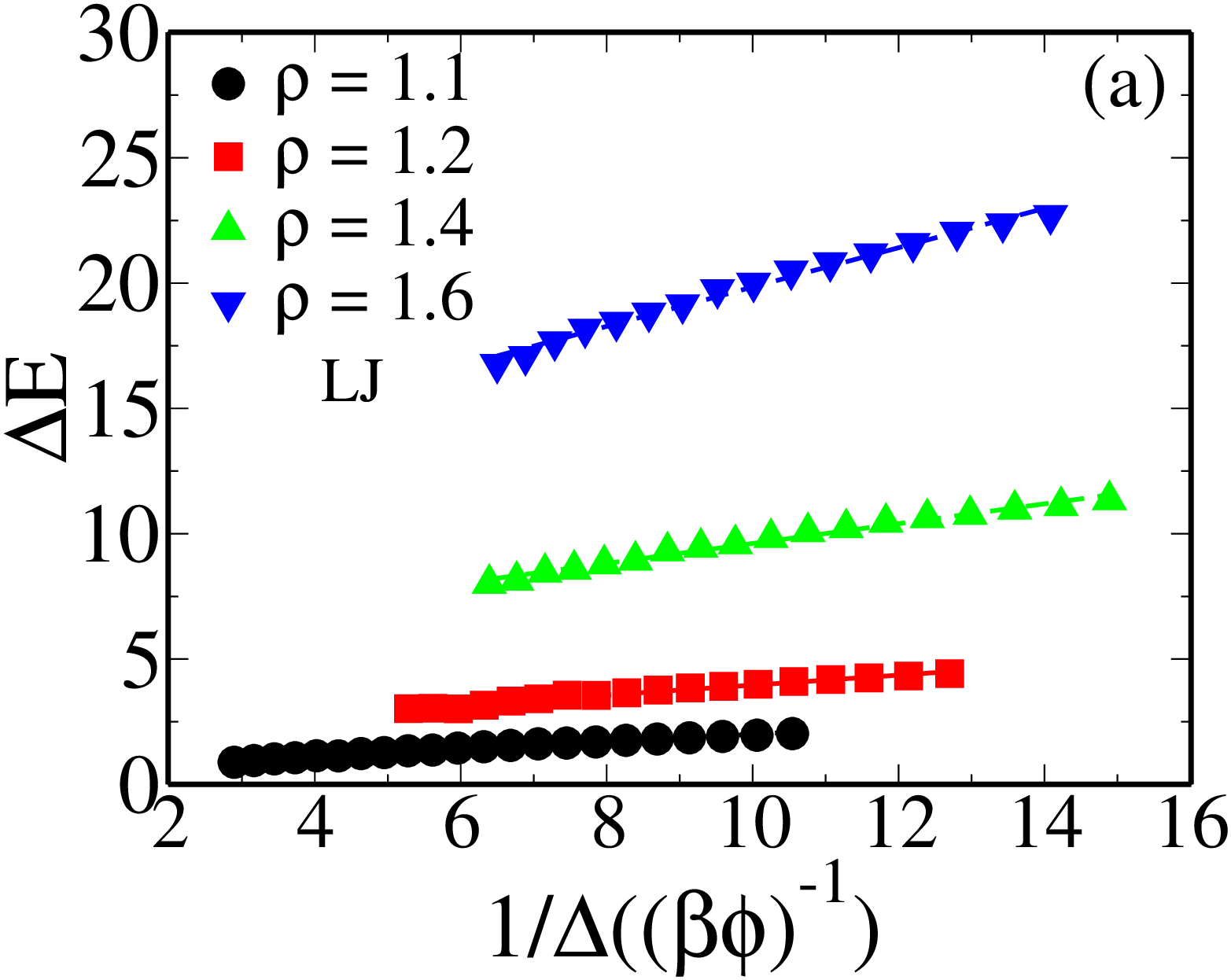}
\end{subfigure}
\hspace{0.2cm}
\begin{subfigure}{0.225\textwidth}
\includegraphics[width=0.98\linewidth]{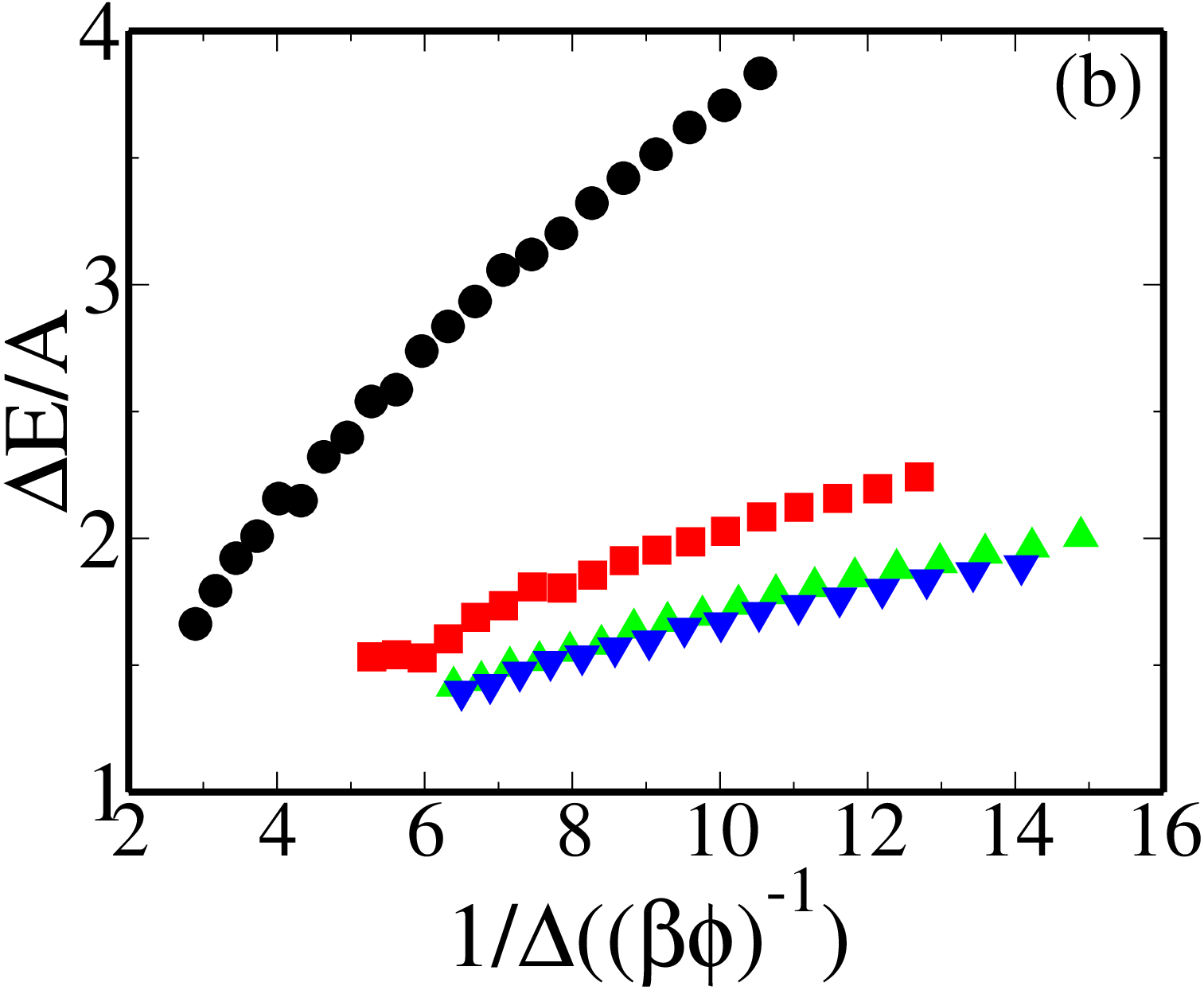}
\end{subfigure}
\vspace{0.2cm}
\begin{subfigure}{0.225\textwidth}
\includegraphics[width=0.98\linewidth]{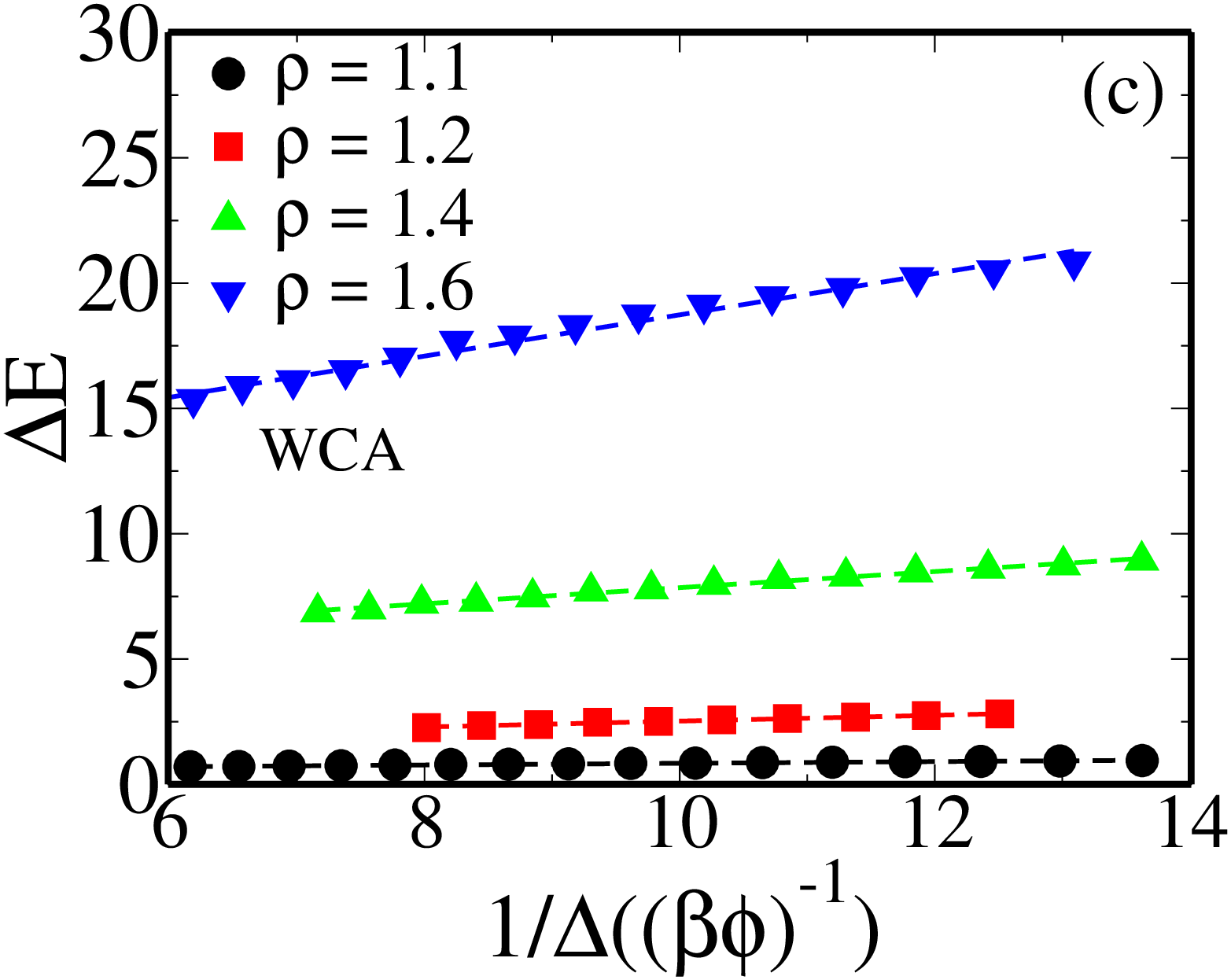}
\end{subfigure}
\hspace{0.2cm}
\begin{subfigure}{0.225\textwidth}
\includegraphics[width=0.98\linewidth]{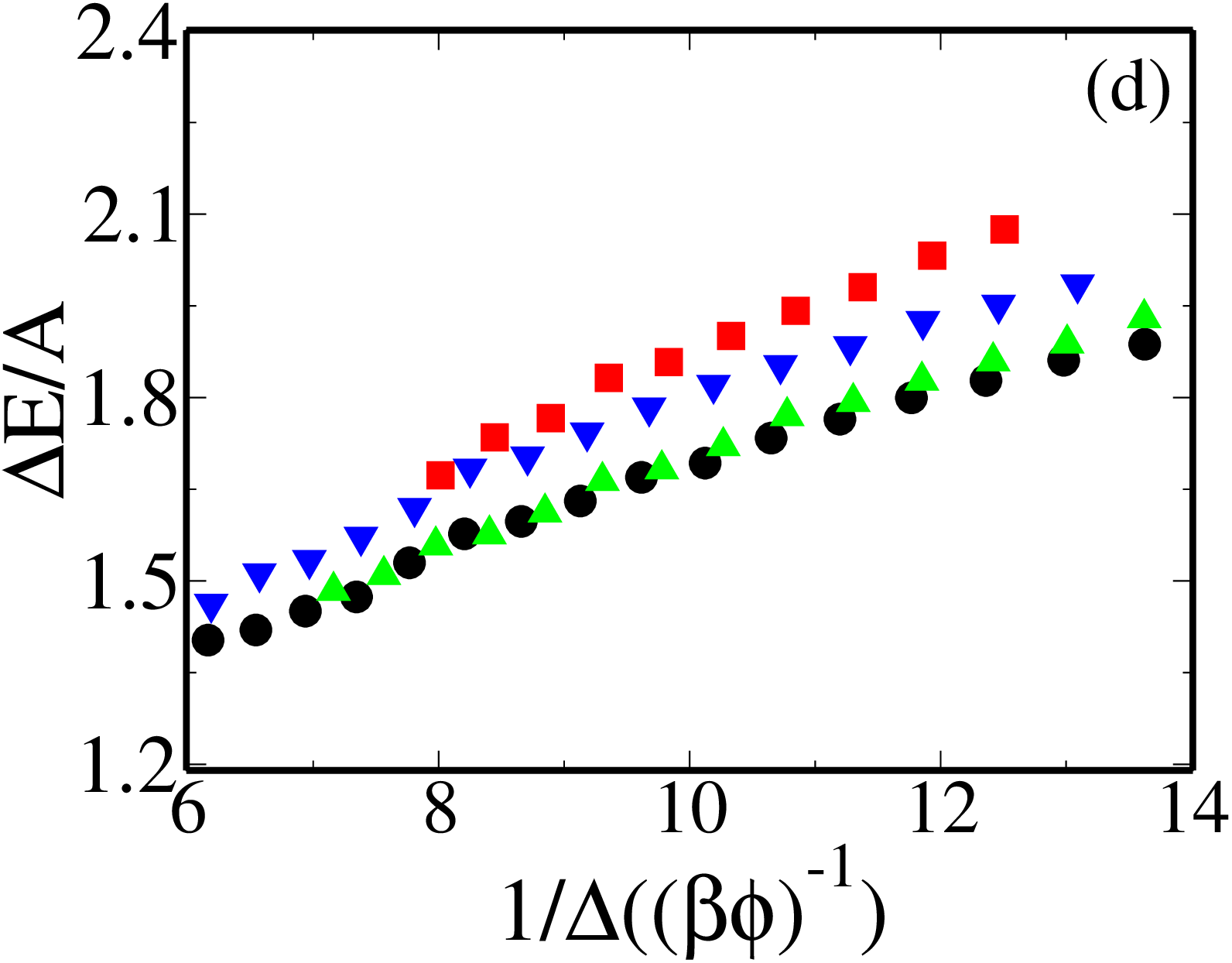}
\end{subfigure}
\vspace{0.2cm}
\begin{subfigure}{0.225\textwidth}
\includegraphics[width=0.98\linewidth]{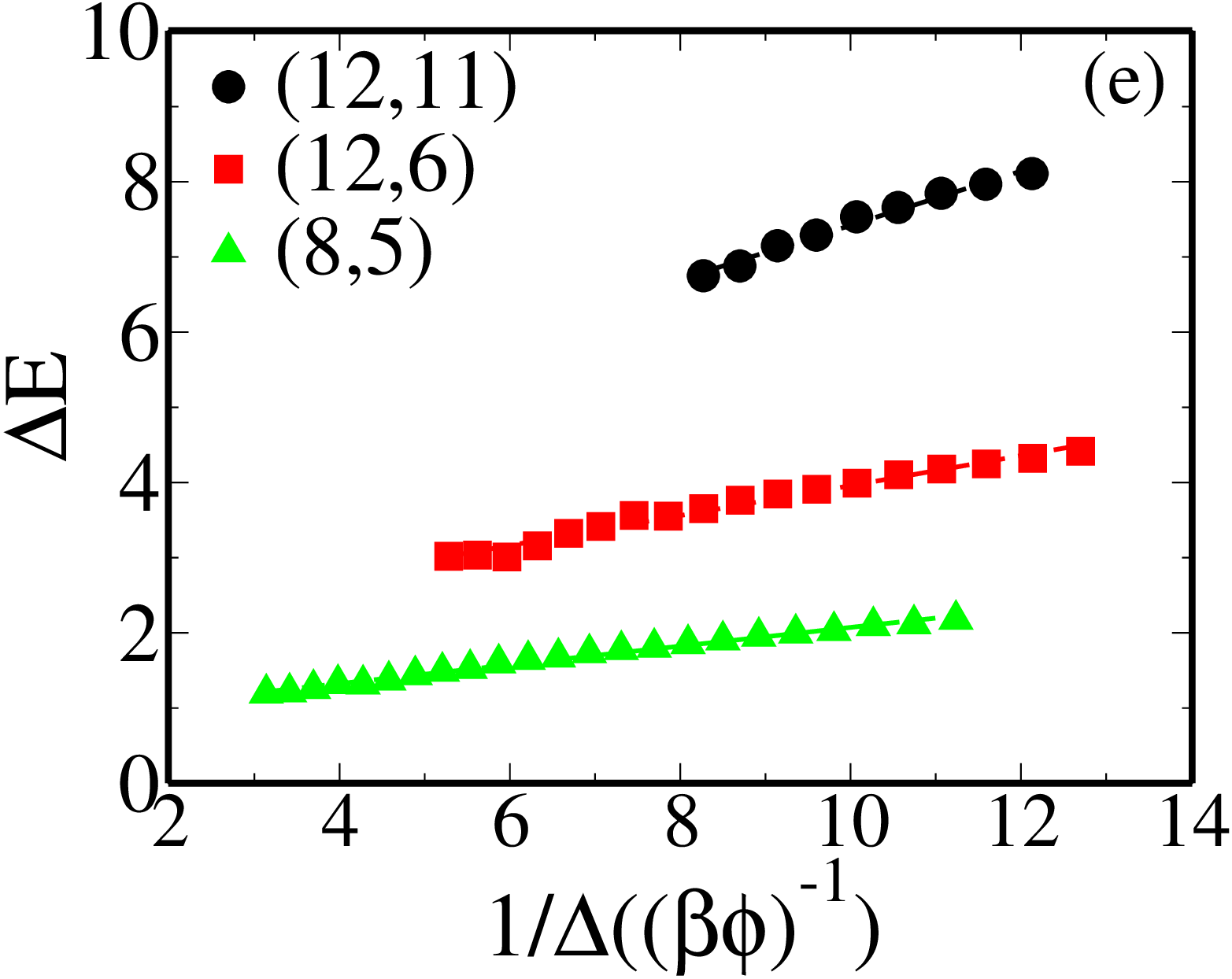}
\end{subfigure}
\hspace{0.2cm}
\begin{subfigure}{0.225\textwidth}
\includegraphics[width=0.98\linewidth]{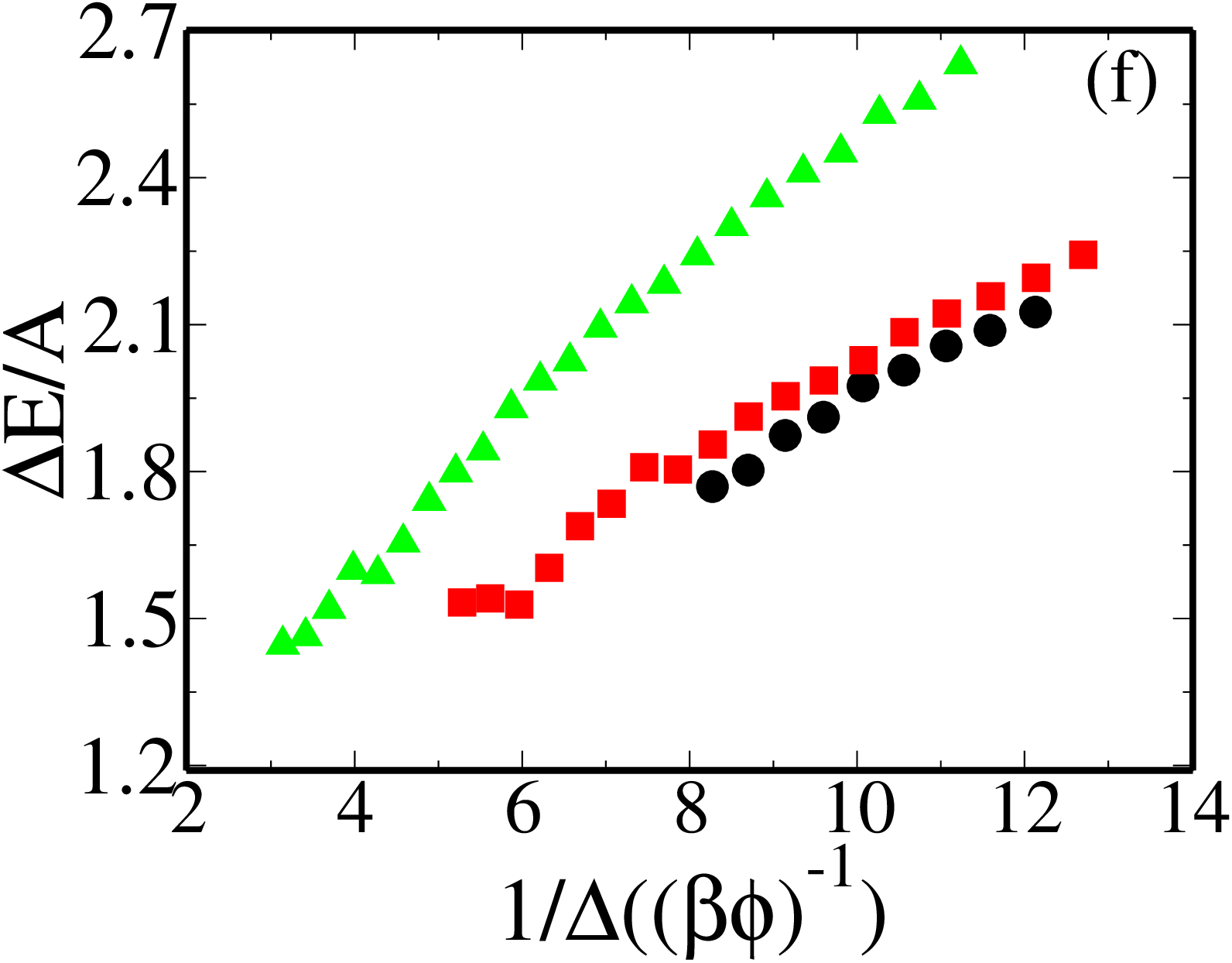}
\end{subfigure}
\hspace{0.2cm}
\caption{\textit{\textbf{Left panels:} Microscopic free energy barrier, $\Delta E$, plotted as a function of inverse microscopic structural order parameter, $1/\Delta((\beta\phi)^{-1})$, for three sets of systems: (a)LJ and (c)WCA, each examined at four different densities, and (e) modified LJ $(q,p)$ system. 
\textbf{Right panels:} Scaled microscopic free energy barrier, $\Delta E/A$, plotted against $1/\Delta((\beta\phi)^{-1})$ for the same systems shown in the left panels. The colour coding in the right panels matches that of the left panels.
}}
\label{Fig_deltaE_micro}
\end{figure}
Next we study the validity of the same analysis for the microscopic $\Delta E$. In Fig.\ref{Fig_micro}(left panels) we plot the $\Delta E$ vs $1/\Delta((\beta\phi)^{-1})$. Notably, unlike at the macroscopic level where the free energy barrier increases more than linearly with the structural order parameter(SOP) (left panels of Fig.\ref{Fig_macro}), at the microscopic level, $\Delta E$ as a function of SOP (left panels of Fig.\ref{Fig_micro}) exhibits a linear behaviour. A similar linear trend is also seen in the $\Delta E$ vs. modified SOP plot (left panels of Fig.\ref{Fig_deltaE_micro}). In the calculation of the microscopic free energy barrier, it is assumed that the dynamics is given by $P_{R}^{-1}$ (see Appendix II). The fact that $\Delta E^{ma}$ grows more strongly compared to $\Delta E$ can be attributed to the fact that $P_{R}^{-1}$ is related to a shorter time dynamics which does not have the information of the cooperativity in the dynamics thus grows slower than $\tau_{\alpha}$, an observation previously reported by Tah et al\cite{Tah2022_jcp}. However, as discussed before, the approximate behaviour at the macroscopic and microscopic levels remains similar. For the LJ and the WCA system, we find that the rate of growth of the free energy barriers follows the fragility, whereas for the modified LJ $(q,p)$ system, the behaviour is the reverse. As discussed in the case of the macroscopic free energy barrier, to do a meaningful comparison of the free energy barrier of different systems, we need to scale it by its temperature independent value. We first fit the $\Delta E$ to a linear form $\Delta E=A+B.\frac{1}{\Delta((\beta\phi)^{-1})}$ (Fig.\ref{Fig_deltaE_micro} (left panels)). We find that the value of $A$ not only increases with the depth of the potential as in the case of the LJ $(q,p)$ system, but it also increases when the density of the system is increased, keeping the interaction potential the same. We next plot in Fig.\ref{Fig_deltaE_micro} (right panels) the scaled energy $\Delta E/A$. According to our previous analysis, $B/A(micro)$, the slope of the scaled microscopic free energy barrier should be equal to $\gamma' T_{VFT}$. However, as shown in Fig.\ref{Fig_slope}(a), that does not seem to be the case. In the inset of Fig.\ref{Fig_slope}(a), we also plot $B/A(micro)$ against $K_{VFT}$, and again no consistent trend is observed. 
This behaviour is in stark contrast to that of the macroscopic counterpart. The slope of the microscopic free energy barrier varies significantly across systems and does not follow a universal trend. For instance, in the modified LJ $(q,p)$ system $B/A(micro)$ appears to track both $\gamma' T_{VFT}$ and $K_{VFT}$. In contrast, the WCA systems, which exhibit the largest variation in fragility, show almost no change in the $B/A(micro)$. Interestingly, for the LJ systems, the slope of the scaled microscopic free energy barrier varies inversely with both $\gamma T_{VFT}$ and $K_{VFT}$. 
\begin{figure}
\vspace{0.2cm}
\centering
\begin{subfigure}{0.4\textwidth}
\includegraphics[width=0.98\linewidth]{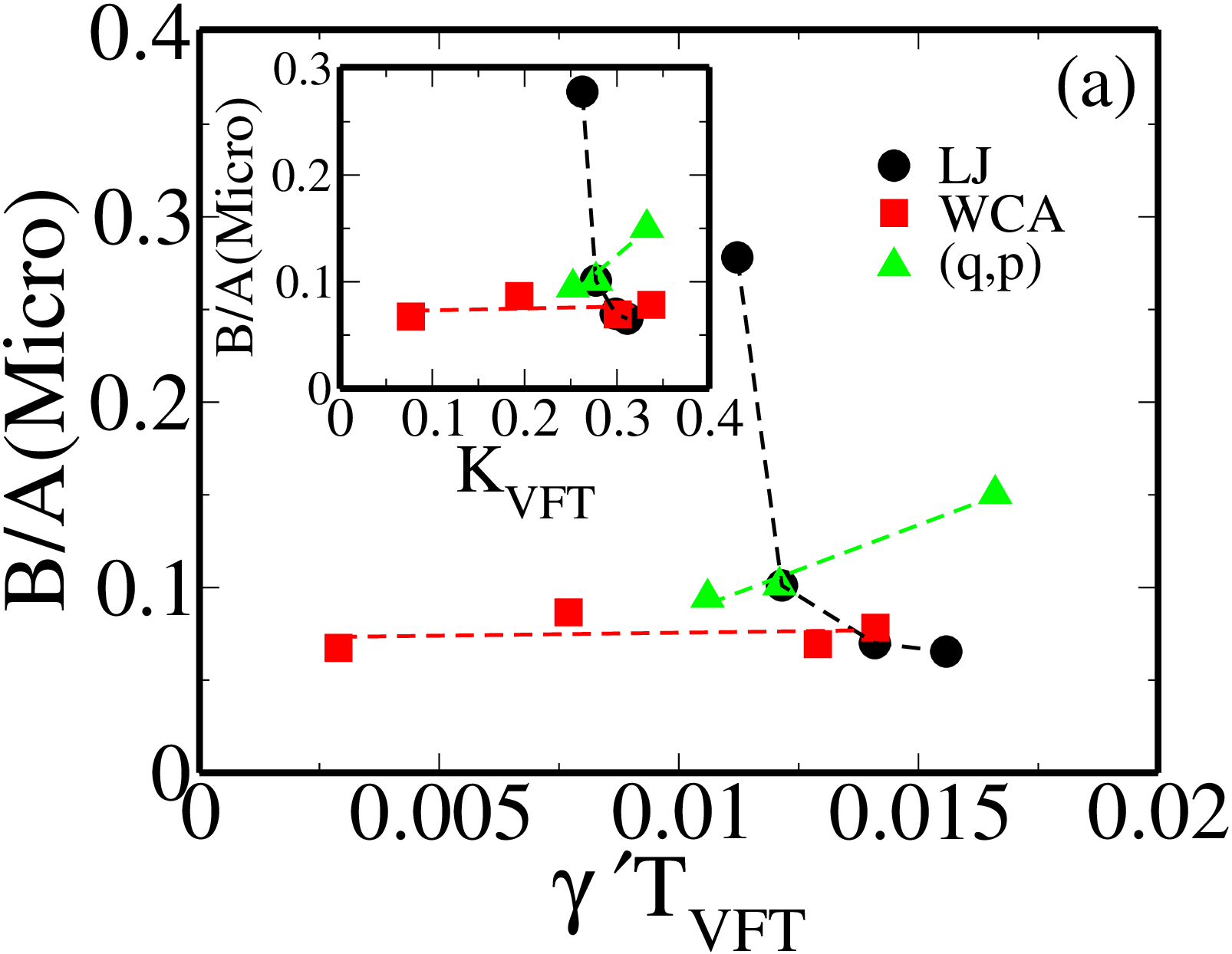}
\end{subfigure}
\vspace{0.2cm}
\begin{subfigure}{0.4\textwidth}
\includegraphics[width=0.98\linewidth]{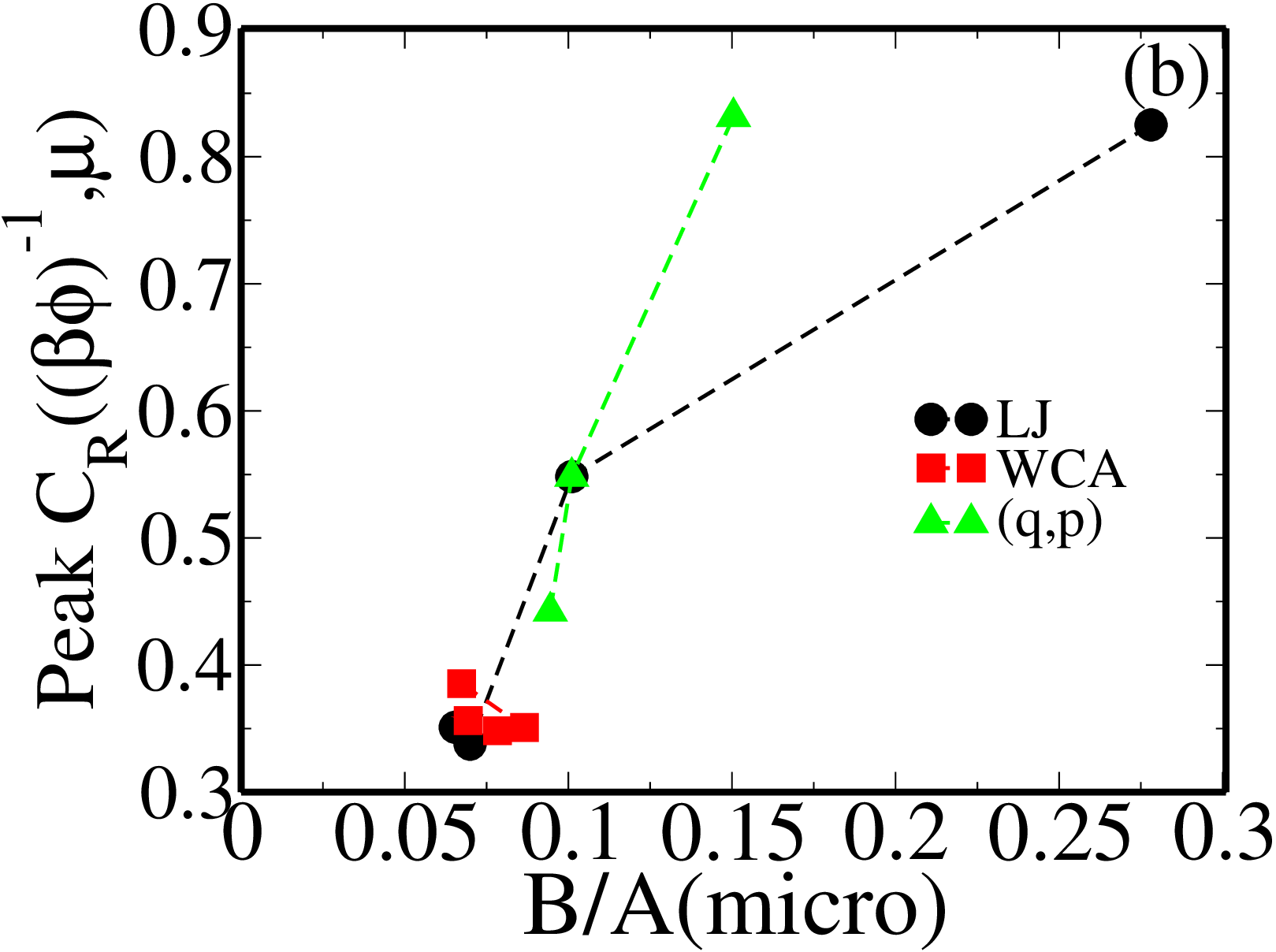}
\end{subfigure}
\vspace{0.2cm}
\caption{\textit{(a)The main panel presents the scaled slope of the microscopic free energy barrier plot, $B/A(micro)$, obtained from the slope of the $\Delta E/A$ versus $1/\Delta((\beta\phi)^{-1})$ plot (right panels of Fig.\ref{Fig_deltaE_micro}), plotted against $\gamma^{'} T_{\mathrm{VFT}}$ for three sets of systems: LJ, WCA, and modified LJ $(q,p)$ system. The data exhibit distinct behaviours across the systems. 
\textbf{Inset:} $B/A(micro)$ plotted against $K_{\mathrm{VFT}}$ shows a similar trend to the main plot, consistent with the linear relationship between $\gamma T_{\mathrm{VFT}}$ and $K_{\mathrm{VFT}}$ (demonstrated in Fig.\ref{Fig_fragility}). (b) The peak value of the Spearman rank correlation between the coarse grained SOP and the dynamics,
$C_{R}((\beta\phi)^{-1},\mu)$, obtained from Fig.\ref{Fig_correlation}, plotted against $B/A(micro)$.}}
\label{Fig_slope}
\end{figure}

Thus, although the macroscopic and microscopic free energy barriers appear to exhibit qualitatively similar variation with the SOP, a more detailed analysis reveals key differences. The scaled slope of the free energy barrier as a function of SOP quantifies how sensitively the barrier and therefore, the dynamics responds to structural changes. A higher slope indicates that even small variations in SOP lead to significant changes in the free energy barrier and dynamics, reflecting a stronger coupling between structure and dynamics.

Accordingly, the macroscopic and microscopic free energy barriers imply different strengths of structure-dynamics correlation. This discrepancy may originate from the nature of the analyses: macroscopic approaches average over many particles, thereby smoothing out local fluctuations and heterogeneities. In contrast, microscopic analysis preserves particle-level detail and captures spatially localised structural variations and dynamic heterogeneity. As a result, macroscopic models may overlook subtle but crucial features that govern dynamic behaviour.

To assess which perspective more accurately captures the structure-dynamics relationship, we next employ a complementary and direct measure of structure-dynamics correlation. The correlation between SOP and particle mobility, obtained from isoconfigurational ensemble simulations.

\subsection{Structure-Dynamics correlation via isoconfigurational study}
As discussed before, in order to apply an independent method to study structure-dynamics correlation, we perform the study using isoconfiguration runs. This is a powerful technique which was proposed by Harrowell and co-workers \cite{Cooper2008_nature} to extract the effect of any structural order parameter on the dynamics. 
To enable a meaningful comparison across systems with different densities, we select temperatures such that the $\alpha$-relaxation time is approximately the same for each system. The working temperatures used for comparing different densities are listed in Table~\ref{tab:workingtemp} in Appendix IV. We then compute the Spearman rank correlation between the SOP at the initial time and the particle mobility measured at different times(Fig.\ref{Fig_correlation}). As observed earlier by us \cite{Sharma2023_jcp} and also discussed in detail in the literature \cite{Tong2018_prx,Tong2019_Nature,Tong2020_prl,Mei2022_localvolume,yethiraj_pnas2025,Tanaka2025_review}, the correlation between an order parameter and the dynamics when the order parameter is calculated at a single particle level is high at short times and then drops at longer times. However, the correlation with coarse grained order parameter increases at long times. The coarse graining of a parameter $X$ for the $i^{th}$ particle, when coarse grained over a length scale $L$, is defined as $\Bar{X}_{i}(L) = \sum_{j}X_{j}P(r_{j}-r_{i})/\sum_{j}P(r_{j}-r_{i})$, where $P(x)$ = $exp(-x/L)$, under the hypothesis that the influence of the neighbourhood decays exponentially with distance. Studies have shown that there is an optimal value of the coarse graining length where the correlation is maximum \cite{Tong2018_prx,Tong2019_Nature,Tong2020_prl,Tanaka2025_review,Sharma2023_jcp, sanket_lengthscale}. The SOP is coarse-grained over this length, $L=4$ for LJ, and modified LJ $(q,p)$ system and $L=3$ for WCA systems in such a way that it yields maximum correlation. Note that, as shown in the left panels of Fig.\ref{Fig_partau} in Appendix IV, although the value of the correlation changes with $L$, the nature of the plot remains similar.
\begin{figure}
\centering
\begin{subfigure}{0.4\textwidth}
\includegraphics[width=0.98\linewidth]{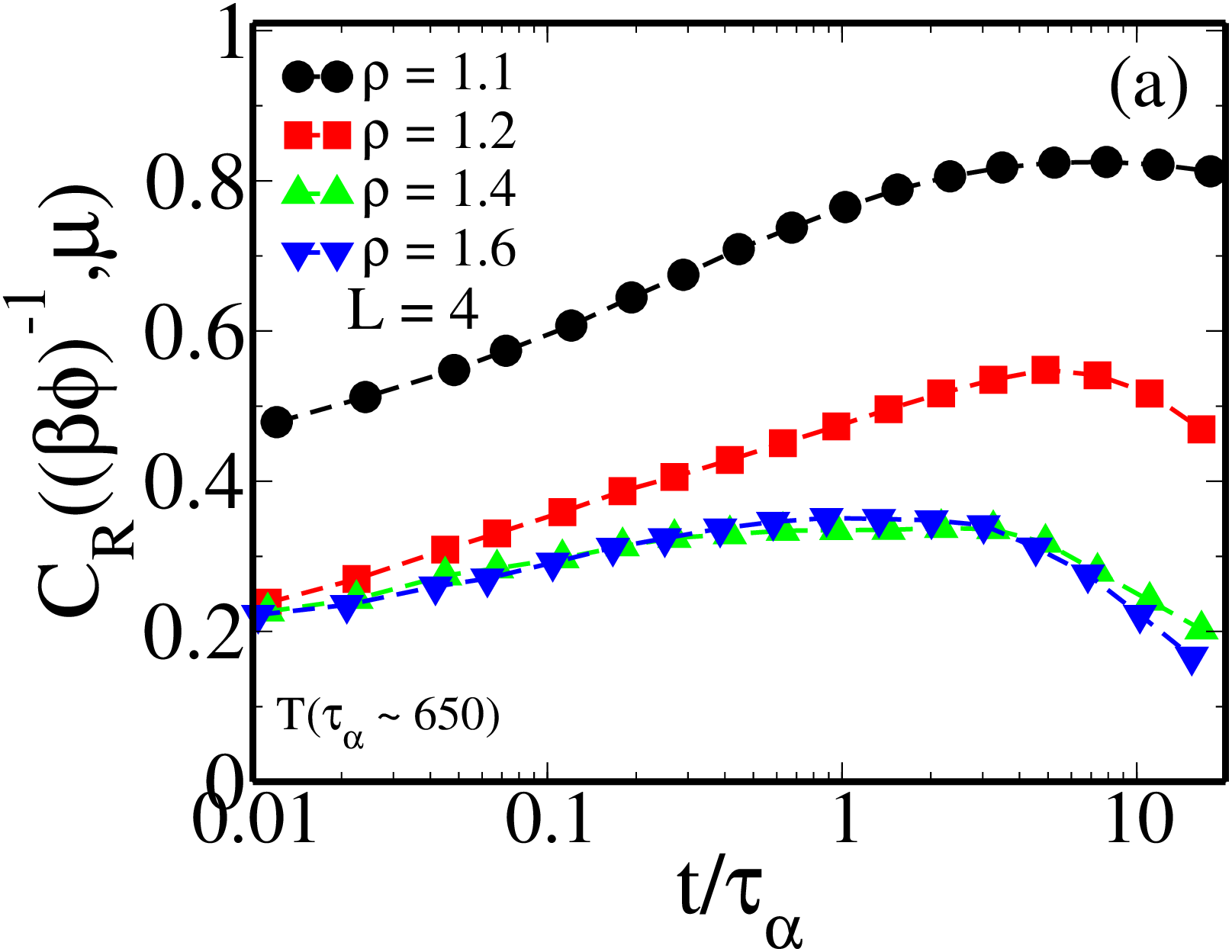}
\end{subfigure}
\vspace{0.2cm}
\begin{subfigure}{0.4\textwidth}
\includegraphics[width=0.98\linewidth]{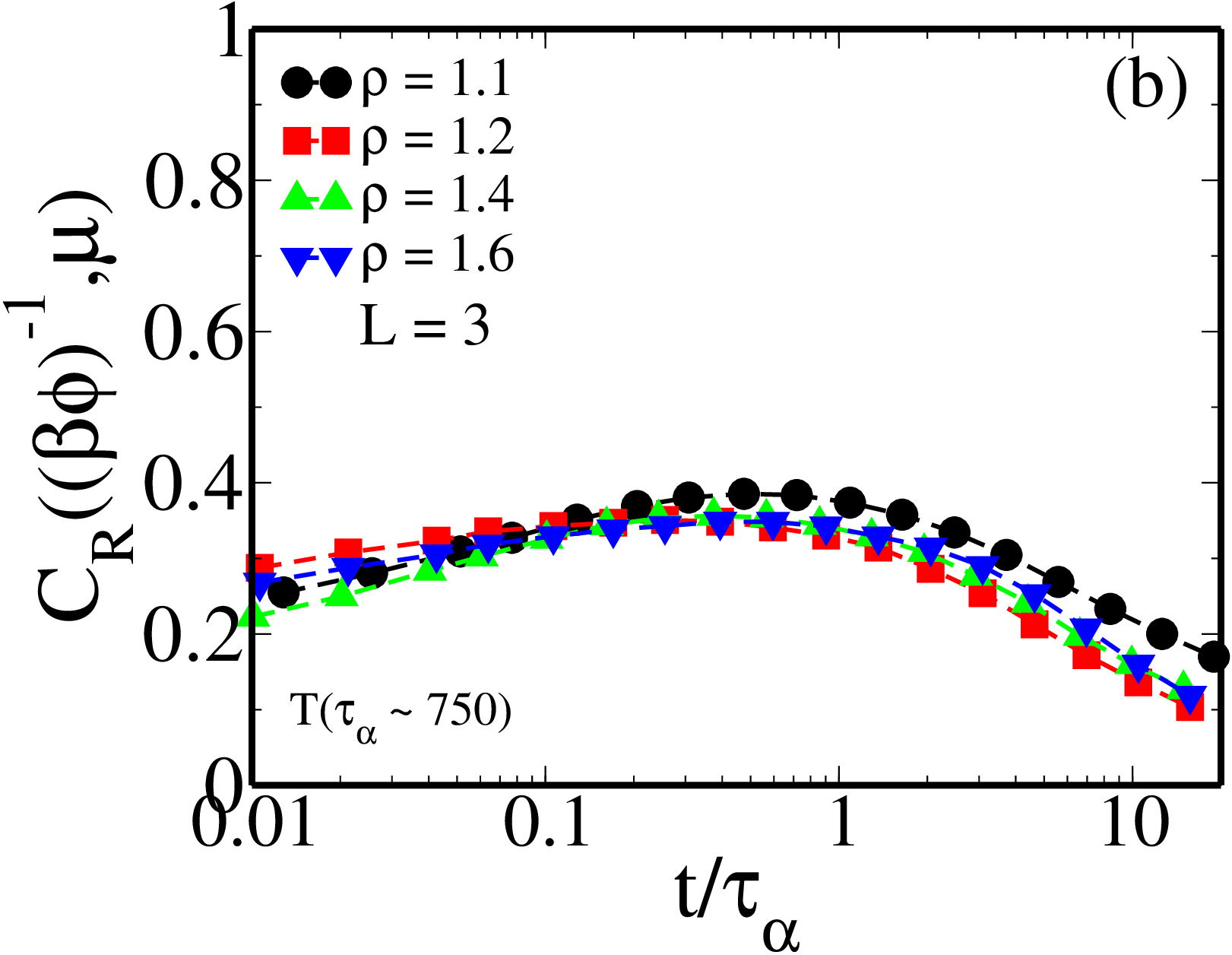}
\end{subfigure}
\vspace{0.2cm}
\begin{subfigure}{0.4\textwidth}
\includegraphics[width=0.98\linewidth]{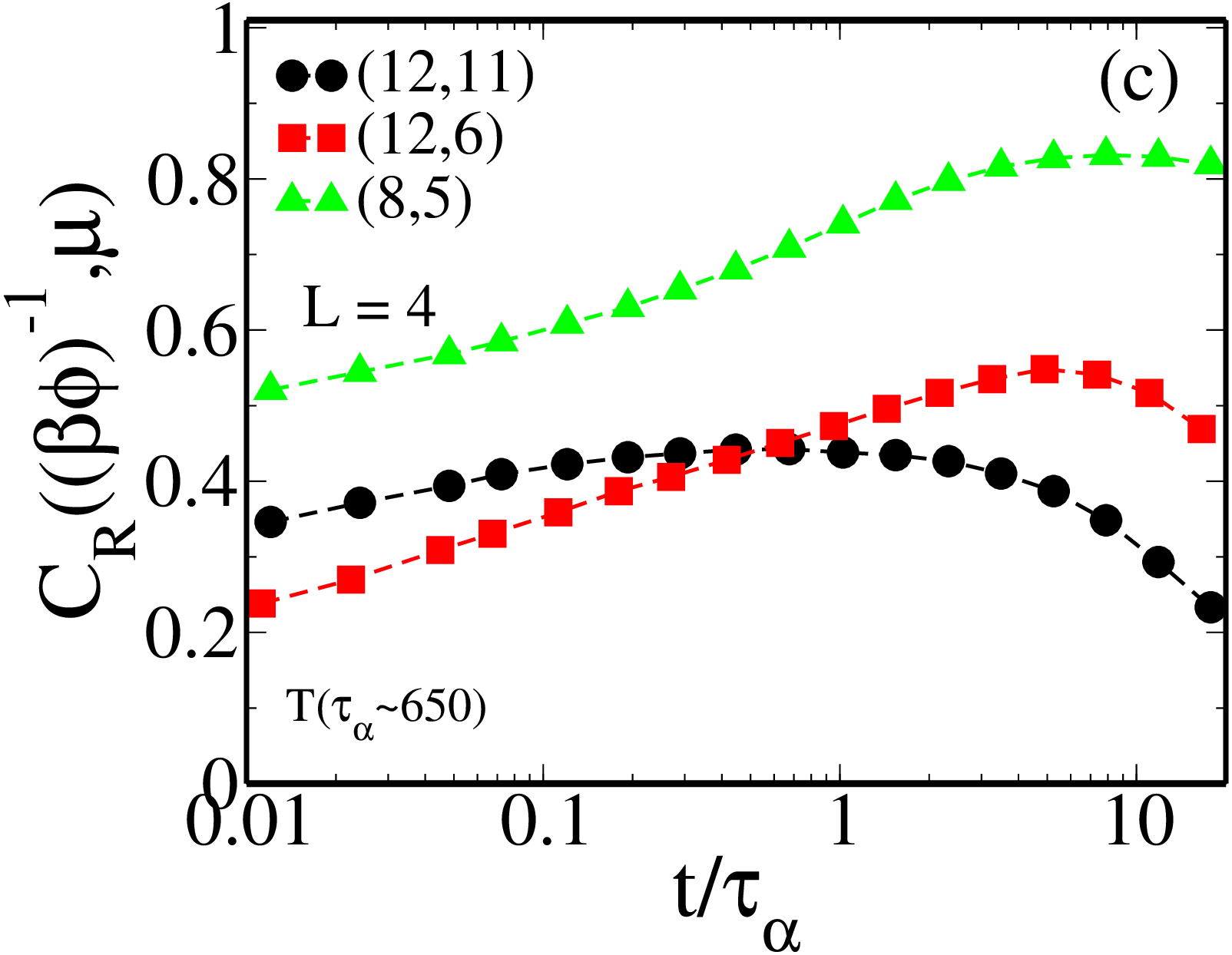}
\end{subfigure}
\caption{\textit{Spearman rank correlation,$C_{R}((\beta\phi)^{-1},\mu)$, between the coarse-grained structural order parameter, $(\beta\phi)^{-1}$, and particle mobility, $\mu$, is plotted as a function of time scaled by the $\alpha$-relaxation time ($t/\tau_{\alpha}$) for three sets of systems:
(a) LJ system at four densities: with coarse-graining length $L = 4$;
(b) WCA system at the four densities with $L = 3$;
(c) modified LJ $(q,p)$ system with $L = 4$.}}
\label{Fig_correlation}
\end{figure}
Our analysis reveals that the structure-dynamics correlation is not always related to fragility. However, we find a strong relationship between the structure–dynamics correlation and the slope of the scaled microscopic free energy barrier across all three sets of systems when we compare the plots in Fig.\ref{Fig_correlation} and the right panels of Fig.\ref{Fig_deltaE_micro}: 
i) For the LJ system, both the slope and the structure–dynamics correlation decrease with increasing fragility.
ii) For the WCA system, despite exhibiting the largest variation in fragility, both the slope and the structure–dynamics correlation remain nearly constant and appear independent of fragility.
iii) For the modified LJ $(q,p)$ system, both the slope and the structure–dynamics correlation increase with fragility. Fig.\ref{Fig_slope}(b) shows the peak value of the structure-dynamics correlation $C_R$ against the scaled microscopic free energy barrier $B/A(micro)$, which quantitatively underscores this observation.

These findings allow us to conclude that the slope of the scaled microscopic free energy barrier is closely linked to the structure-dynamics correlation. In contrast, the slope of the scaled macroscopic free energy barrier, given by 
$\gamma T_{VFT}$ does follow fragility but not the structure-dynamics correlation. This implies that the structure–dynamics correlation is not trivially connected to fragility.
However, phenomenologically, fragility has been linked to dynamic heterogeneity, with more fragile systems typically exhibiting stronger heterogeneity in dynamics \cite{Sastry2001_nature,Ediger2000_Physchem}. This raises an important and open question: What fundamental mechanism governs the system dependent trends in structure-dynamics correlation, and how is this related to dynamic heterogeneity?

\subsection{Dynamic heterogeneity and structure-dynamics correlation}
\begin{figure}
\centering
\begin{subfigure}{0.22\textwidth}
\includegraphics[width=0.98\linewidth]{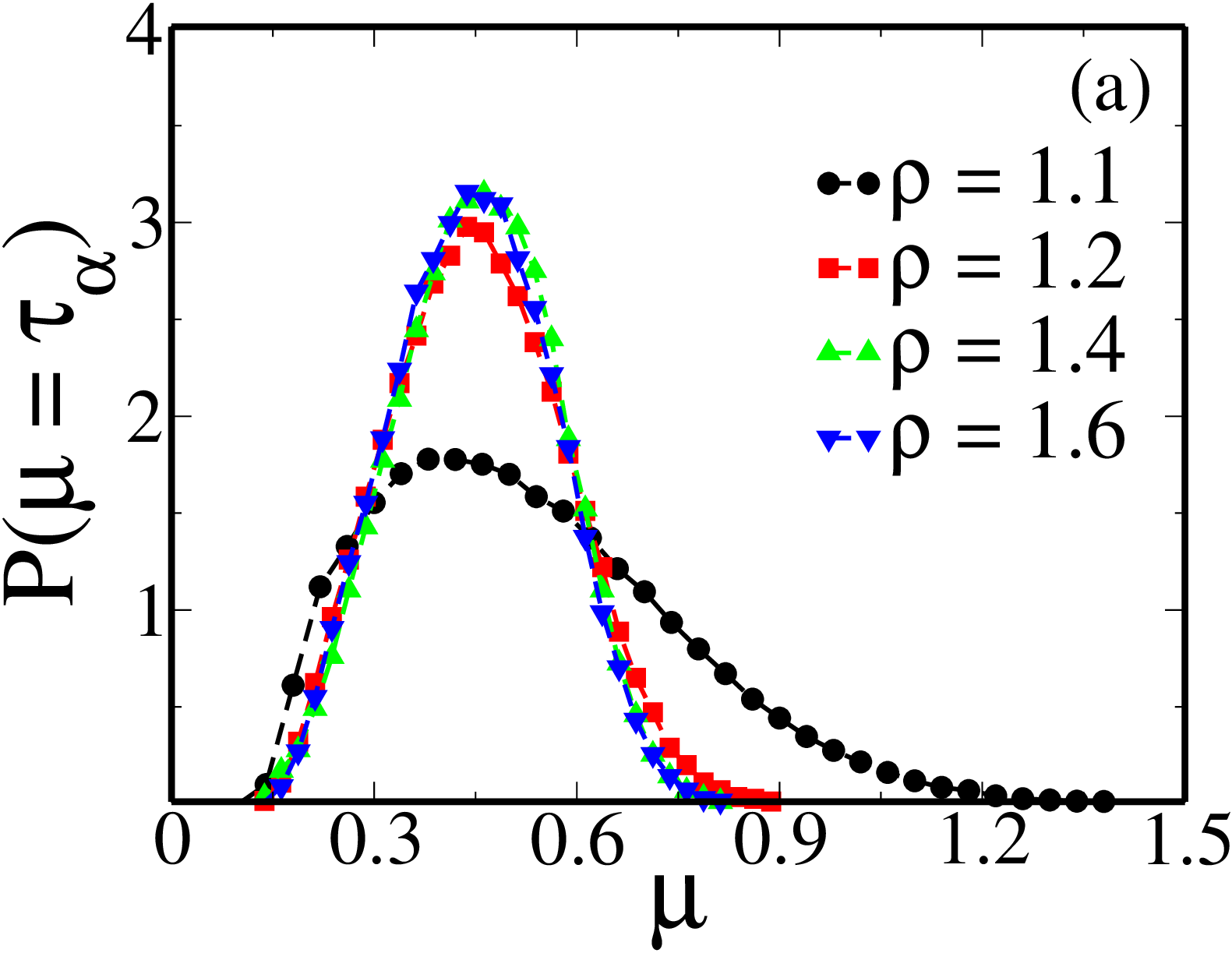}
\end{subfigure}
\hspace{0.2cm}
\begin{subfigure}{0.225\textwidth}
\includegraphics[width=0.98\linewidth]{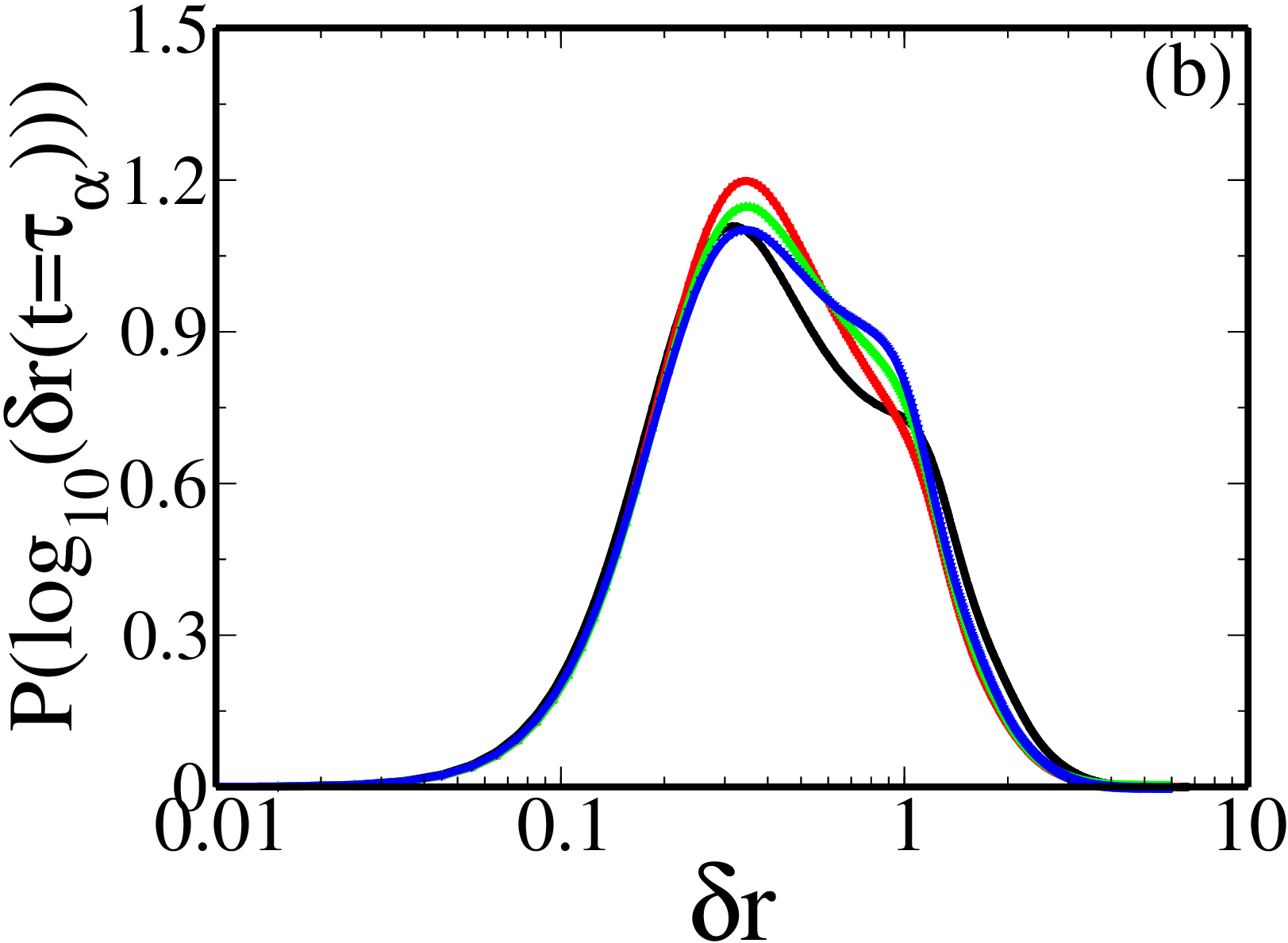}
\end{subfigure}
\vspace{0.2cm}
\begin{subfigure}{0.225\textwidth}
\includegraphics[width=0.98\linewidth]{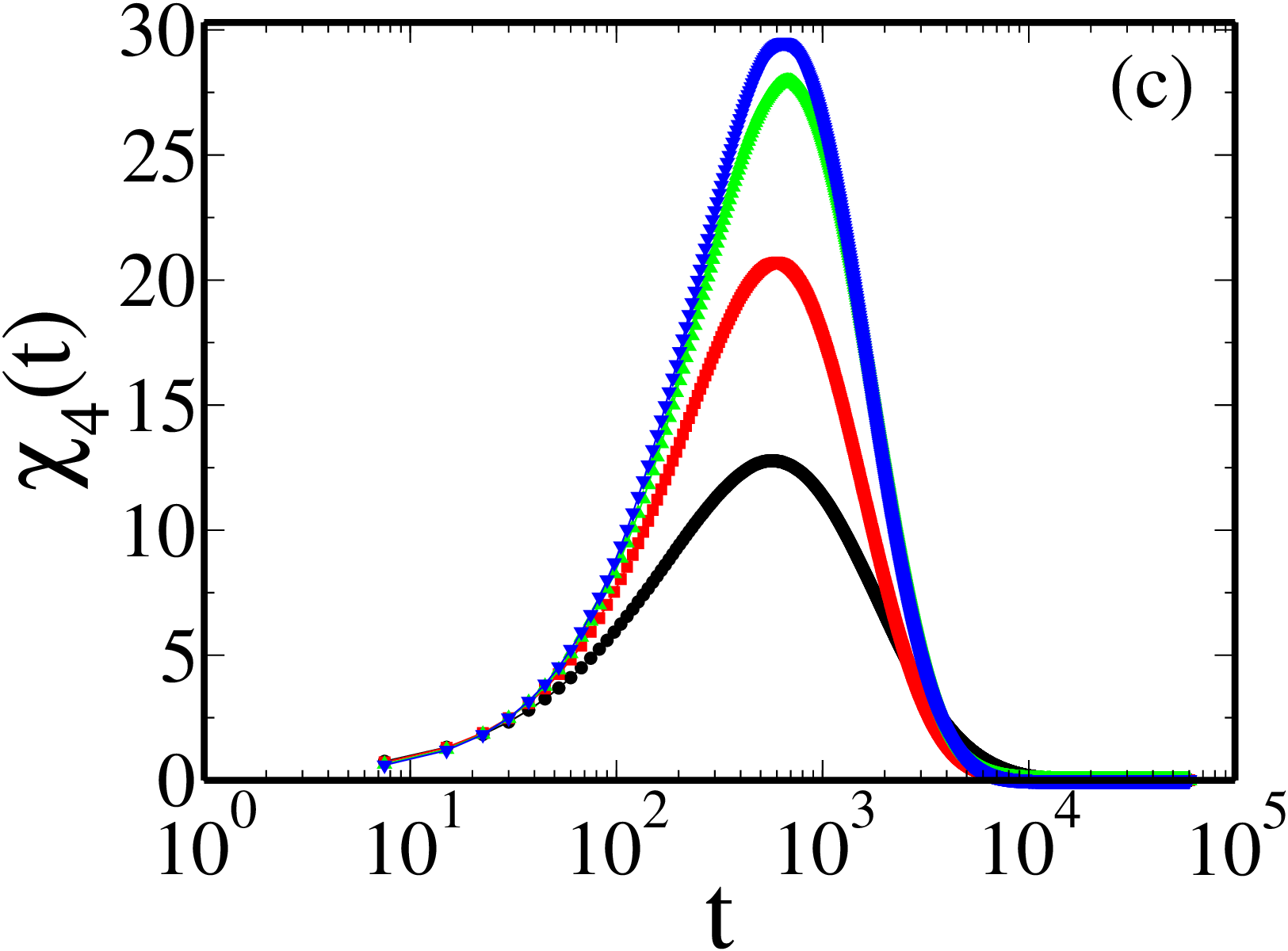}
\end{subfigure}
\hspace{0.2cm}
\begin{subfigure}{0.225\textwidth}
\includegraphics[width=0.98\linewidth]{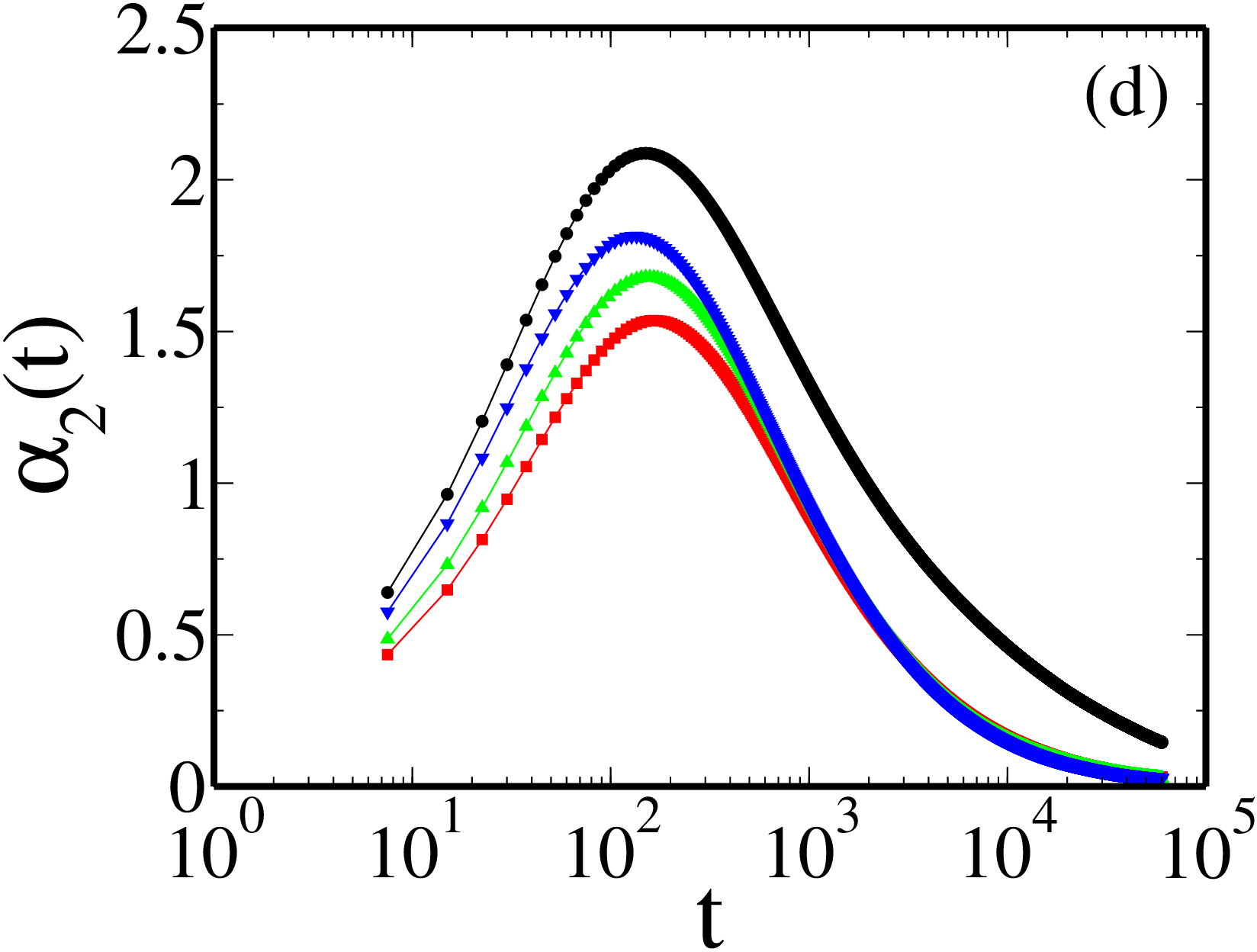}
\end{subfigure}
\vspace{0.2cm}
\caption{\textit{Different measures of heterogeneity for the LJ systems (a) Distribution of particle mobilities at the $\alpha$-relaxation time, obtained from isoconfigurational ensemble simulations.
(b) Logarithmic distribution of single-particle displacements, $P(\log_{10}(\delta r(t=\tau_{\alpha})))$, derived from standard simulation trajectories.
(c) Four-point susceptibility, $\chi_4(t)$, plotted as a function of time.
(d) Non-Gaussian parameter, $\alpha_2(t)$, plotted as a function of time.
The system exhibiting the maximum structure–dynamics correlation ($\rho = 1.1$) is characterized by a broader $P(\mu)$, a bimodal $P(\log_{10}(\delta r(t=\tau_{\alpha})))$, and a higher peak in $\alpha_2(t)$. The color coding in (b),(c),(d) matches that of (a). 
}}
\label{Fig_hetero_lj}
\end{figure}
\begin{figure}
\centering
\begin{subfigure}{0.22\textwidth}
\includegraphics[width=0.98\linewidth]{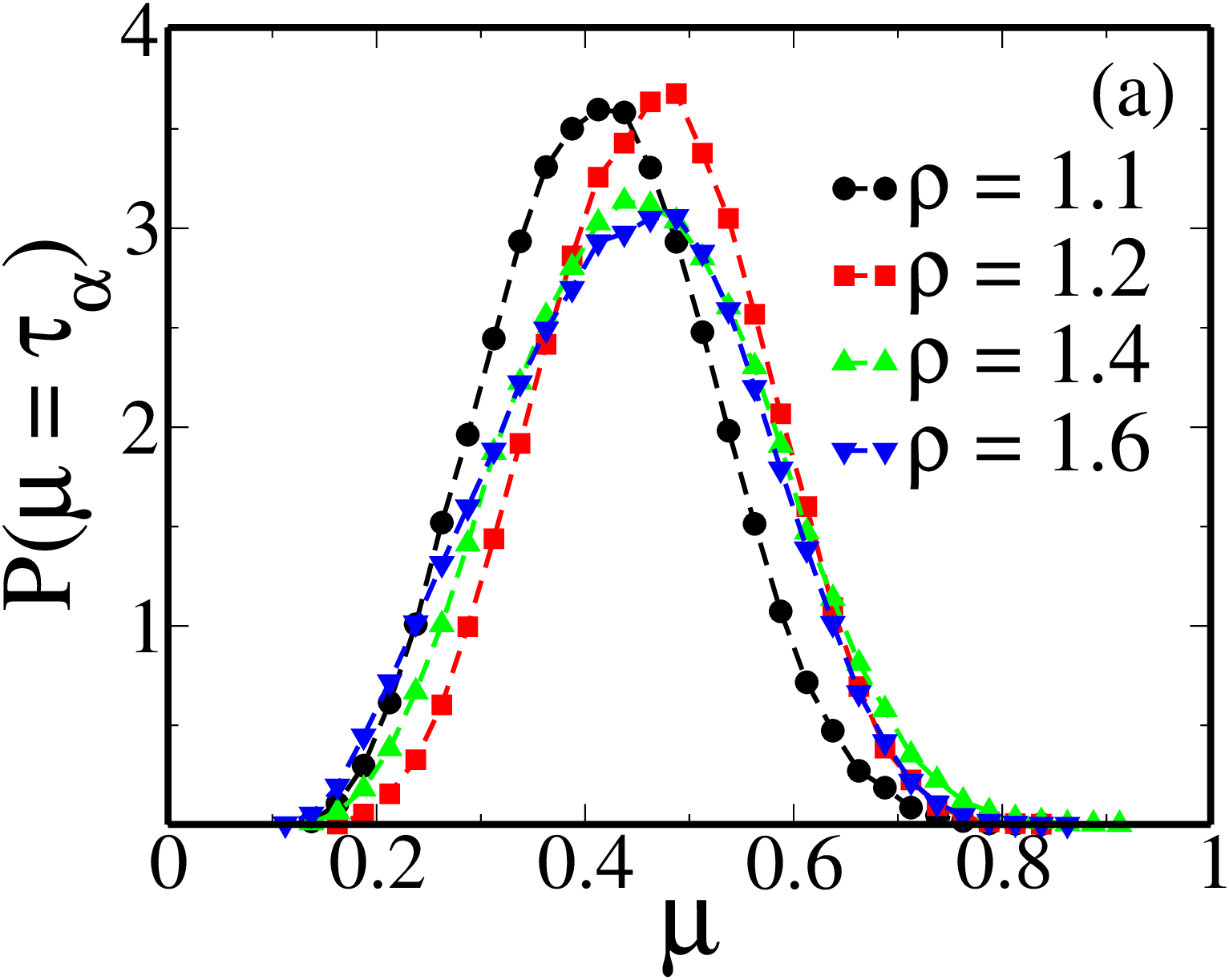}
\end{subfigure}
\hspace{0.2cm}
\begin{subfigure}{0.225\textwidth}
\includegraphics[width=0.98\linewidth]{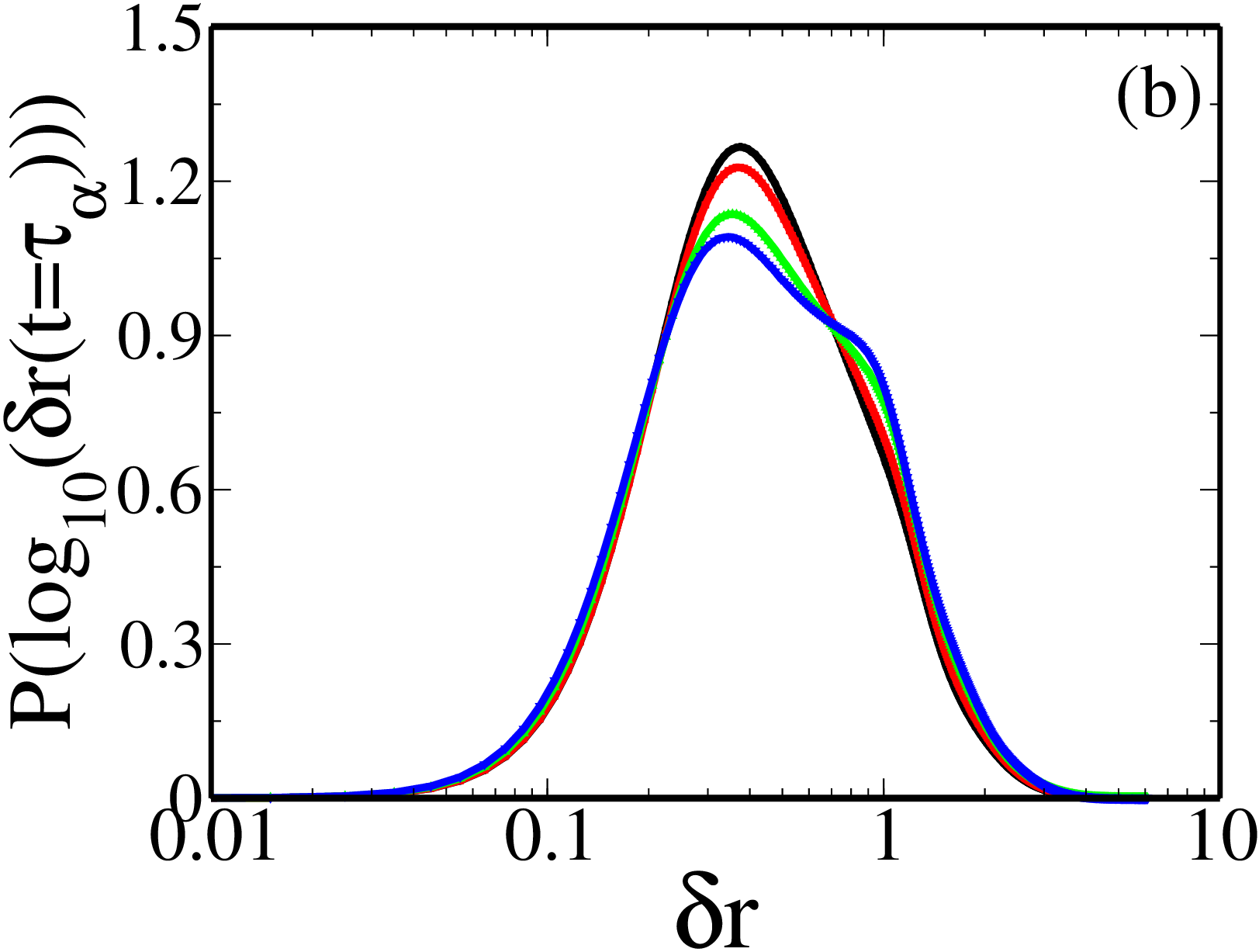}
\end{subfigure}
\vspace{0.2cm}
\begin{subfigure}{0.225\textwidth}
\includegraphics[width=0.98\linewidth]{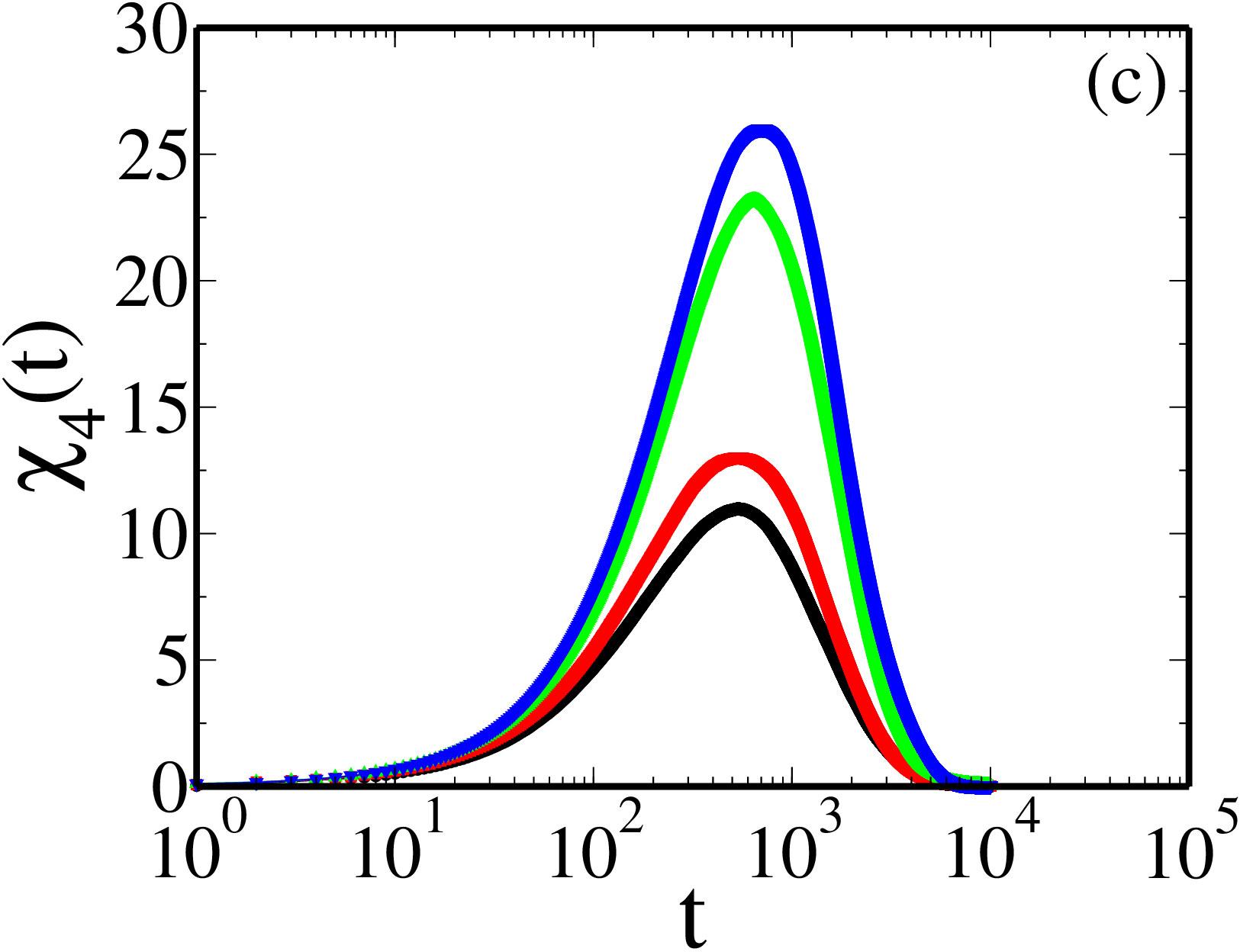}
\end{subfigure}
\hspace{0.2cm}
\begin{subfigure}{0.225\textwidth}
\includegraphics[width=0.98\linewidth]{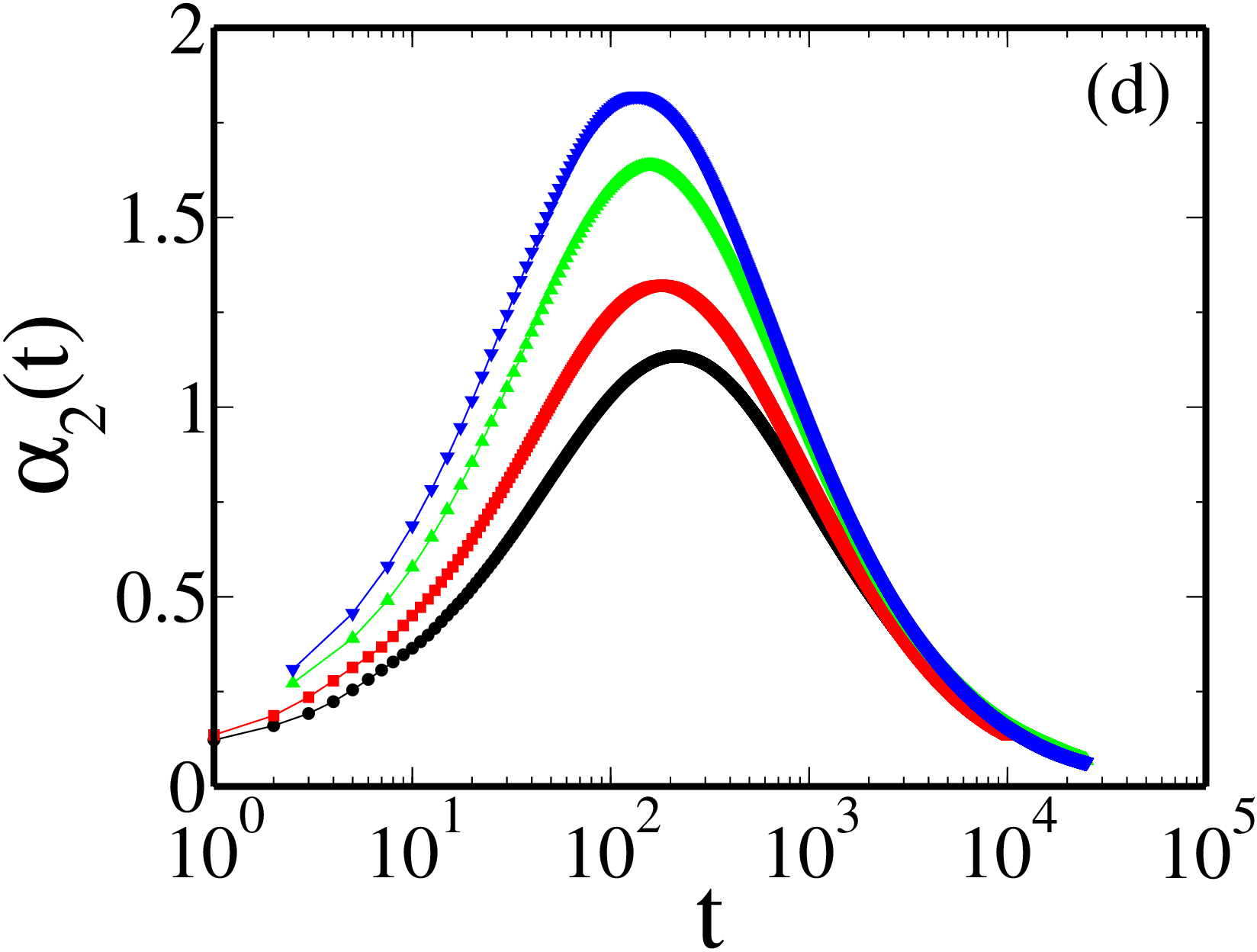}
\end{subfigure}
\hspace{0.2cm}
\caption{\textit{Different measures of heterogeneity for the WCA systems {(a) Distribution of particle mobilities at the $\alpha$-relaxation time, obtained from isoconfigurational ensemble simulations.
(b) Logarithmic distribution of single-particle displacements, $P(\log_{10}(\delta r(t=\tau_{\alpha})))$, derived from standard simulation trajectories.
(c) Four-point susceptibility, $\chi_4(t)$, plotted as a function of time.
(d) Non-Gaussian parameter, $\alpha_2(t)$, plotted as a function of time.
Across all densities, the system exhibits consistent trends in dynamic heterogeneity metrics, with structure–dynamics correlations remaining similar. The colour coding in (b),(c),(d) matches that of (a). 
 }}}
\label{Fig_hetero_wca}
\end{figure}
\begin{figure}
\centering
\begin{subfigure}{0.22\textwidth}
\includegraphics[width=0.98\linewidth]{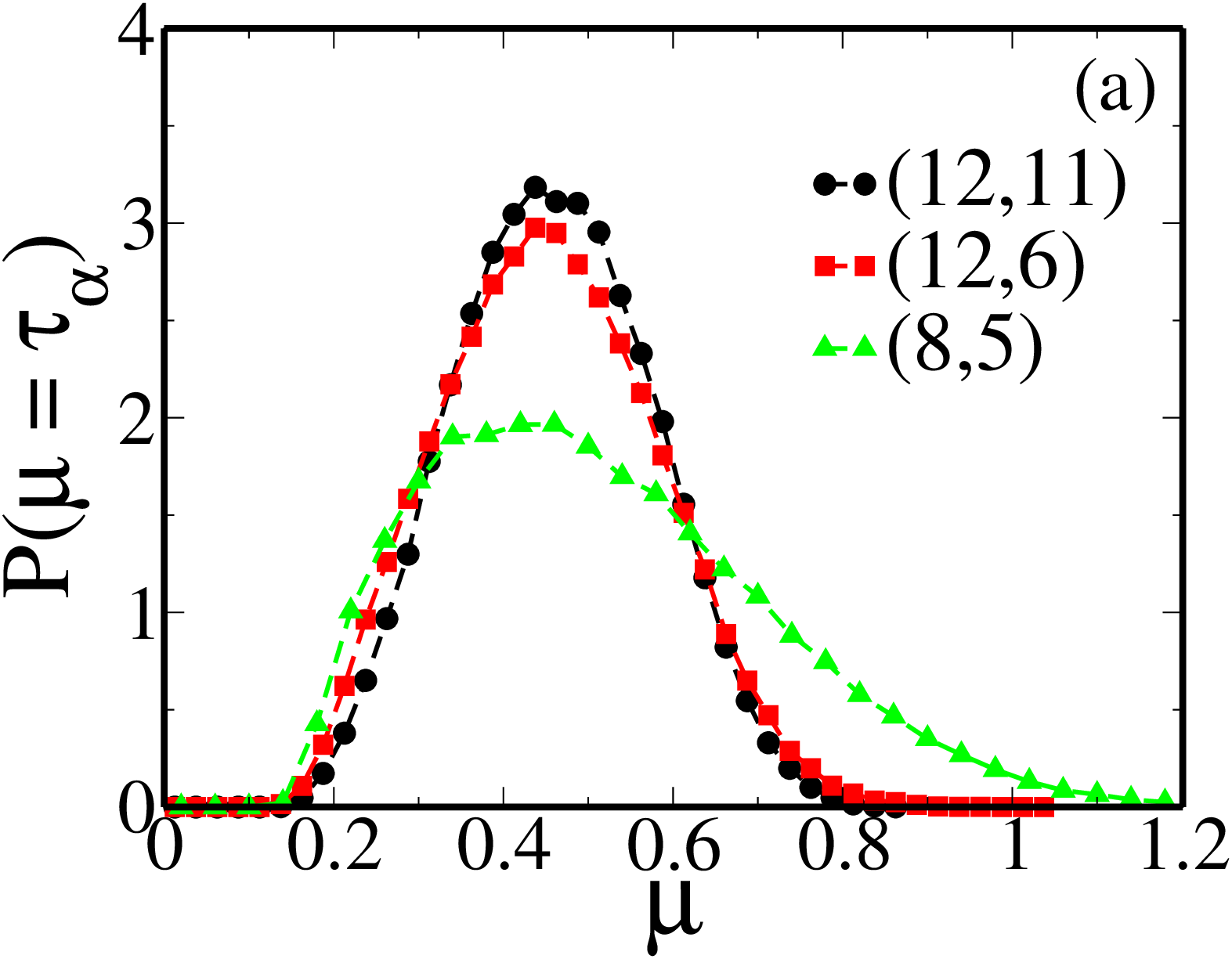}
\end{subfigure}
\hspace{0.2cm}
\begin{subfigure}{0.225\textwidth}
\includegraphics[width=0.98\linewidth]{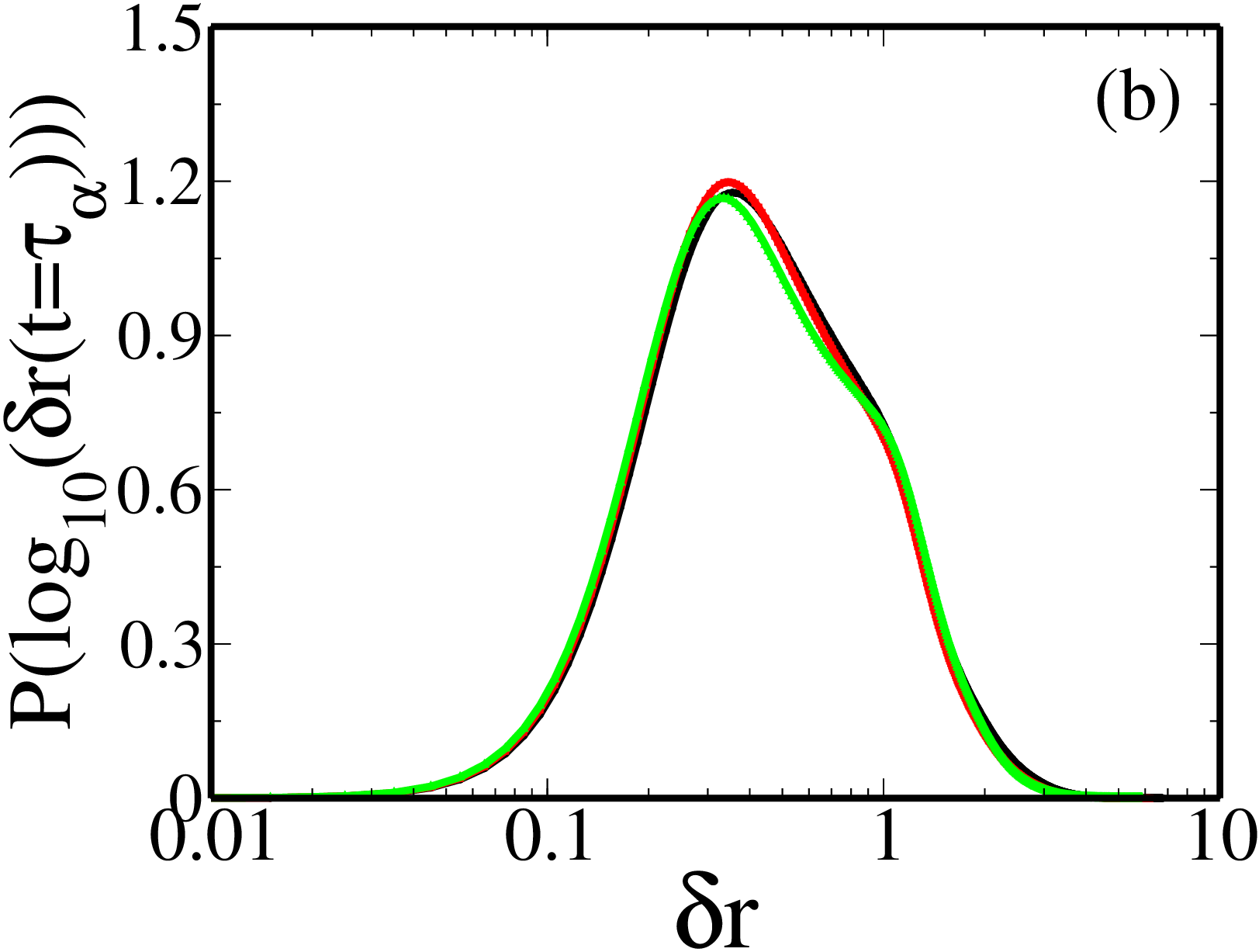}
\end{subfigure}
\vspace{0.2cm}
\begin{subfigure}{0.225\textwidth}
\includegraphics[width=0.98\linewidth]{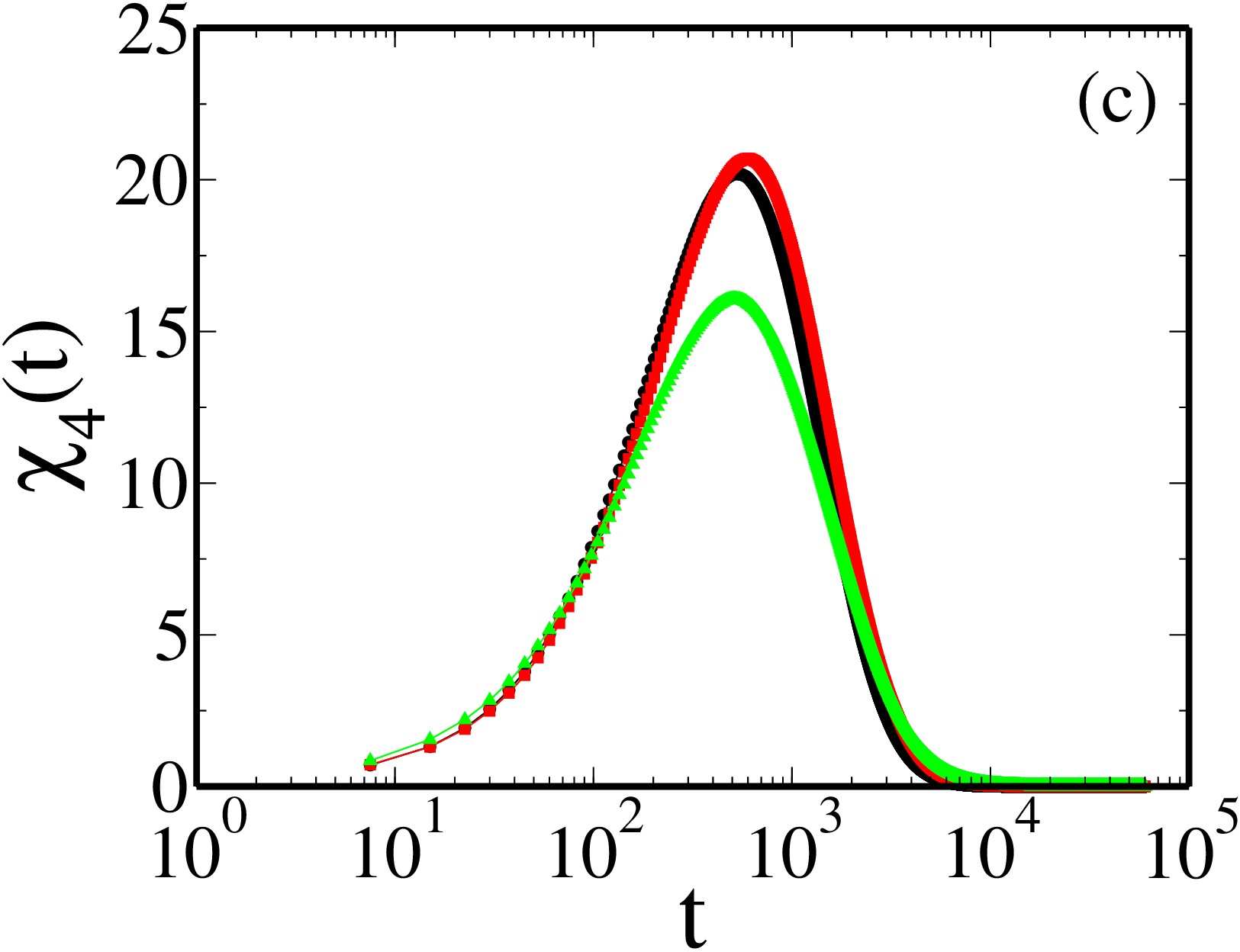}
\end{subfigure}
\hspace{0.2cm}
\begin{subfigure}{0.225\textwidth}
\includegraphics[width=0.98\linewidth]{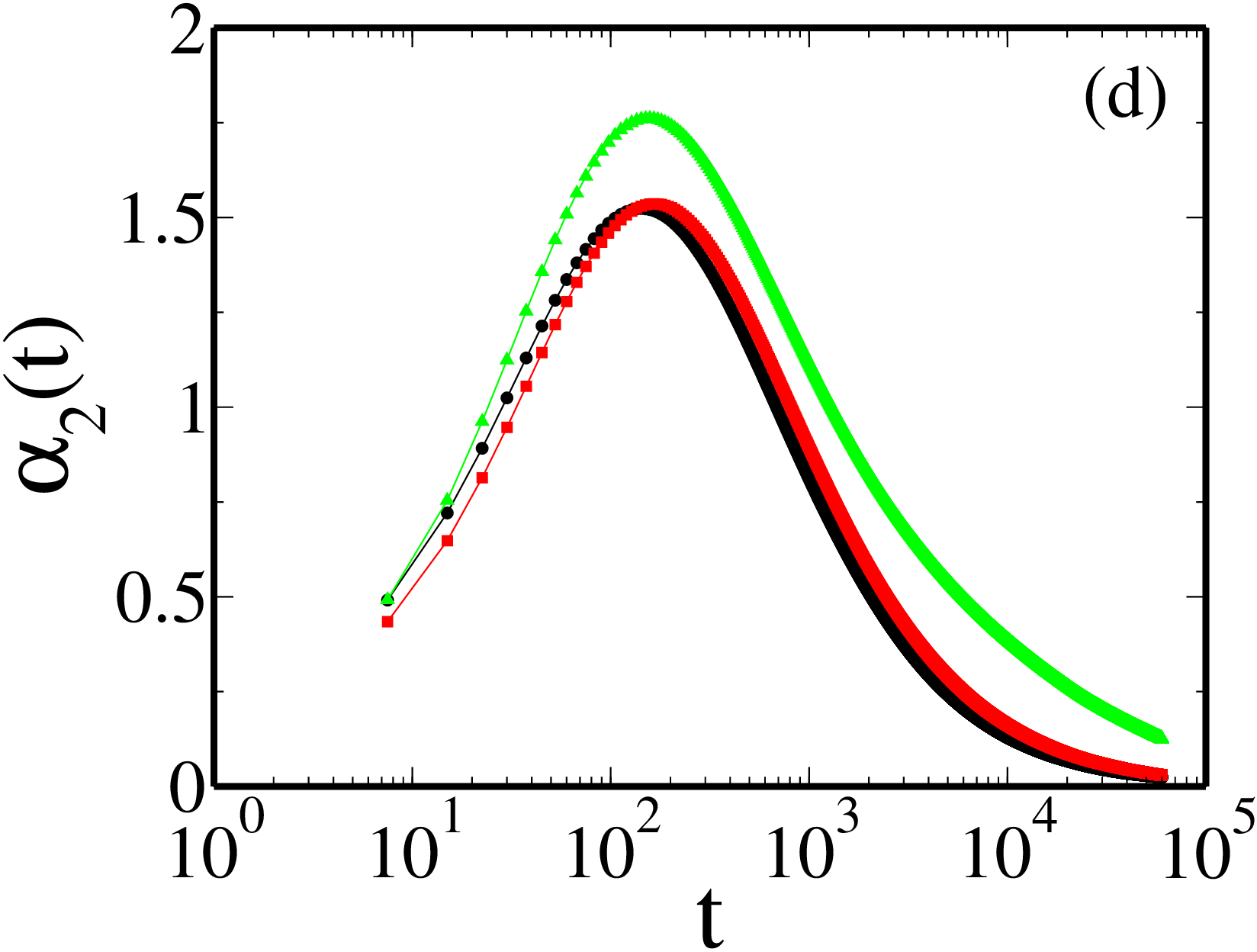}
\end{subfigure}
\vspace{0.2cm}
\caption{\textit{Different measures of heterogeneity for the modified LJ $(q,p)$ system. (a) Distribution of particle mobilities at the $\alpha$-relaxation time, obtained from isoconfigurational ensemble simulations.
(b) Logarithmic distribution of single-particle displacements, $P(\log_{10}(\delta r(t=\tau_{\alpha})))$, derived from standard simulation trajectories.
(c) Four-point susceptibility, $\chi_4(t)$, plotted as a function of time.
(d) Non-Gaussian parameter, $\alpha_2(t)$, plotted as a function of time.
The system exhibiting the maximum structure–dynamics correlation ($(8,5)$) is characterized by a broader $P(\mu$), a bimodal $P(\log_{10}(\delta r(t=\tau_{\alpha})))$, and a higher peak in $\alpha_2(t)$. The colour coding in (b),(c),(d) matches that of (a). }}
\label{Fig_hetero_ksys}
\end{figure}
\begin{figure*}
\centering
\begin{subfigure}{0.31\textwidth}
\includegraphics[width=0.98\linewidth]{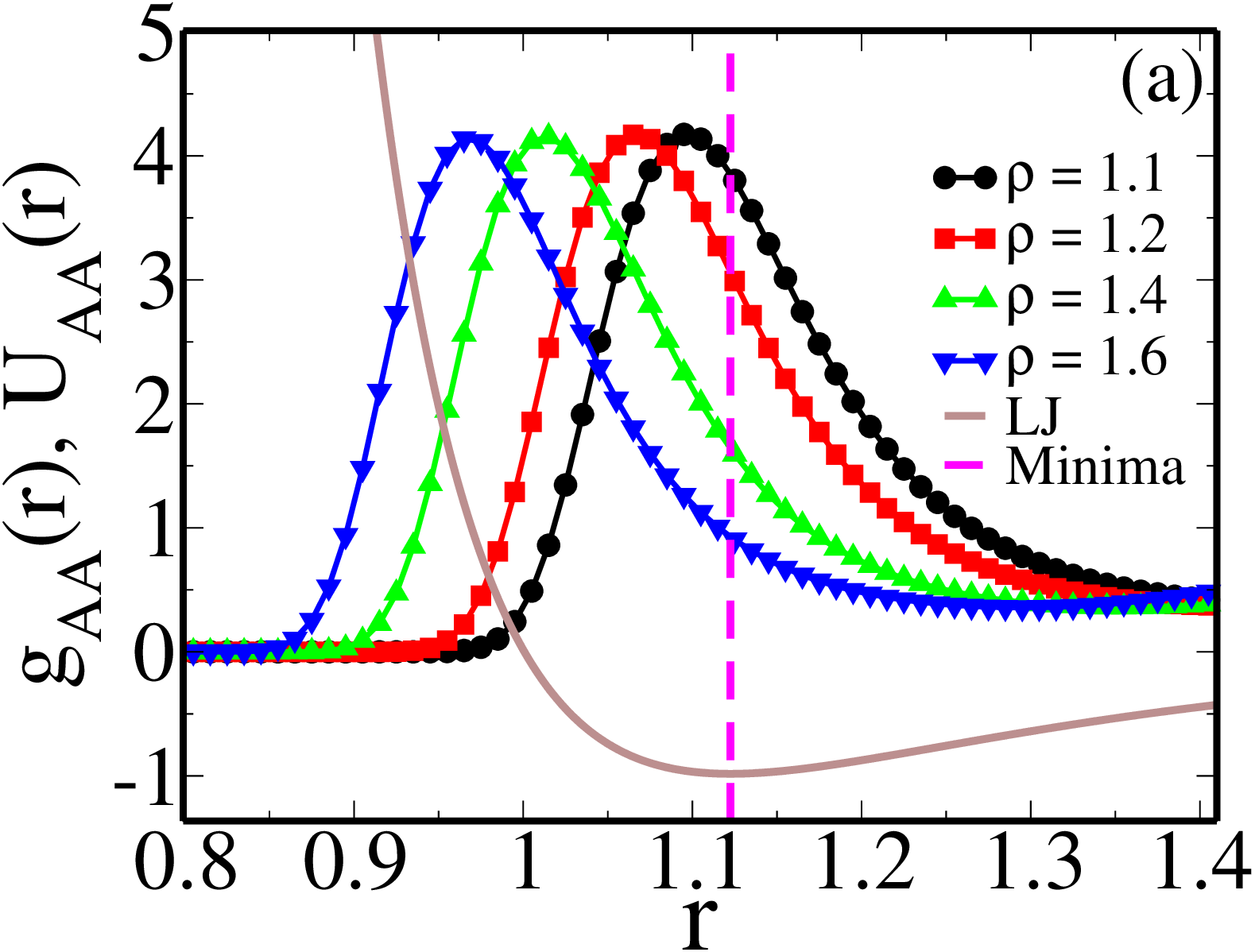}
\end{subfigure}
\hspace{0.2cm}
\begin{subfigure}{0.31\textwidth}
\includegraphics[width=0.98\linewidth]{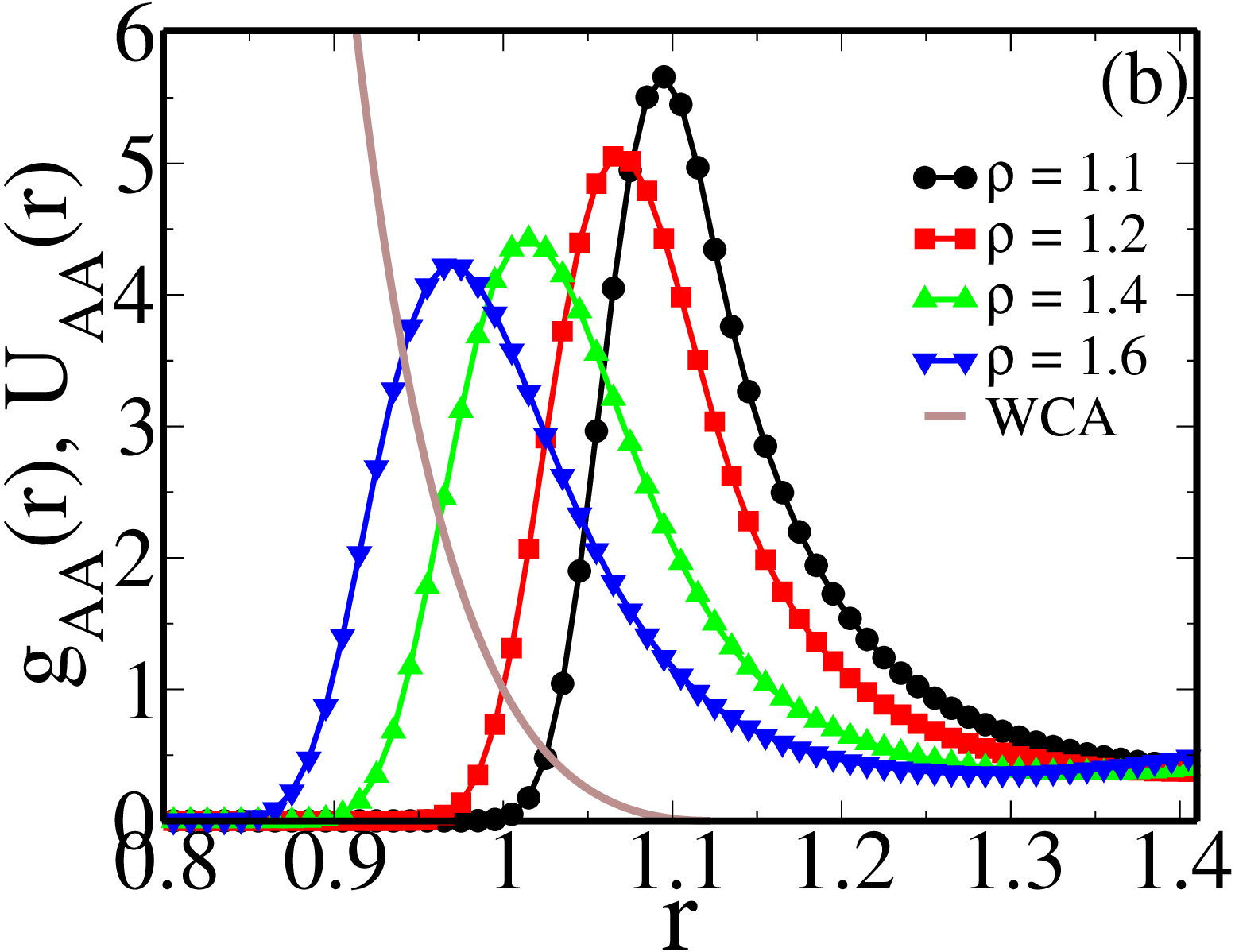}
\end{subfigure}
\hspace{0.2cm}
\begin{subfigure}{0.31\textwidth}
\includegraphics[width=0.98\linewidth]{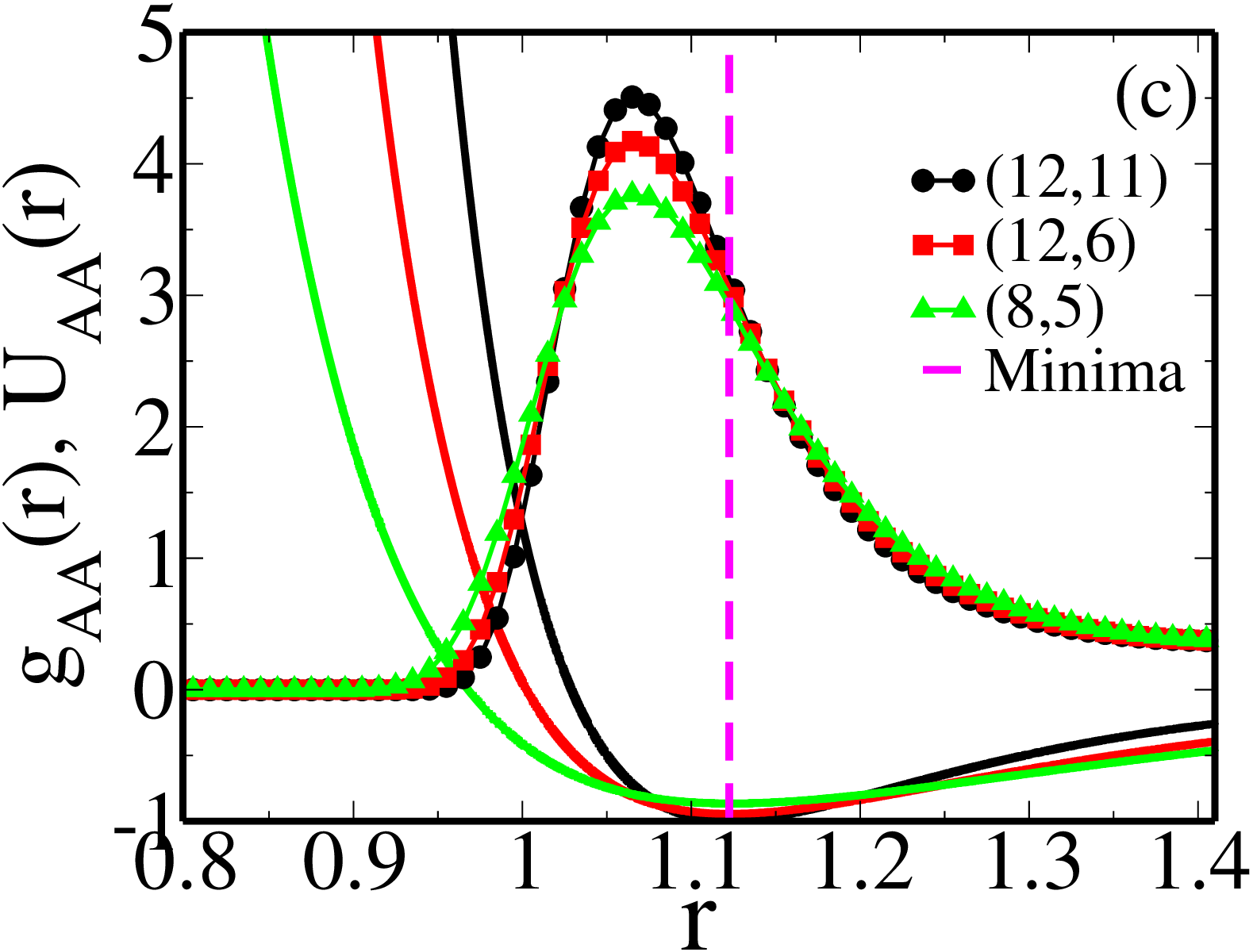}
\end{subfigure}
\vspace{0.2cm}
\begin{subfigure}{0.31\textwidth}
\includegraphics[width=0.98\linewidth]{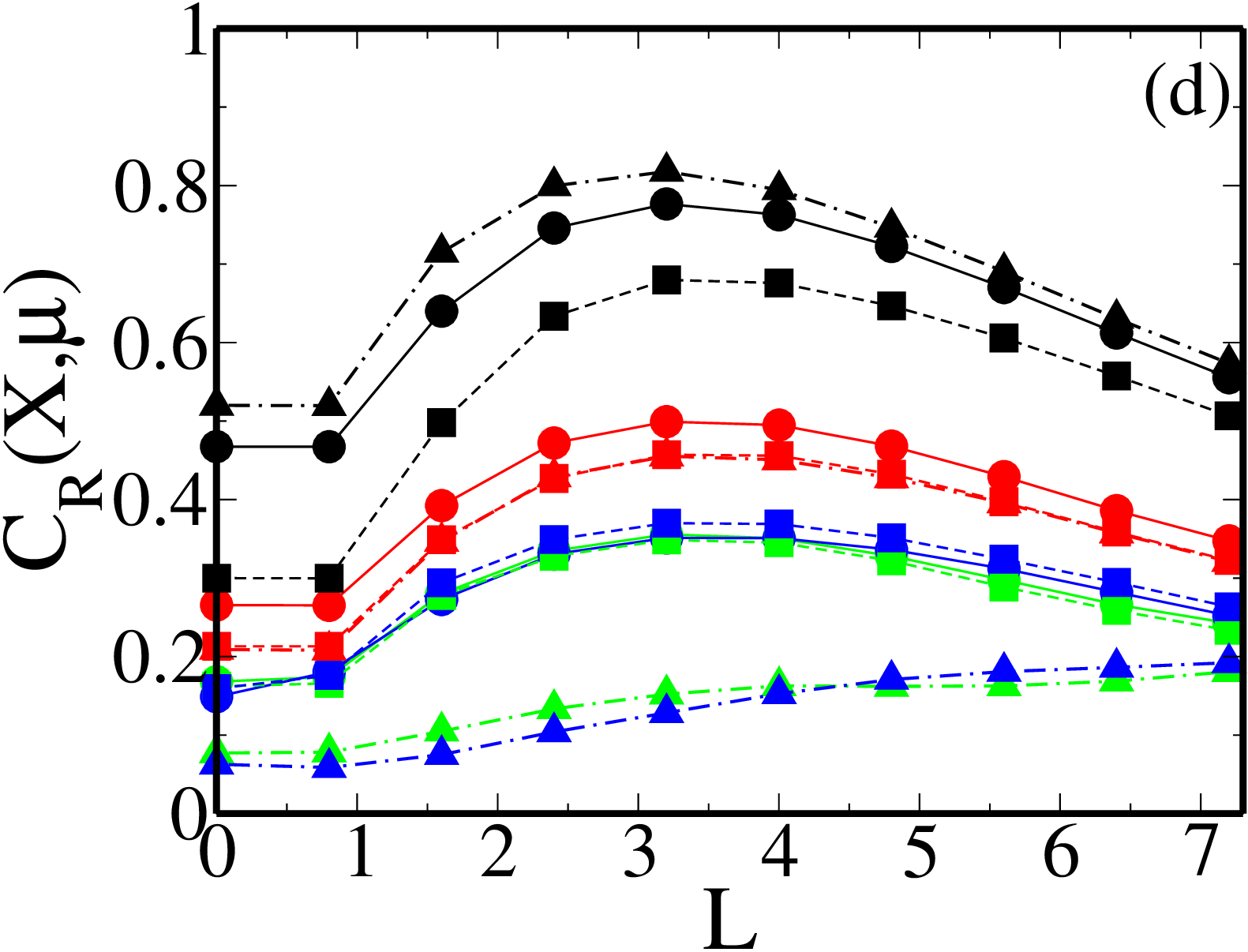}
\end{subfigure}
\hspace{0.2cm}
\begin{subfigure}{0.31\textwidth}
\includegraphics[width=0.98\linewidth]{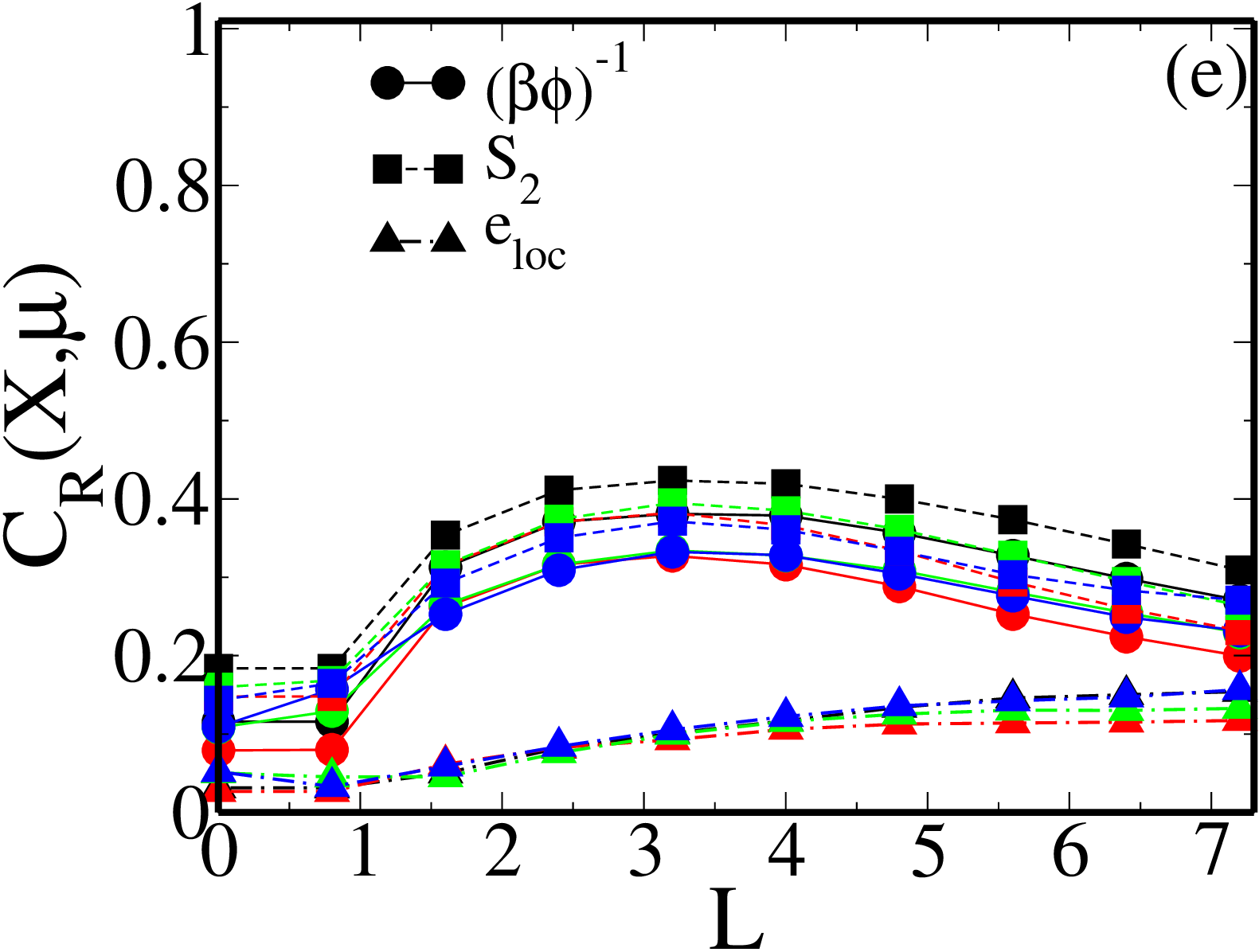}
\end{subfigure}
\hspace{0.2cm}
\begin{subfigure}{0.31\textwidth}
\includegraphics[width=0.98\linewidth]{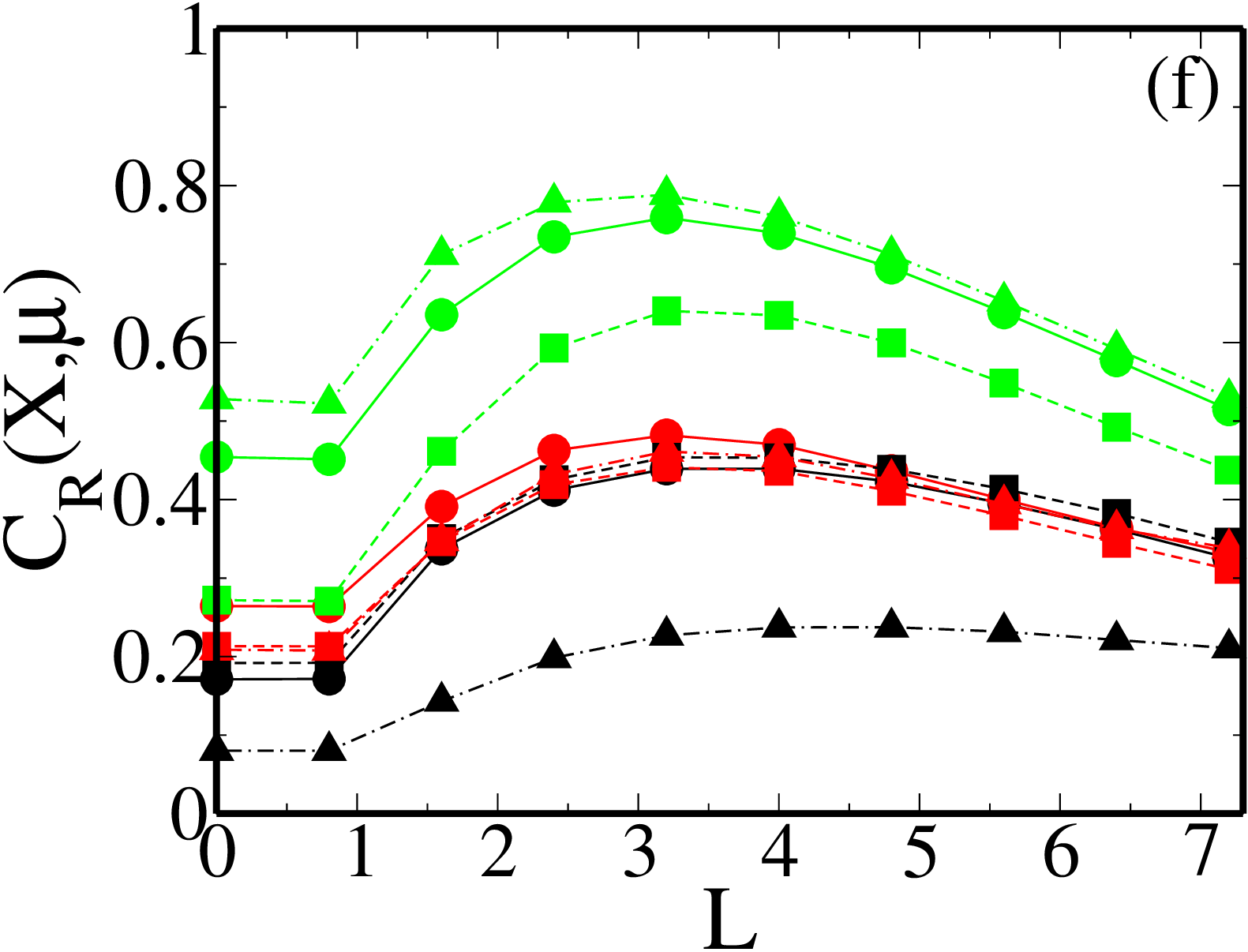}
\end{subfigure}
\vspace{0.2cm}
\caption{\textit{\textbf{Top panels:} Partial radial distribution function, $g_{AA}(r)$, and interparticle potential, $U_{AA}(r)$, plotted as functions of $r$ for three sets of systems: (a)LJ and (b)WCA, each examined at four different densities, and (c) modified LJ $(q,p)$ system.
\textbf{Bottom panels:} Spearman rank correlation,$C_{R}((\beta\phi)^{-1},X)$, between particle mobility, $\mu$, and coarse-grained structural parameters $X$—where $X$ represents SOP, $(\beta\phi)^{-1}$ (circles), two-body entropy, $S_2$ (squares), and local energy, $e_{\text{loc}}$ (triangles)—plotted as a function of coarse-graining length, $L$, for the same systems shown in top panels. The colour coding in the bottom panels matches that of the top panels.
}
}
\label{Fig_gr_potn}
\end{figure*}
In the literature, dynamical heterogeneity is characterised using several complementary approaches. In this study, we employ the logarithm of the distribution of single-particle displacements at the $\alpha$- relaxation time, expressed as, $P(log_{10}(\delta r(t=\tau_{\alpha}))=ln(10)4\pi \delta r^{3}G_{s}(\delta r,t)$ which is related to the self-part of the van Hove correlation function, $G_{s}(\delta r,t)$ \cite{Flenner2005_pre}. In addition, we analyse two commonly used dynamical measures: the four-point dynamic susceptibility $\chi_4$,\cite{Flenner2014_prl}, and the non-Gaussian parameter $\alpha_2$ \cite{Rahman1964_review}. Systems exhibiting stronger dynamical heterogeneity typically show a pronounced long tail in, $G_{s}(\delta r,t)$ and a double peak structure in $P(log_{10}(\delta r(t=\tau_{\alpha}))$, and both the $\chi_{4}$ and the $\alpha_{2}$ are found to have higher values \cite{Barrat2010_soft,Lacevic2003_jcp, Flenner2014_prl,Kob1997_prl}. 
To further probe dynamic variability, we also examine the distribution of particle mobility $\mu$ obtained from isoconfigurational ensemble simulations as $\mu^{j}(t) = \frac{1}{N_{\text{IC}}} \sum_{i=1}^{N_{\text{IC}}} \left| \mathbf{r}_{i}^{j}(t) - \mathbf{r}_{i}^{j}(0) \right|$, where $\mu^j(t)$ is the average displacement of the $j^\text{th}$ particle over $N_{\text{IC}}$ isoconfigurational trajectories at time $t$. The mobility distribution, $P(\mu)$ is not identical but similar to the $4\pi \delta r^{2} G_{s}(\delta r,t)$. The former represents a averaged measure of motion over many velocity realizations and thus suppresses some of the heterogeneity captured in the latter. Along with the distribution of $\mu$, we also calculate the distribution of the $\alpha$- relaxation time from the single particle overlap function, where the overlap function is averaged over the different velocity realizations. In Fig.\ref{Fig_hetero_lj} we plot the functions for the LJ system, in Fig.\ref{Fig_hetero_wca} we plot the same for the WCA system and in Fig.\ref{Fig_hetero_ksys} we plot them for the modified LJ $(q,p)$ system. 

Our analysis reveals that different measures of dynamic heterogeneity do not necessarily track each other. Interestingly, the systems exhibiting the maximum structure–dynamics correlation, LJ at $\rho = 1.1 $ and $(8,5)$ are both characterized by broader mobility distributions (Fig.\ref{Fig_hetero_lj}(a) and Fig.\ref{Fig_hetero_ksys}(a)), broader distribution of $P(\tau_{\alpha})$ (Fig.\ref{Fig_suspt}(a) and Fig.\ref{Fig_suspt}(e) in the Appendix IV), prominent two-peaked structure of $P(log_{10}(\delta r(t=\tau_{\alpha}))$ (Fig.\ref{Fig_hetero_lj}(b) and Fig.\ref{Fig_hetero_ksys}(b)) and a higher $\alpha_2$ value
(Fig.\ref{Fig_hetero_lj}(d) and Fig.\ref{Fig_hetero_ksys}(d)). However both these systems show low values of $\chi_{4}$ (Fig.\ref{Fig_hetero_lj}(c) and Fig.\ref{Fig_hetero_ksys}(c)). Interestingly, the LJ at $\rho=1.1$ within its set is the least fragile system but the $(8,5)$ within its set is the most fragile system. 
For the LJ system, except for $\chi_{4}$, no other functions follow fragility, and it is just the reverse for the modified LJ $(q,p)$ system, where $\chi_{4}$ does not follow fragility, but all the other functions do. For the WCA system where the change in fragility is the widest, the distribution of mobility $P(\mu)$ (Fig.\ref{Fig_hetero_wca}(a)) and the $\alpha$ relaxation time $P(\tau_{\alpha})$ (Fig.\ref{Fig_suspt}(c) in the Appendix) are quite similar for all the densities however the other functions, $P(log_{10}(\delta r(t=\tau_{\alpha}))$, $\alpha_{2}$ and $\chi_{4}$ follow fragility.  

An important point to note is that, although both the mobility distribution and the single-particle overlap function are derived from the same set of isoconfigurational runs, they are not equivalent. The mobility distribution characterises the actual displacement of individual particles, whereas the overlap function quantifies the decorrelation of a particle's position, based on whether its displacement exceeds a threshold distance `a'. However, we find that they always correlate. 

Another measure of dynamic heterogeneity, the four-point dynamic susceptibility $\chi_{4}$ is directly related to the fluctuations in the overlap function (Section \ref{section_relaxation}). In the Appendix (see right panels of Fig.\ref{Fig_suspt}), we plot $\chi_{4}$ as obtained from isoconfigurational ensemble simulations. Interestingly not the value but the nature of $\chi_{4}$ obtained from the isoconfigurational run matches with that obtained from regular MD simulation studies which implies that isoconfigurational runs can capture the correlated particle motion. However when compared with the distribution of particle-level relaxation times, $P(\tau_{\alpha})$(see left panels of Fig.\ref{Fig_suspt}), we find that a broader distribution of $P(\tau_{\alpha})$ does not necessarily correspond to larger values of $\chi_{4}$. This distinction is crucial: while $P(\tau_{\alpha})$ reflects the heterogeneity in relaxation times across individual particles, $\chi_{4}$ captures how correlated are the dynamics of the particles. Our analysis clearly shows that these two aspects of dynamic heterogeneity are not always aligned.

Our results indicate that while fragility generally correlates with $\chi_{4}$, which captures the spatial extent of correlated regions, it does not always align with other measures of dynamic heterogeneity that reflect the width of the distribution of individual particle mobility. In contrast, the structure–dynamics correlation is robustly linked to the width of the mobility distribution and does not depend on the spatial correlation length and, consequently, the fragility.

\subsection{Enthalpy vs entropy contribution}

In this section, our analysis will shed some light on why the structure-dynamics correlation and distribution of mobility for the two systems (LJ at $\rho=1.1$ and $(8,5)$) exhibit similar behaviour, despite having different fragility values. In our previous study involving some of us, we had shown that for an attractive system, the enthalpy, $e_{loc}$, plays a dominant role, and for a repulsive system, the entropy, $S_{2}$, plays a dominant role in determining the dynamics, and our order parameter can pick up both the contributions \cite{Sharma2023_jcp}. We do a similar analysis for the systems studied here.

In Fig.\ref{Fig_gr_potn}(top panels), we plot the overlay between the radial distribution function and the potential for the three sets of systems. We also plot the correlations between different SOPs and the dynamics for the corresponding systems in Fig.\ref{Fig_gr_potn}(bottom panels).

We find that for both the LJ and WCA systems, as density increases, the radial distribution function shifts toward lower values of $r$ (Fig.\ref{Fig_gr_potn}(a) and Fig.\ref{Fig_gr_potn}(b)). In the LJ system, this shift implies that at low density, particles explore the attractive part of the potential, while at high density, they predominantly sample the repulsive part. This transition is reflected in the correlation between local energy, $e_{loc}$, and dynamics (Fig.\ref{Fig_gr_potn}(a)). At $\rho = 1.1$, the system lies in the enthalpy-dominated regime, and as density increases, the correlation between $e_{loc}$ and the dynamics drops sharply. Interestingly, in the high-density regime, where the repulsive part of the potential dominates, the correlation between entropy ($S_2$) and dynamics becomes strong. Beyond a certain density, however, this correlation plateaus and shows little further variation.

For the WCA system, due to the nature of the potential, particles explore only the repulsive regime at all densities. Consequently, enthalpy exhibits almost no correlation with the dynamics. In contrast, entropy shows a moderate correlation with dynamics, which remains largely unchanged across densities.

In the modified LJ $(q,p)$ system, increasing the softness of the potential results in a broader RDF, allowing particles in softer systems to access more of the attractive part of the potential. As a result, in the $(8,5)$ system, enthalpy plays a dominant role in controlling the dynamics. In contrast, in the $(12,11)$ system, the correlation between $e_{\text{loc}}$ and dynamics is weak. We also find that as the potential becomes stiffer, the systems become increasingly entropy-driven. Notably, our SOP is capable of capturing both entropic and enthalpic contributions to the dynamics.

In Fig.\ref{Fig_slope} (b) where we plot the peak values of the Spearman rank correlation between the SOP and dynamics against the $B/A(micro)$, we find that all the systems which are strongly entropy driven, the WCA systems and LJ systems at $\rho=1.4$ and $\rho=1.6$ are all clustered together in the left hand corner of the plot and the enthalpy driven systems are on the right hand corner.

\section{Conclusion}
\label{conclusion_section}
The main objective of this study is to investigate the relationship between structure–dynamics correlation and fragility in supercooled liquids. Fragility is often associated with enhanced dynamic heterogeneity \cite{Sastry2001_nature, Ediger2000_Physchem}. Thus, if structural features significantly influence dynamics, one would expect systems with higher fragility to also exhibit stronger structure–dynamics correlation. In a previous study involving some of us, it was shown that the slope of the macroscopic free energy barrier, obtained from average dynamics as a function of the structural order parameter (SOP), correlates with fragility \cite{Nandi2021_prl}. A more recent study demonstrated that the slope of the microscopic free energy barrier, when plotted against a machine-learned softness parameter, also increases with fragility \cite{Tah2022_jcp}. Physically, the slope of the free energy barrier versus SOP quantifies how sensitively the barrier, and thus the dynamics, responds to structural changes: a higher slope indicates that even small variations in SOP lead to significant changes in the barrier, reflecting stronger structure–dynamics coupling.

To explore this connection in a general and systematic manner, we consider three distinct sets of model systems: (i) the attractive Lennard-Jones (LJ) system, (ii) the purely repulsive Weeks–Chandler–Andersen (WCA) system, where fragility is tuned by varying density, and (iii) the modified LJ $(q,p)$ system, where fragility is modulated by changing the softness of the interaction potential. This diversity enables us to decouple effects such as density and interaction range from the fragility itself.

Our SOP is observed to vary linearly with temperature, which allows the relaxation time to be fitted in terms of the SOP using a Vogel-Fulcher-Tammann (VFT) form\cite{GarciaColin1989_vft}. The fragility extracted from this SOP-based VFT expression is consistent with the temperature-based fragility. However, a key insight from this analysis is that the slope of the free energy barrier as a function of SOP does not necessarily scale with fragility, as often assumed \cite{Nandi2017_prl,Tah2022_jcp}. We also find that the barrier contains a temperature/SOP independent term, which sets the high-temperature limit of the activation energy and affects the slope when unaccounted for. At the macroscopic level, this scaled slope does align with fragility trends. However, at the microscopic level, the relationship is not universal. Notably, an independent and complementary analysis based on isoconfigurational ensemble simulations reveals that the structure–dynamics correlation correlates well with the microscopic scaled slope, reinforcing the idea that high structure–dynamics correlation does not necessarily coincide with high fragility.

We further observe that structure–dynamics correlation is strongest in systems where the dynamics are enthalpy driven. The systems with the highest correlation are the LJ system at $\rho=1.1$ and the $(8,5)$ system; the former is the least fragile within its class, while the latter is the most fragile. Despite this contrast, both systems show broad mobility distributions, bimodal displacement profiles, and high values of the non-Gaussian parameter.

This raises the question of whether enthalpy dominance is intrinsically linked to broad mobility distributions and a form of heterogeneity. Interestingly, both the LJ at $\rho=1.1$ and the $(8,5)$ systems lie close to the spinodal line \cite{Sastry2000_prl_press, Sengupta2011_jcp}(see Appendix V), a regime known to amplify dynamic heterogeneity due to mechanical instability and growing fluctuations \cite{Sastry2000_prl_press}. Although the mechanisms underlying heterogeneity near the spinodal and in supercooled liquids differ, both are characterised by spatially heterogeneous mobility. In systems like $(8,5)$, high fragility and spinodal proximity may act in tandem to produce pronounced dynamical heterogeneity.

In conclusion, our findings reveal that fragility and structure–dynamics correlation do not always correlate. We would like to clarify an important distinction regarding different measures of dynamic heterogeneity. The peak of the four-point susceptibility, $\chi_{4}$, captures the spatial extent of correlated regions in the dynamics and is commonly used to quantify heterogeneity that grows with fragility. In contrast, other indicators of dynamic heterogeneity, such as the distribution of particle mobility or the non-Gaussian parameter, reflect the contrast in individual particle mobilities, which can be substantial even when $\chi_{4}$ is small. Our results show that the structure–dynamics correlation is more strongly associated with this contrast in mobility than the spatial correlation length captured by $\chi_{4}$. Consequently, systems with high structure–dynamics correlation do not necessarily exhibit high fragility or large $\chi_{4}$ values. This distinction highlights that different aspects of dynamic heterogeneity need not vary together and that structure–dynamics correlation can be strong even in systems with low fragility.\\
\\
\textbf{Appendix I: Structural order parameter at macroscopic and microscopic levels}\\
\\
In a binary system, at the microscopic level, the effective caging potential experienced by particle \( i \) of type $u$ is given by
\begin{equation}
\beta\phi^i_u = -\rho \int \textbf{dr} \sum_{v} x_v C_{uv}^i(r)\, g_{uv}^i(r),
\label{eq:micro_sop}
\end{equation}
where \( \rho \) is the number density, \( C_{uv}^i(r) \) is the local direct correlation function, and \( g_{uv}^i(r) \) is the local radial distribution function for particle \( i \). To maintain consistency and avoid unphysical contributions in the microscopic calculations, particularly at small \( r \), we adopt an approximation for the direct correlation function. This has been discussed in detail in earlier studies \cite{Sharma2023_jcp, Patel2023_jcp}. Specifically, we replace the HNC expression with a simplified form:
\begin{equation}
C_{uv}^{\text{approx}}(r) = g_{uv}(r) - 1,
\end{equation}
and use this same approximation consistently in both macroscopic and microscopic calculations. 

We work with the absolute value of the caging potential, and the SOP, which is the inverse of the depth of the caging potential, is given by, 
\begin{equation}
SOP = (\beta \phi^i_u)^{-1} = 1/\Big(\rho \int \textbf{dr} \sum_{v} x_v C_{uv}^i(r)\, g_{uv}^i(r)\Big)
\label{eq:micro_sop}
\end{equation}
This gives rise to a distribution of the SOP \cite{Sharma2022_pre}. 
The average of this distribution can be written as,
\begin{equation}
\begin{split}
&\langle (\beta\phi^{i}_u)^{-1}\rangle = \left\langle 1/\Big(\rho\int \textbf{dr}\, \sum_{v} x_v C^{i}_{uv}(r) g_{uv}^i(r)\Big) \right\rangle_i\\
&= 1/\Big(\rho\int \textbf{dr}\, \sum_{v} \langle C^{i}_{uv}(r) g_{uv}^i(r) \rangle x_v\Big)
\end{split}
\end{equation}
We present the temperature dependence of the SOP for all three sets of systems studied (LJ, WCA, and modified LJ $(q,p)$ system). 
In the right panels of Fig.\ref{Fig_micro} we show that both $1/(\beta\Phi)$ and $<1/(\beta\phi)>$ exhibit a linear dependence on temperature across all systems. This observation is consistent with earlier studies~\cite{Nandi2021_prl,Sharma2022_pre}. Although the macroscopic and microscopic SOP have the same functional forms, it is important to note that since the SOP is a nonlinear function of the radial distribution function, the average of the microscopic SOP is quantitatively not the same as the macroscopic SOP(Eq.\ref{eq:macro_sop}). Although $\langle g^{i}_{uv}(r) \rangle=g_{uv}(r)$, $\langle C^{i}_{uv}(r) g^{i}_{uv}(r) \rangle
\ne C_{uv}(r) g_{uv}(r)$, i.e., the average of a product is not equal to the product of averages. As a result, the macroscopic SOP \((\beta \Phi)^{-1} \) and microscopic average SOP \( \langle(\beta\phi)^{-1} \rangle \) have different values but interestingly similar trend as seen in right panels of Fig.\ref{Fig_micro}.\\
\\
\textbf{Appendix II: Identification of fast particles and calculation of microscopic activation energy }\\
\\
To identify dynamically fast particles and calculate their microscopic activation energy, we follow the method introduced by Candelier et al.~\cite{Candelier2010_prl,Smessaert2013_pre}. This approach allows us to detect cage rearrangements by analysing particle trajectories in a fixed time window. 
For each particle $i$ at time $t$, a hopping propensity $p_{\text{hop}}(i,t)$ is computed over a window $W = [t_1, t_2]$ using the expression:
\begin{equation}
p_{\text{hop}}(i,t) = \sqrt{
  \left\langle (\vec{r}_i - \left\langle \vec{r}_i \right\rangle_U)^2 \right\rangle_V
  \left\langle (\vec{r}_i - \left\langle \vec{r}_i \right\rangle_V)^2 \right\rangle_U},
  \label{phop}
\end{equation}
where the time intervals $U = [t - \Delta t / 2, t]$ and $V = [t, t + \Delta t / 2]$ represent regions before and after time $t$, respectively. In our calculations, we use $\Delta t = 15\tau$. A low $p_{\text{hop}}$ value indicates the particle is localised within a cage, whereas a high value signifies a cage jump.
To determine a threshold $p_c$ for identifying rearranging particles, we use the value of mean square displacement (MSD) at which the non-Gaussian parameter $\alpha_2(t)$ attains its maximum\cite{Landes2020_pre}. The formal definition of $\alpha_2(t)$ is provided in Section\ref{section_alpha2}, and the corresponding time is used to set the relevant rearrangement scale.
Any particle $i$ with $p_{\text{hop}}(i,t) > p_c$ is categorised as dynamically fast and said to undergo a rearrangement at time $t$.
We then analyse the correlation between structure and dynamics using the probability of rearrangement, $P_{R}$, as a function of the inverse SOP. We denote this probability as $P_{R}((\beta\phi))$, where $\beta\phi$ represents the inverse of the SOP for a given particle. In Fig.\ref{Fig_calculation}(a), we plot $P_{R}(\beta\phi)$ versus $(\beta\phi)$ for different temperatures. At high temperatures, the curves appear flat, indicating that rearrangement is nearly independent of local structure. However, at lower temperatures, a clear dependence emerges—particles with smaller $\beta\phi$ values are significantly more likely to rearrange.
\begin{figure}
\centering
\begin{subfigure}{0.225\textwidth}
\includegraphics[width=0.98\linewidth]{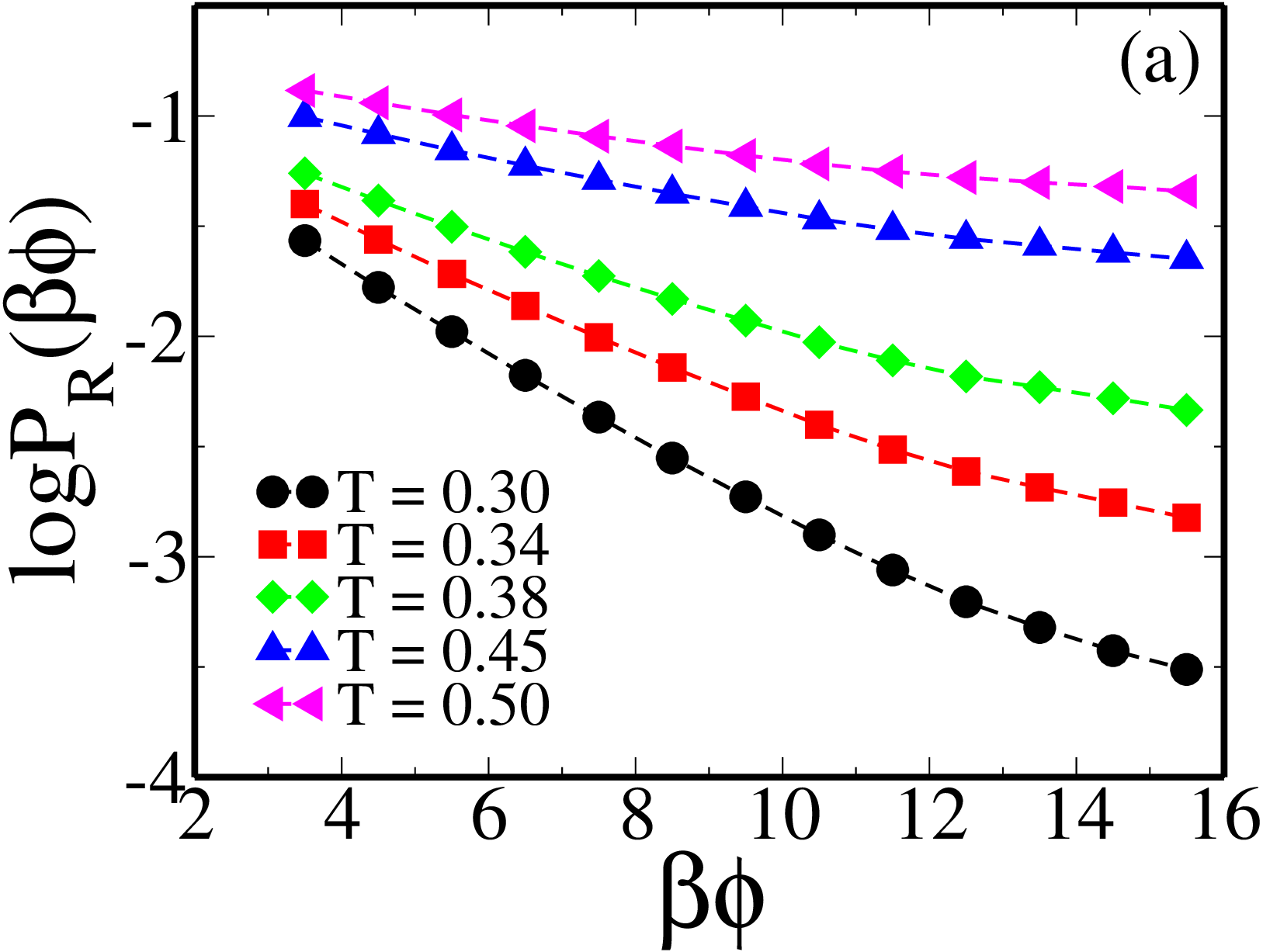}
\end{subfigure}
\hspace{0.2cm}
\begin{subfigure}{0.225\textwidth}
\includegraphics[width=0.98\linewidth]{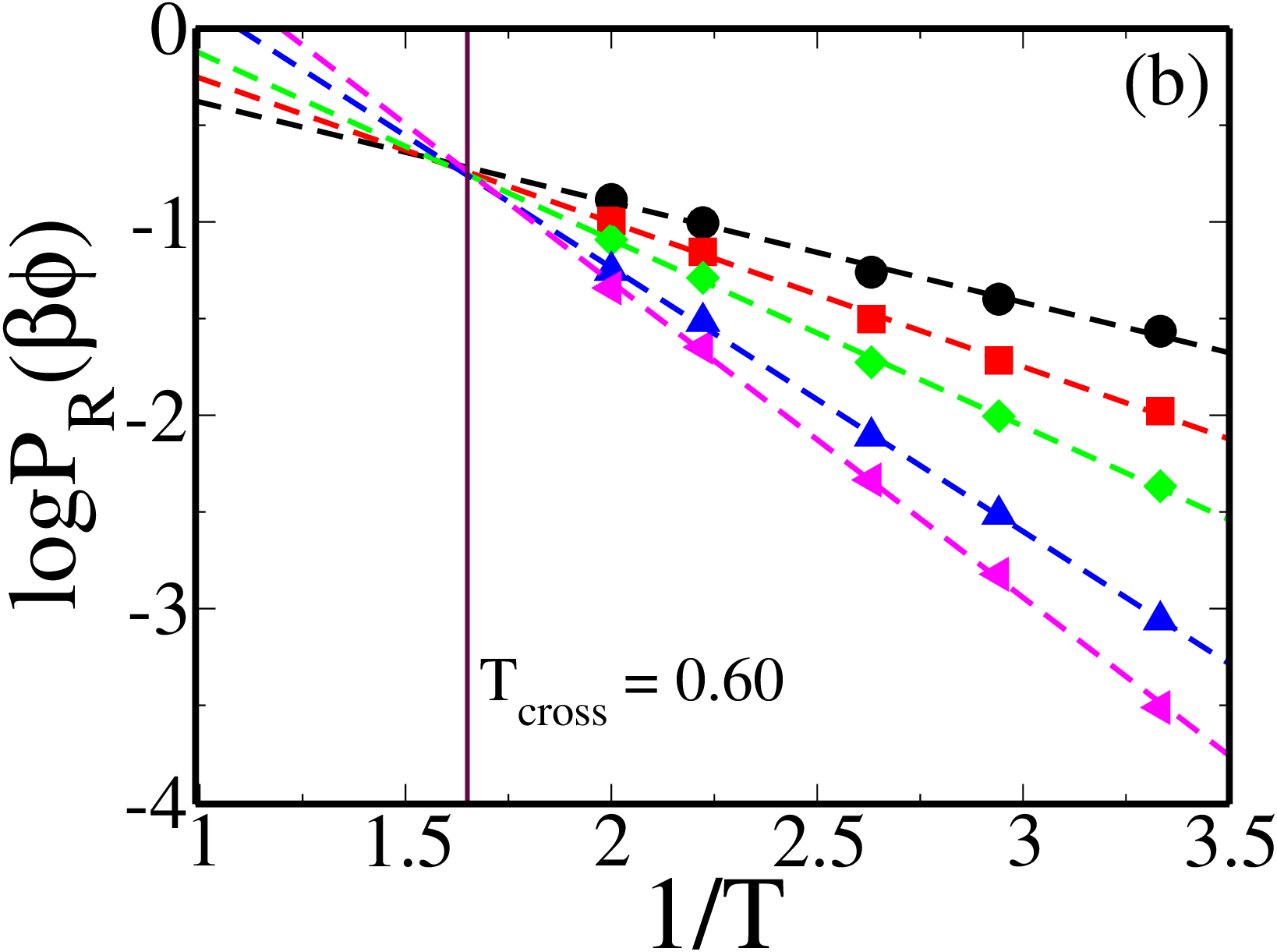}
\end{subfigure}
\vspace{0.2cm}
\caption{\textit{(a) Plot of $\log P_{R}$ as a function of inverse structural order parameter, $\beta\phi$, at various temperatures.(b) Plot of $\log P_{R}$ versus inverse temperature $1/T$ for several fixed $(\beta\phi)$ values. Each curve corresponds to a different inverse SOP (i.e., $(\beta\phi)$) value, with black representing low inverse SOP values and magenta representing high inverse SOP values. The vertical solid line indicates the temperature at which all curves intersect, and this cross temperature is the onset temperature for all systems\cite{Schoenholz2015_nature,Sharma2022_pre}.
}
}
\label{Fig_calculation}
\end{figure}

To extract the microscopic activation energy, we analyse the temperature dependence of the rearrangement probability for fixed SOP values. Specifically, we fit $P_{R}(\beta\phi)$ as a function of inverse temperature $1/T$ for several values of $(\beta\phi)$ using the Arrhenius form:
\begin{equation}
  P_{R}(\beta\phi) = P_0(\beta\phi) \exp\left[-\frac{\Delta E(\beta\phi)}{T}\right],
\label{prs}
\end{equation}
where $\Delta E(\beta\phi)$ is interpreted as the microscopic activation energy required for rearrangement and depends on the local structure via $(\beta\phi)$. This is shown in Fig.\ref{Fig_calculation}(b), which demonstrates the increasing structural control over dynamics at lower temperatures.
This method thus establishes a direct quantitative link between local structural features and the free energy barrier for particle rearrangements.\\
\\
\textbf{Appendix III: Macroscopic activation energy as a function of temperature}\\
\\
The temperature dependence of the barrier $\Delta E^{ma}=T*ln(\tau_{\alpha}/\tau_{0})$ as predicted by the VFT expression (Eq.\ref{vft_T}) can be written as,
\begin{equation}
 \Delta E^{ma}/T = \frac{T_{VFT}}{K_{VFT}(T-T_{VFT})}
\end{equation}
 \noindent
In the range away from $T_{VFT}$, we can expand $(1 - T_{VFT}/T)$ by Taylor expansion, and using the first-order approximation, we can write
\begin{equation}
 \Delta E^{ma} \approx \frac{T_{VFT}}{K_{VFT}}+ \frac{T_{VFT}^2}{K_{VFT}}.\frac{1}{T}
\label{deltaE_T}
\end{equation} 
This clearly shows that the slope of $\Delta E^{ma}$ is not directly proportional but inversely proportional to the fragility. For both the LJ and WCA systems, fragility increases with density, accompanied by a corresponding rise in $T_{VFT}$ as the systems at higher densities operate in higher temperature regimes. This trend explains why we observe an increase in the slope, $\frac{T_{VFT}^2}{K_{VFT}}$, with increasing fragility. For the modified LJ $(q,p)$ system the fragility is higher for the $(8,5)$ system which has the lowest value of $T_{VFT}$ and this is the reason even in the $\Delta E^{ma}$ vs $1/T$ plots the $(8,5)$ system has the slowest growth (left panels of Fig.\ref{Fig_Macro_extra}). Interestingly the slope of $1/\Delta E^{ma}$ vs $1/T$ is the fragility (right panels of Fig.\ref{Fig_Macro_extra}),
\begin{equation}
 1/\Delta E^{ma} = \frac{K_{VFT}}{T_{VFT}} - K_{VFT}.\frac{1}{T}
\label{ideltaE_T}
\end{equation}

The slopes of these linear fits yield the fragility, $K_{\mathrm{VFT}}$, which exactly matches the fragility values obtained from the VFT fit of the relaxation time using Eq.\ref{vft_T}. The extracted values are listed in Table~\ref{tab:vftfragility}.
\\
\begin{figure}
\centering
\begin{subfigure}{0.225\textwidth}
\includegraphics[width=0.98\linewidth]{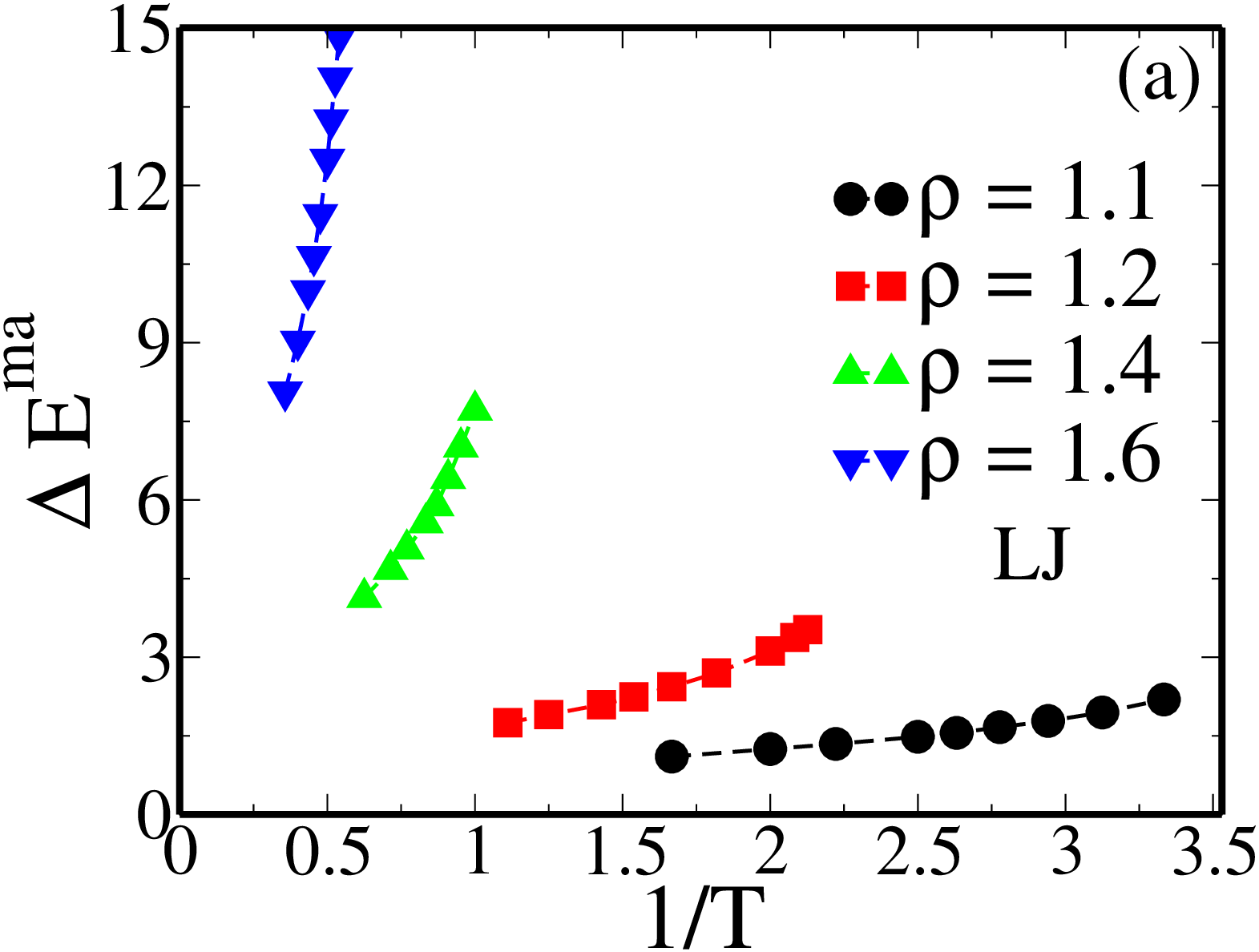}
\end{subfigure}
\hspace{0.2cm}
\begin{subfigure}{0.225\textwidth}
\includegraphics[width=0.98\linewidth]{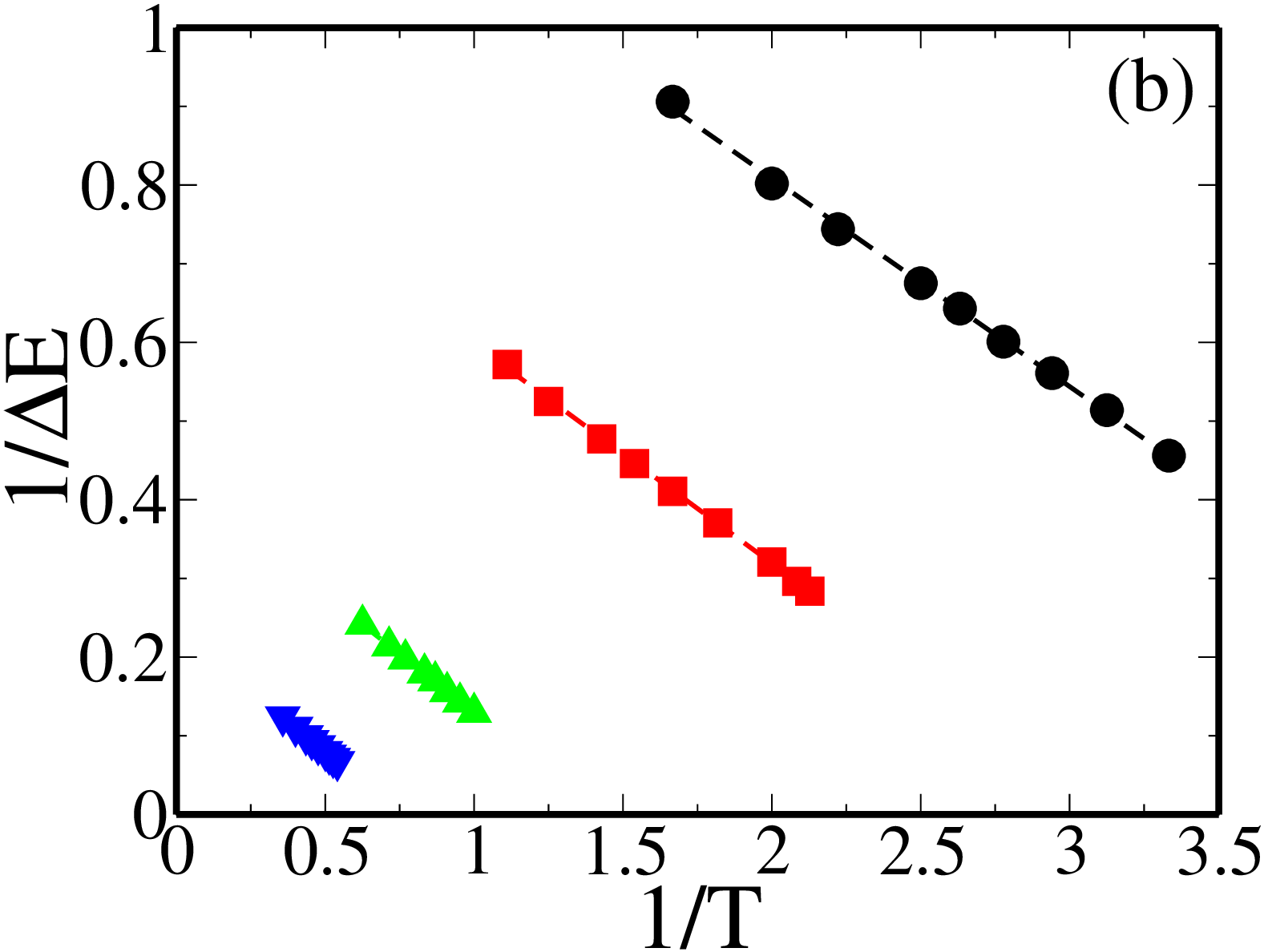}
\end{subfigure}
\vspace{0.2cm}
\begin{subfigure}{0.225\textwidth}
\includegraphics[width=0.98\linewidth]{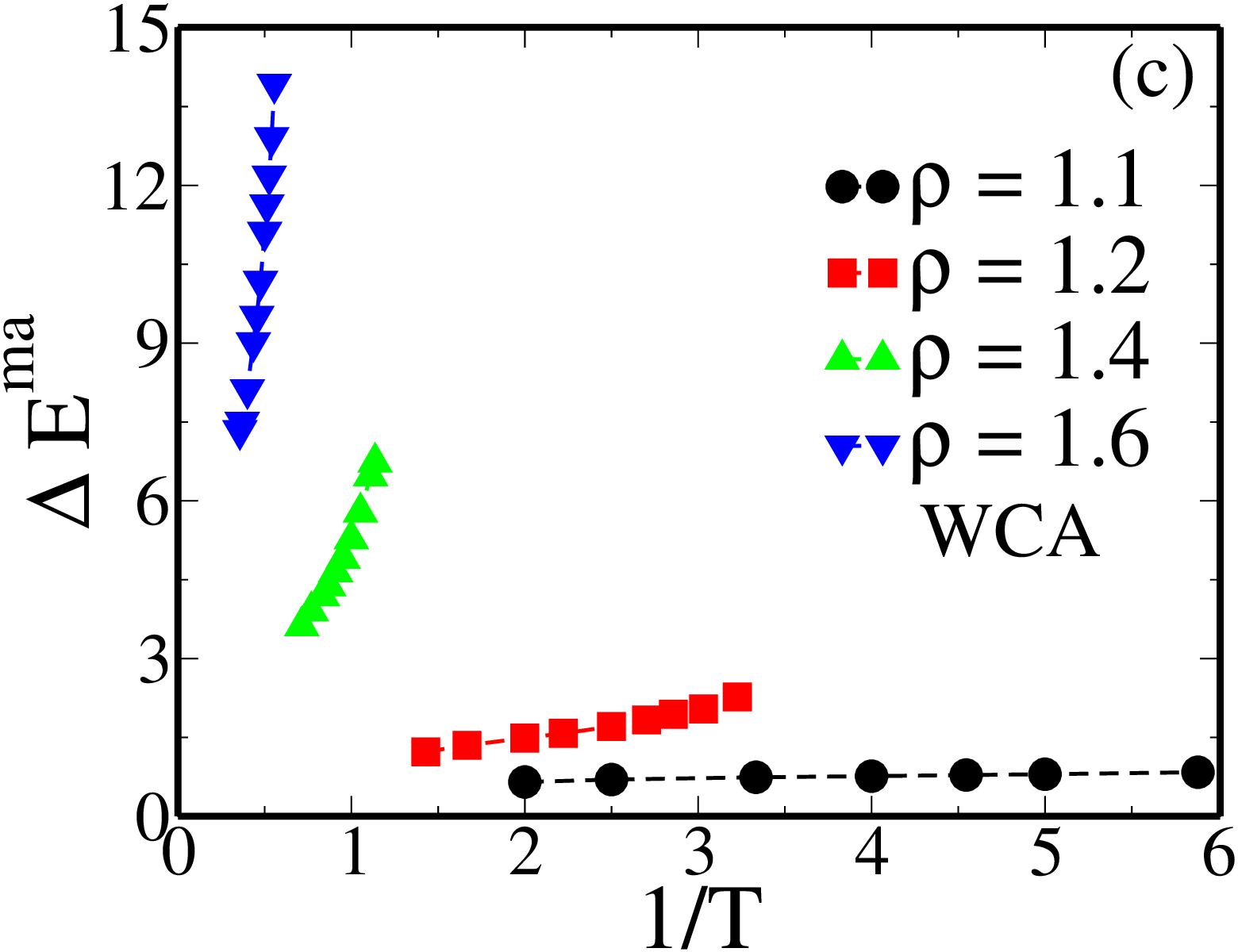}
\end{subfigure}
\hspace{0.2cm}
\begin{subfigure}{0.225\textwidth}
\includegraphics[width=0.98\linewidth]{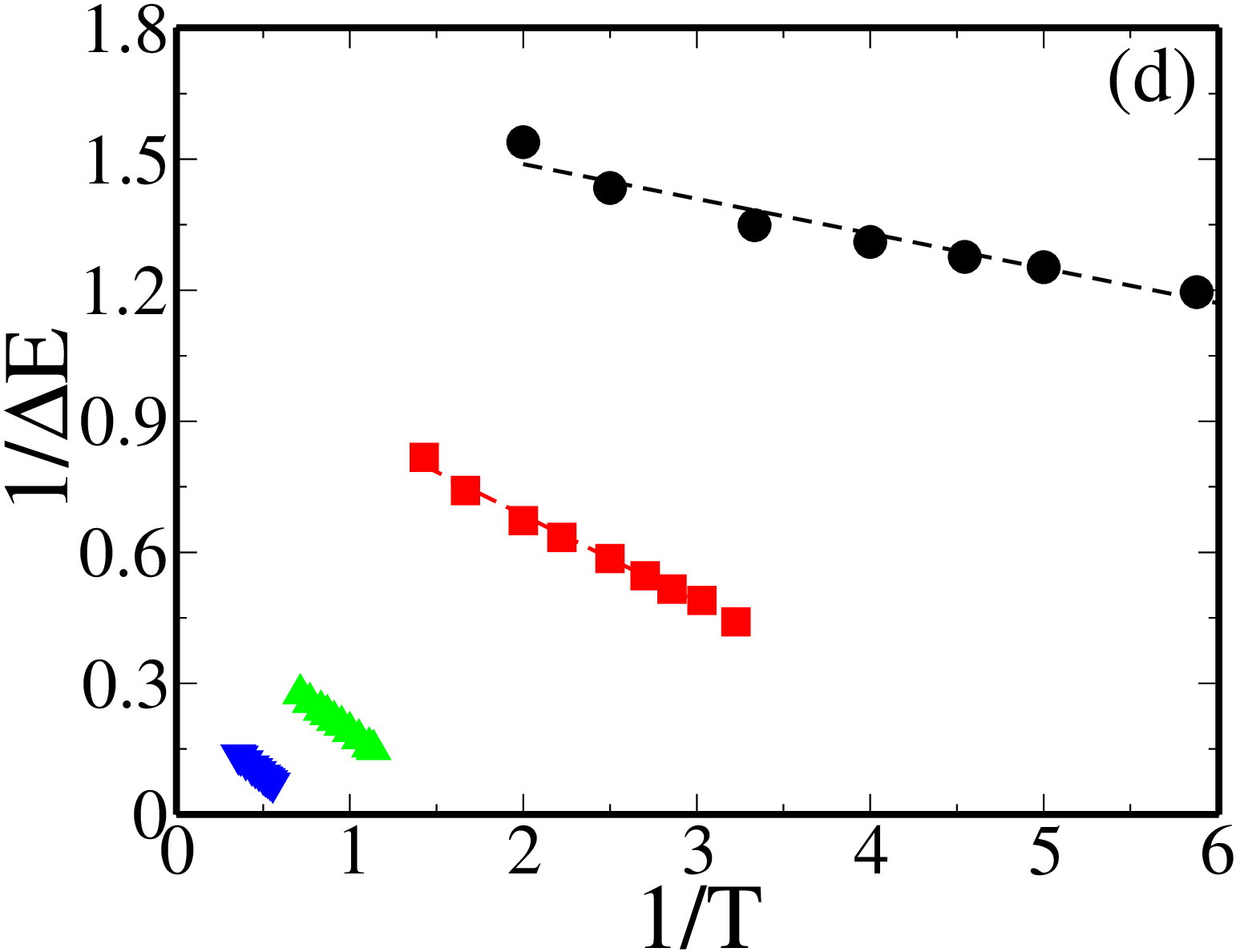}
\end{subfigure}
\vspace{0.2cm}
\begin{subfigure}{0.225\textwidth}
\includegraphics[width=0.98\linewidth]{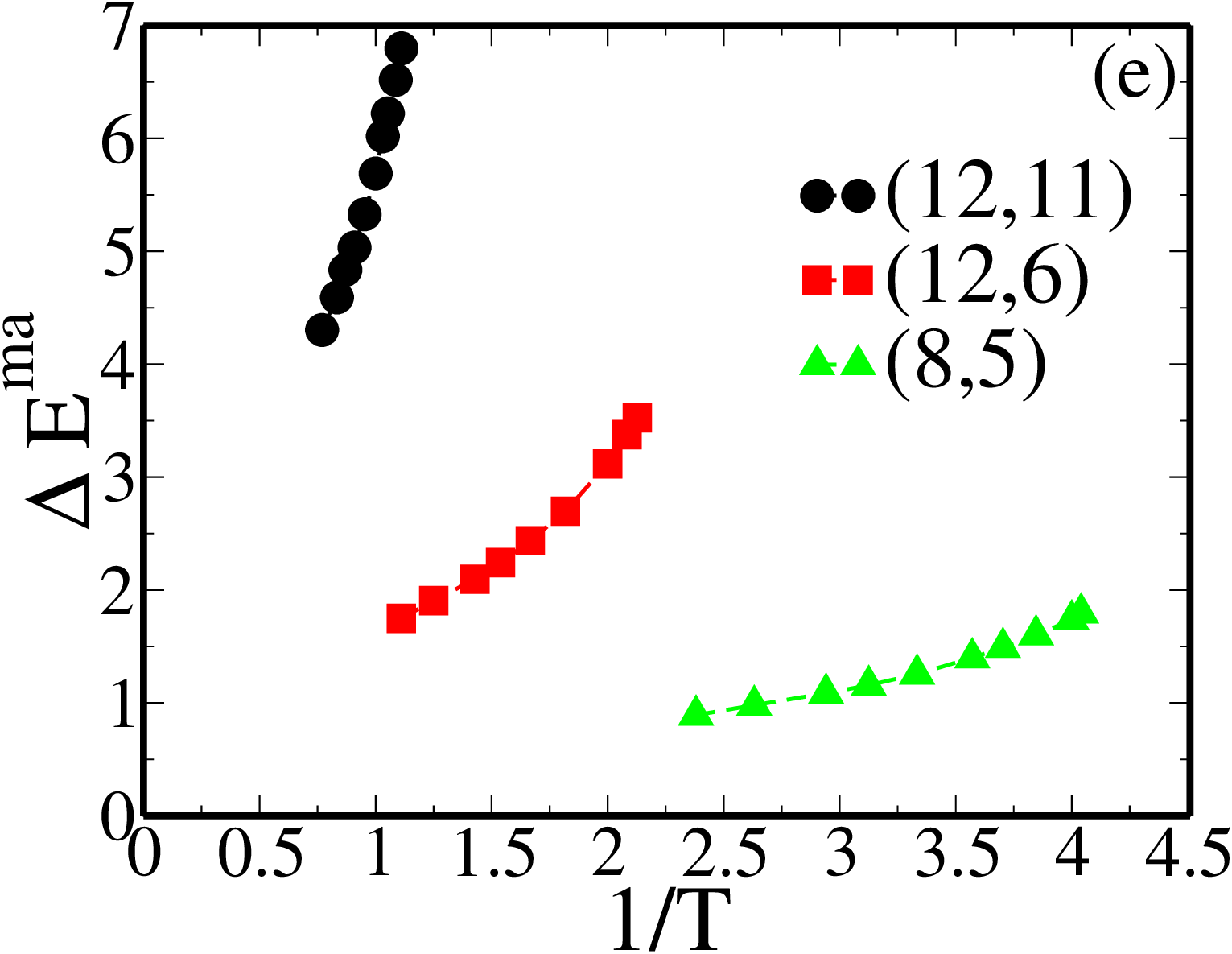}
\end{subfigure}
\hspace{0.2cm}
\begin{subfigure}{0.225\textwidth}
\includegraphics[width=0.98\linewidth]{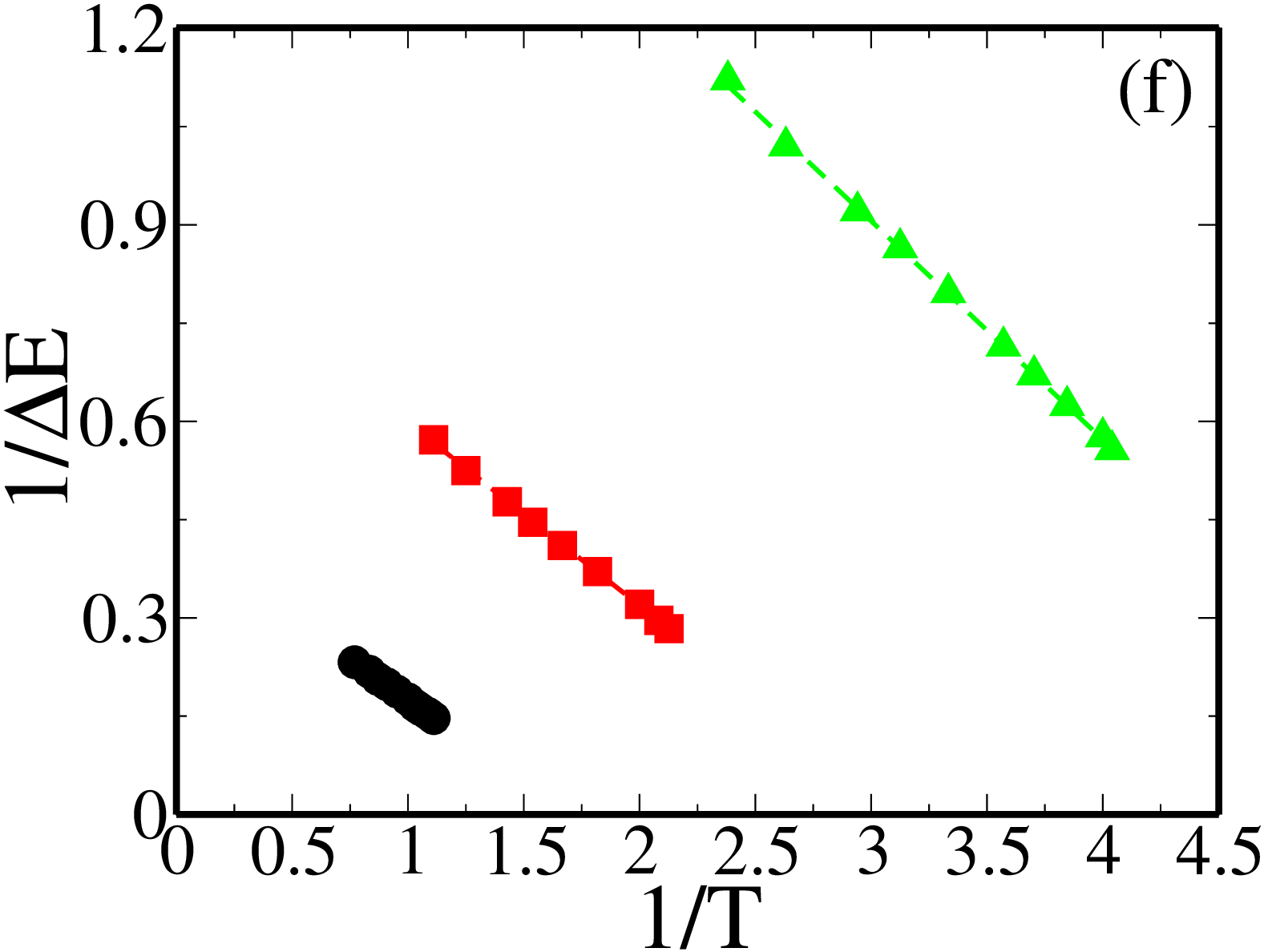}
\end{subfigure}
\vspace{0.2cm}
\caption{\textit{\textbf{Left panels:} Variation of the macroscopic activation energy $\Delta E^{ma}$ with inverse temperature $1/T$ for three sets of systems: (a)LJ and (c)WCA, each examined at four different densities, and (e) modified LJ $(q,p)$ system. \textbf{Right panels:}The inverse activation energy, $1/\Delta E^{ma}$, is plotted against $1/T$ for the same systems. The slope of the resulting linear fits provides the fragility, $K_{\mathrm{VFT}}$ as shown in Eq.\ref{ideltaE_T}. The corresponding values are listed in Table~\ref{tab:vftfragility}. The colour coding in the right panels matches that of the left panels.}}

\label{Fig_Macro_extra}
\end{figure}
\begin{table}[h]
\centering
\caption{\textit{Slope ($B'$) and intercept ($A'$) values from the linear fit of $\frac{1}{\Delta E^{\mathrm{ma}}}$ vs. $1/\Delta((\beta\Phi)^{-1})$ (right panels of Fig.\ref{Fig_macro}), and slope ($B(micro)$) and intercept ($A(micro)$) values from the fit of $\Delta E$ vs. $1/\Delta((\beta\phi)^{-1})$ (left panels of Fig.\ref{Fig_deltaE_micro}). The corresponding values for each system are listed in the table.
}}
\vspace{0.2cm}
\label{tab:slope_intercept}
\begin{tabular}{|c|c|c|c|c|}
\hline
\textbf{System} & $\boldsymbol{A'}$ & $\boldsymbol{B'}$ & $\boldsymbol{A}(micro)$ & $\boldsymbol{B}(micro)$ \\
\hline
LJ, $\rho = 1.1$  & 1.320 & 0.0177 & 0.528 & 0.146\\
\hline
LJ, $\rho = 1.2$  & 0.869 & 0.0123 & 1.968 & 0.198\\
\hline
LJ, $\rho = 1.4$  & 0.426 & 0.0066 & 5.654 & 0.395\\
\hline
LJ, $\rho = 1.6$  & 0.234 & 0.0038 & 11.99 & 0.784\\
\hline
WCA, $\rho = 1.1$ & 1.614 & 0.0057 & 0.501 & 0.033\\
\hline
WCA, $\rho = 1.2$ & 1.067 & 0.0098 & 1.352 & 0.117\\
\hline
WCA, $\rho = 1.4$ & 0.486 & 0.0070 & 4.632 & 0.321\\
\hline
WCA, $\rho = 1.6$ & 0.251 & 0.0039 & 10.50 & 0.822\\
\hline
$(12,11)$         & 0.425 & 0.0052 & 3.815 & 0.360\\
\hline
$(12,6)$          & 0.948 & 0.0143 & 2.030 & 0.196\\
\hline
$(8,5)$             & 1.893 & 0.0368 & 0.827 & 0.124\\
\hline
\end{tabular}
\label{AB}
\end{table}
\\
\textbf{Appendix IV: Structure–dynamics correlation: Isoconfigurational ensemble and rank analysis }\\
\\
To investigate the correlation between structure and dynamics in our system, we use the isoconfigurational ensemble method introduced by Harrowell et al.~\cite{Cooper2008_nature}. In this method, an ensemble of trajectories is generated from the same initial configuration of particles, with initial velocities drawn randomly from the Maxwell-Boltzmann distribution at the corresponding temperature. This approach eliminates trivial variations in dynamics due to different initial momenta, allowing us to assess the role of structural features in determining the dynamics.

In our study, we use 32 different initial configurations, each separated by at least $75\tau_{\alpha}$ to ensure statistical independence. From each configuration, we generate 100 trajectories with randomised initial momenta. The dynamics of each particle are quantified by its mobility, defined as:
\begin{equation}
\mu^{j}(t) = \frac{1}{N_{\text{IC}}} \sum_{i=1}^{N_{\text{IC}}} \left| \mathbf{r}_{i}^{j}(t) - \mathbf{r}_{i}^{j}(0) \right|,
\end{equation}
where $\mu^j(t)$ is the average displacement of the $j^\text{th}$ particle over $N_{\text{IC}}$ isoconfigurational trajectories at time $t$.
We have also calculated the per-particle $\alpha$-relaxation time by averaging over 200 isoconfigurations for each particle.
In the left panels of Fig.\ref{Fig_partau}, we present the distribution of the per-particle $\alpha$-relaxation times. We find that the distribution of the mobility and the relaxation time show similar behaviour. 
\begin{figure}
\centering
\begin{subfigure}{0.225\textwidth}
\includegraphics[width=0.98\linewidth]{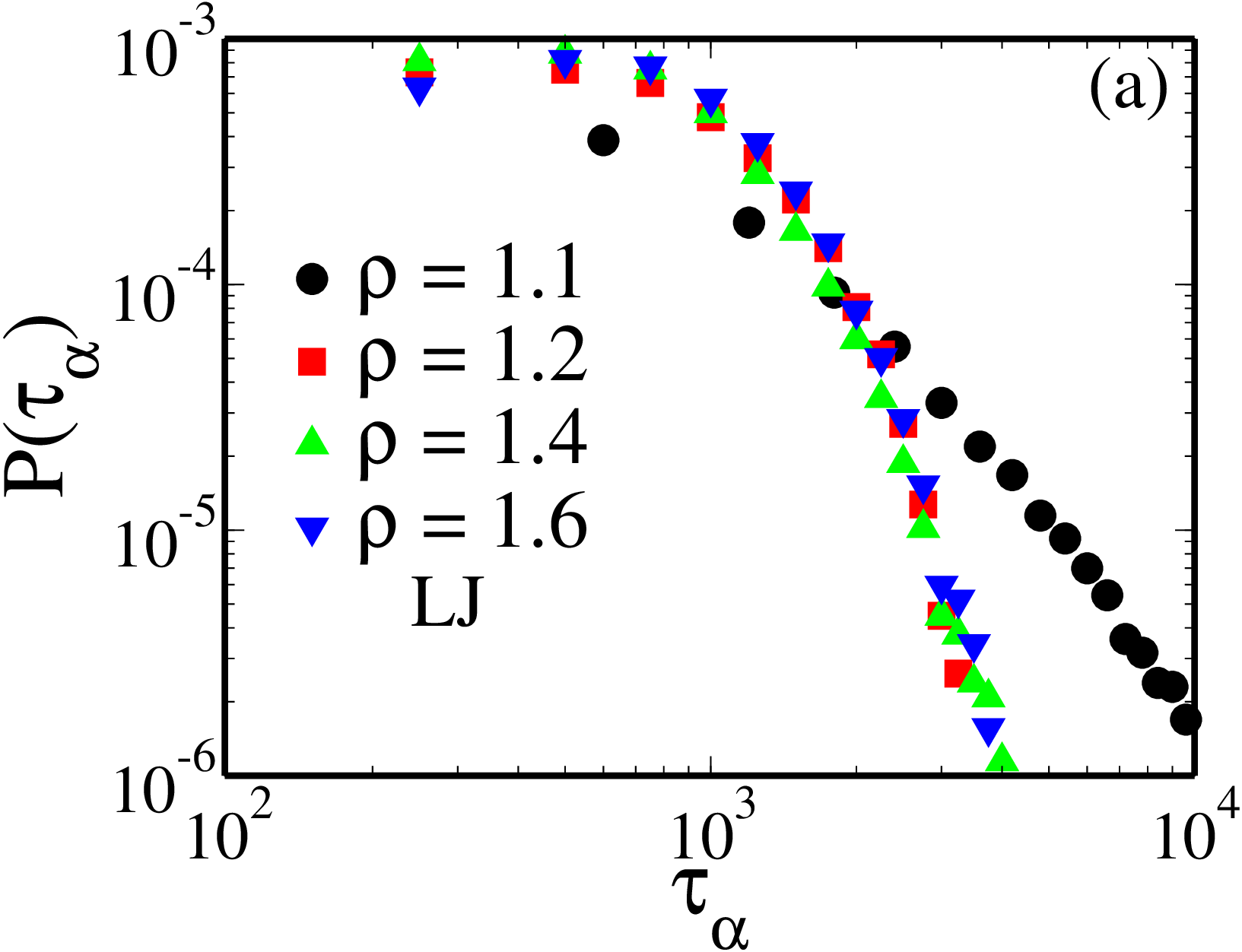}
\end{subfigure}
\hspace{0.2cm}
\begin{subfigure}{0.225\textwidth}
\includegraphics[width=0.98\linewidth]{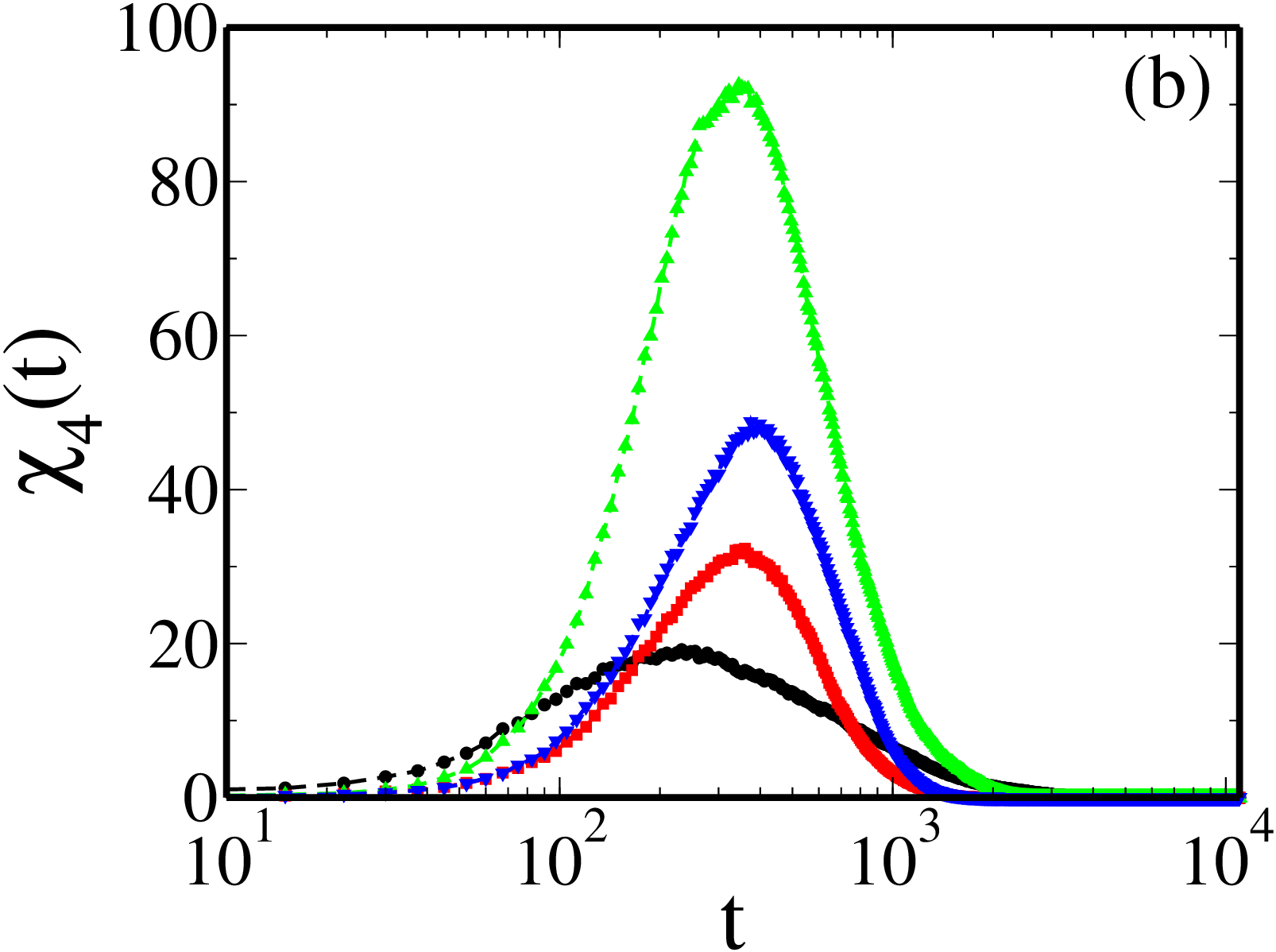}
\end{subfigure}
\vspace{0.2cm}
\begin{subfigure}{0.225\textwidth}
\includegraphics[width=0.98\linewidth]{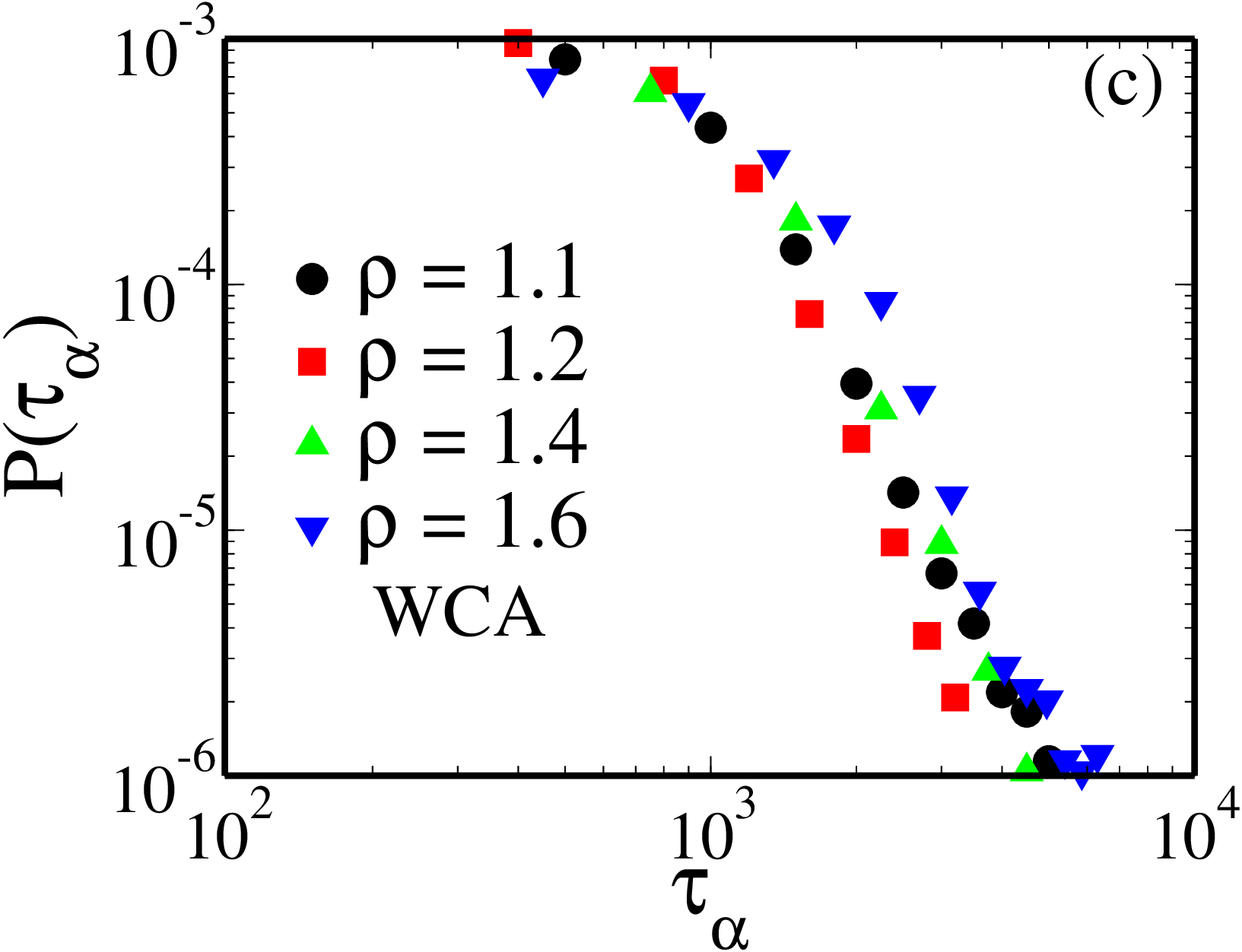}
\end{subfigure}
\hspace{0.2cm}
\begin{subfigure}{0.225\textwidth}
\includegraphics[width=0.98\linewidth]{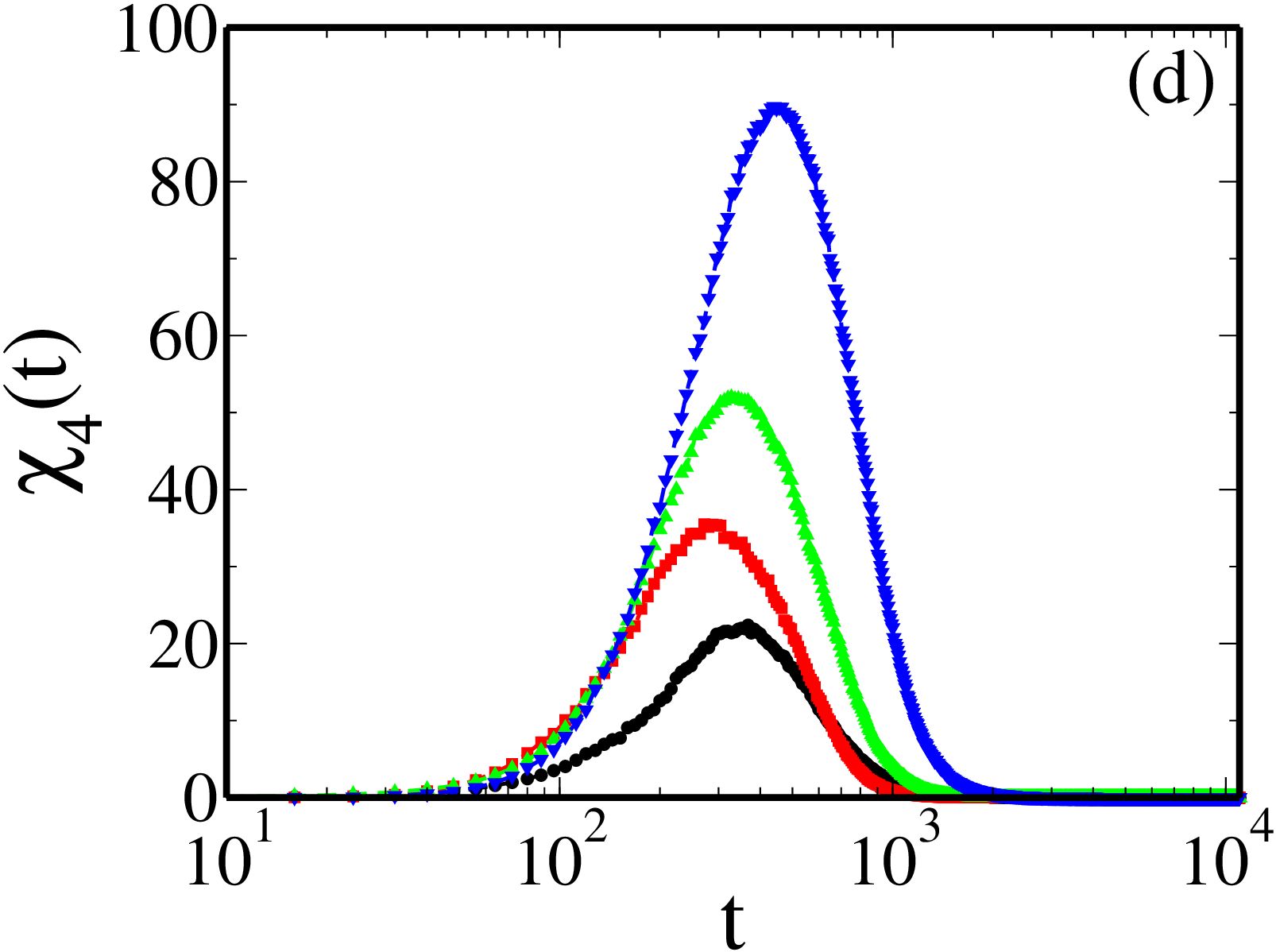}
\end{subfigure}
\vspace{0.2cm}
\begin{subfigure}{0.225\textwidth}
\includegraphics[width=0.98\linewidth]{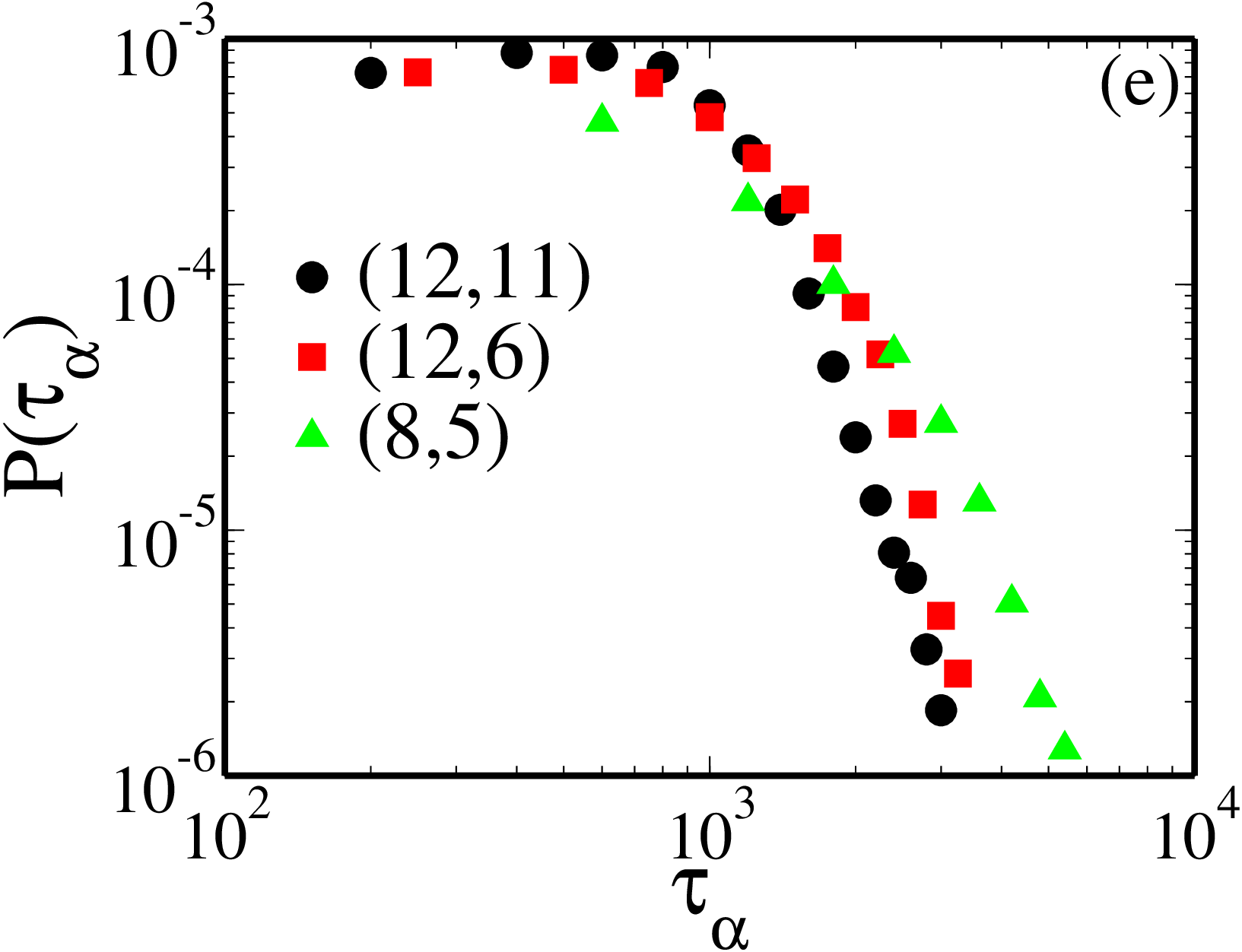}
\end{subfigure}
\hspace{0.2cm}
\begin{subfigure}{0.225\textwidth}
\includegraphics[width=0.98\linewidth]{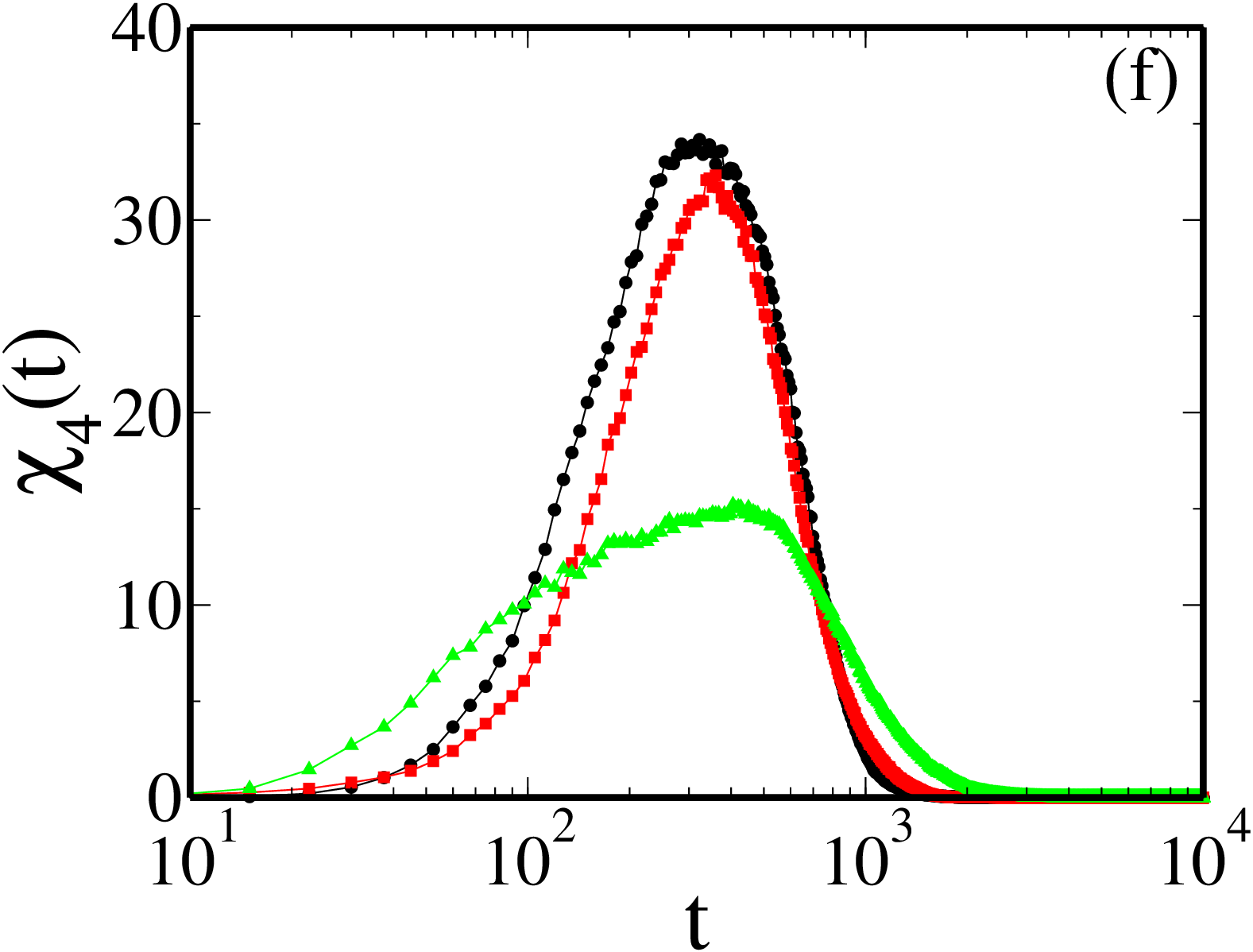}
\end{subfigure}
\vspace{0.2cm}
\caption{\textit{\textbf{Left panels:} Distribution of per-particle $\alpha$-relaxation times obtained from isoconfigurational ensemble simulations for three sets of systems: (a)LJ and (c) WCA, each examined at four different densities, and (e) modified LJ $(q,p)$ system. \textbf{Right panels:} Corresponding four-point susceptibility, $\chi_4(t)$, plotted as a function of time for the same systems shown in the left panels. The colour coding in the right panels matches that of the left panels.
}}
\label{Fig_suspt}
\end{figure}

In right panels of Fig.\ref{Fig_suspt}, we plot $\chi_{4}$ as obtained from isoconfigurational ensemble simulations. Interestingly, not the value but the nature of $\chi_{4}$ obtained from the isoconfigurational run matches with that obtained from regular MD simulation studies, which implies that isoconfigurational runs can capture the correlated particle motion. However, when compared with the distribution of particle-level relaxation times, $P(\tau_{\alpha})$ (see left panels of Fig.\ref{Fig_suspt}), we find that a broader distribution of $P(\tau_{\alpha})$ does not necessarily correspond to larger values of $\chi_{4}$. This distinction is crucial: while $P(\tau_{\alpha})$ reflects the heterogeneity in relaxation times across individual particles, $\chi_{4}$ captures how correlated the dynamics of the particles are. Our analysis clearly shows that these two aspects of dynamic heterogeneity are not always aligned.

To quantify the correlation between structural parameters (such as the SOP) and dynamics (such as mobility), we use the Spearman rank correlation coefficient. Given $m$ data points $\{X_i\}$ and $\{Y_i\}$, the Spearman coefficient is defined as:
\begin{equation}
C_{R}(X,Y) = 1 - \frac{6 \sum_{i=1}^{m} d_i^2}{m(m^2 -1)},
\label{rank_corrl}
\end{equation}
where $d_i = R(X_i) - R(Y_i)$ is the difference in ranks of the $i^\text{th}$ values of variables $X$ and $Y$.

This correlation coefficient is used to assess how well the SOP of particles in the initial configuration correlates with their mobility at later times. A strong (positive or negative) correlation indicates that the structure of a particle's local environment significantly influences its dynamic behaviour.

To demonstrate that the qualitative nature of the structure-dynamics correlation for different systems is not specific to any particular choice of $L$, we show in the left panels of Fig.\ref{Fig_partau} the same correlations as presented in Fig.\ref{Fig_correlation} but calculated at a fixed $L = 1$. The similar trends observed across densities confirm that our findings are robust and do not depend sensitively on the choice of $L$. The right panels of Fig.\ref{Fig_partau} shows the Spearman rank correlation, $C_{R}((\beta\phi)^{-1},X)$, between the SOP, $(\beta\phi)^{-1}$, and both mobility, $\mu$, calculated at $\alpha$-relaxation time and per-particle level $\alpha$-relaxation time. Both correlations exhibit similar trends. \\

\begin{figure}
\centering
\vspace{0.2cm}
\begin{subfigure}{0.225\textwidth}
\includegraphics[width=0.98\linewidth]{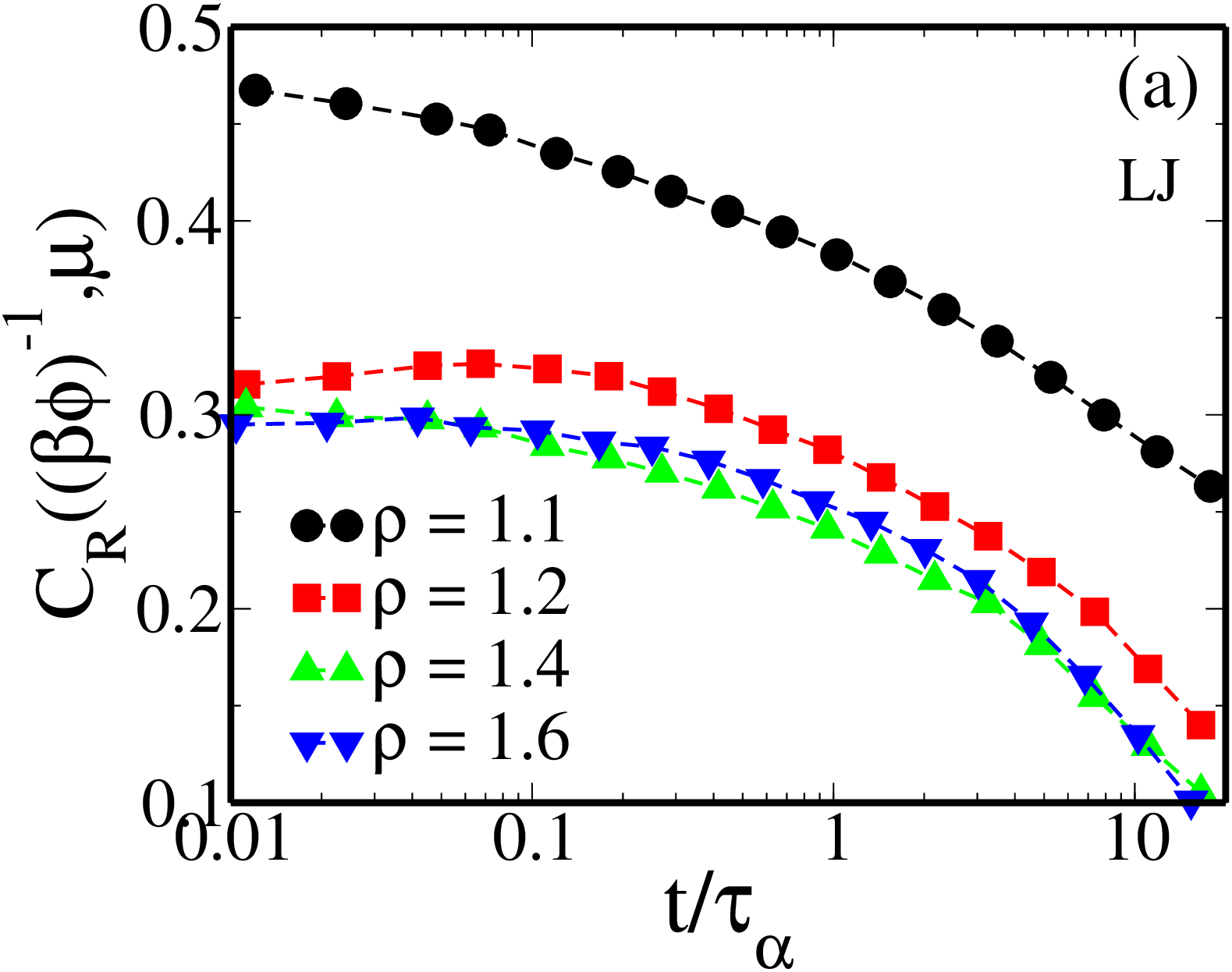}
\end{subfigure}
\hspace{0.2cm}
\begin{subfigure}{0.232\textwidth}
\includegraphics[width=0.98\linewidth]{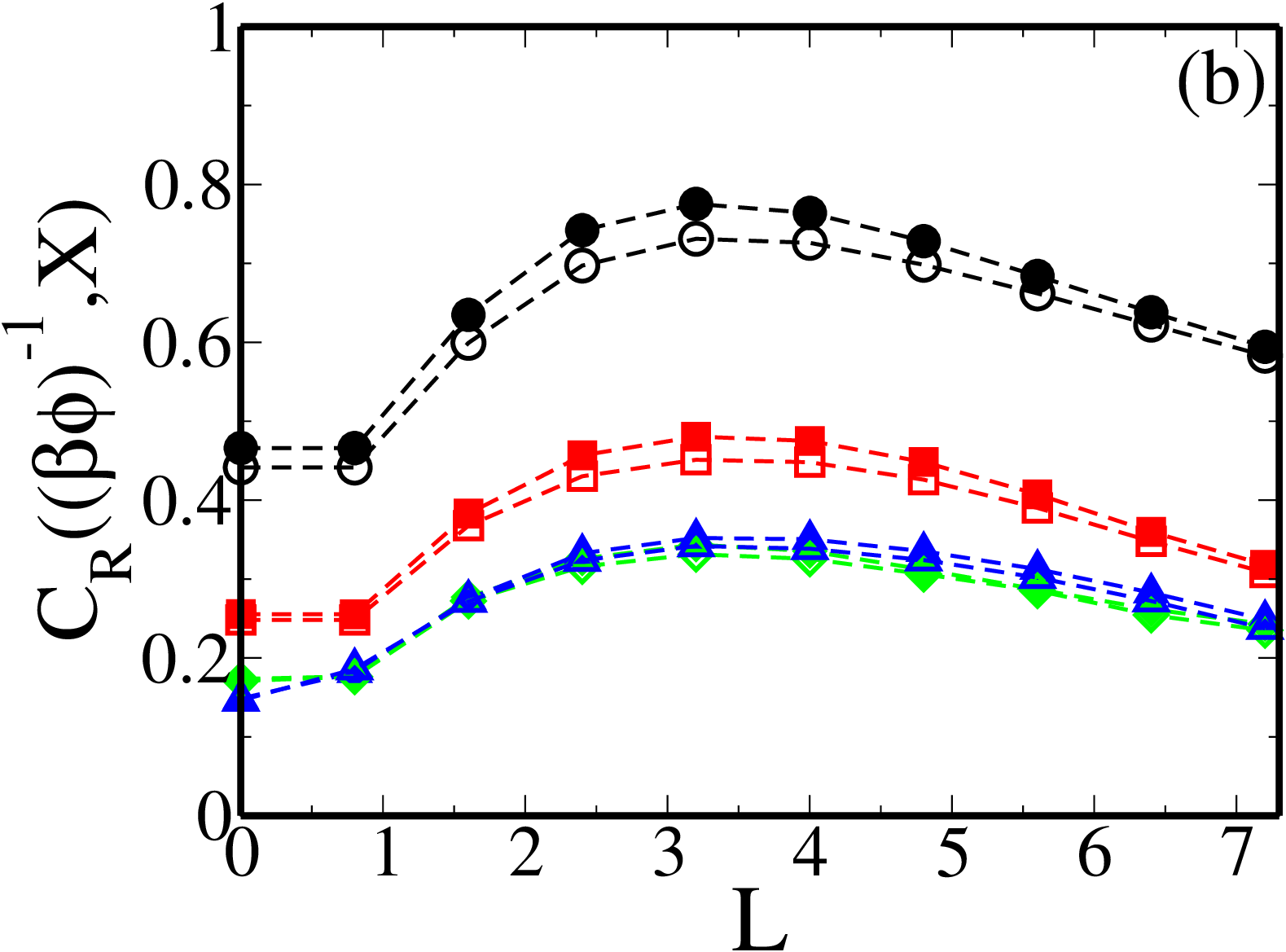}
\end{subfigure}
\vspace{0.2cm}
\begin{subfigure}{0.225\textwidth}
\includegraphics[width=0.98\linewidth]{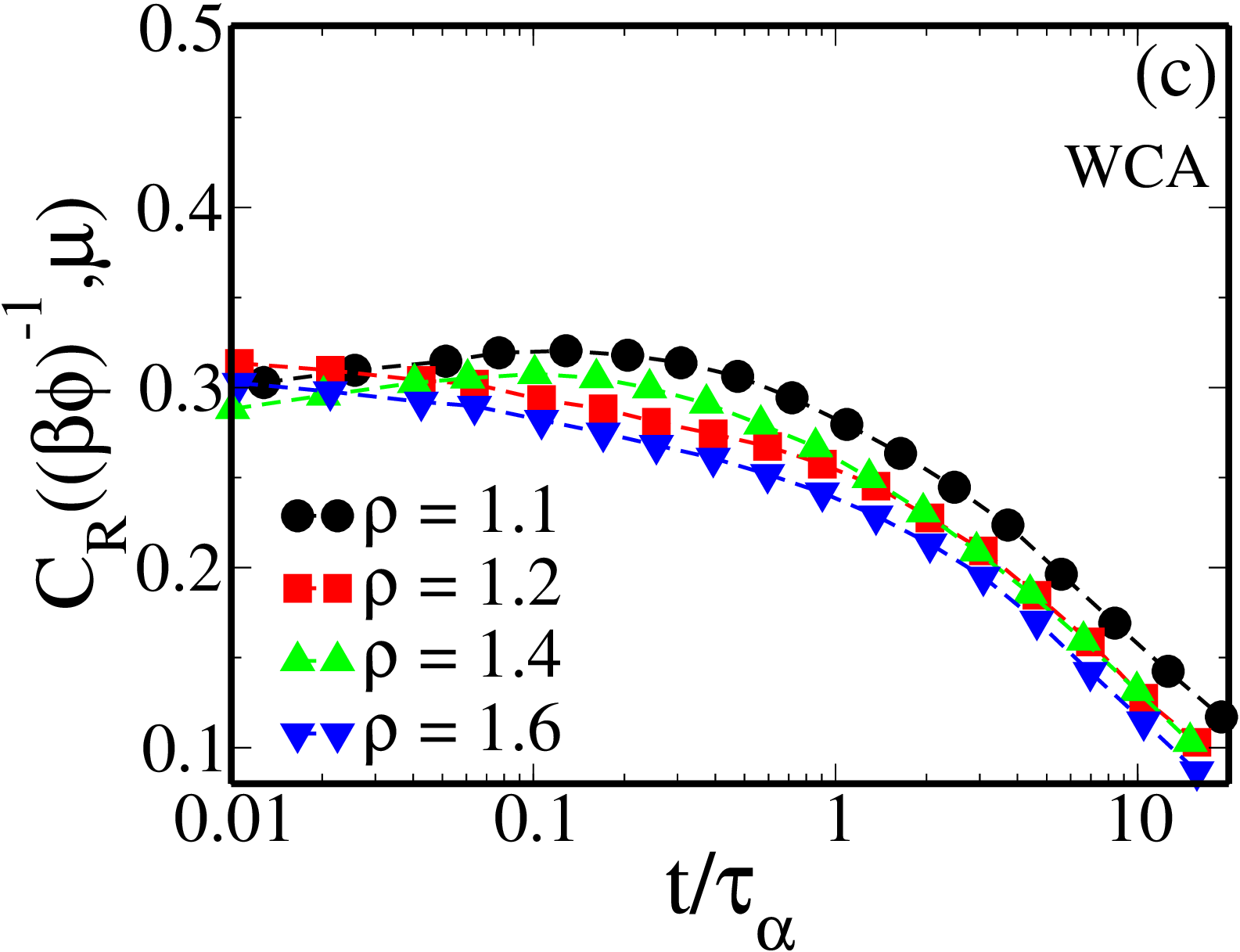}
\end{subfigure}
\hspace{0.2cm}
\begin{subfigure}{0.232\textwidth}
\includegraphics[width=0.98\linewidth]{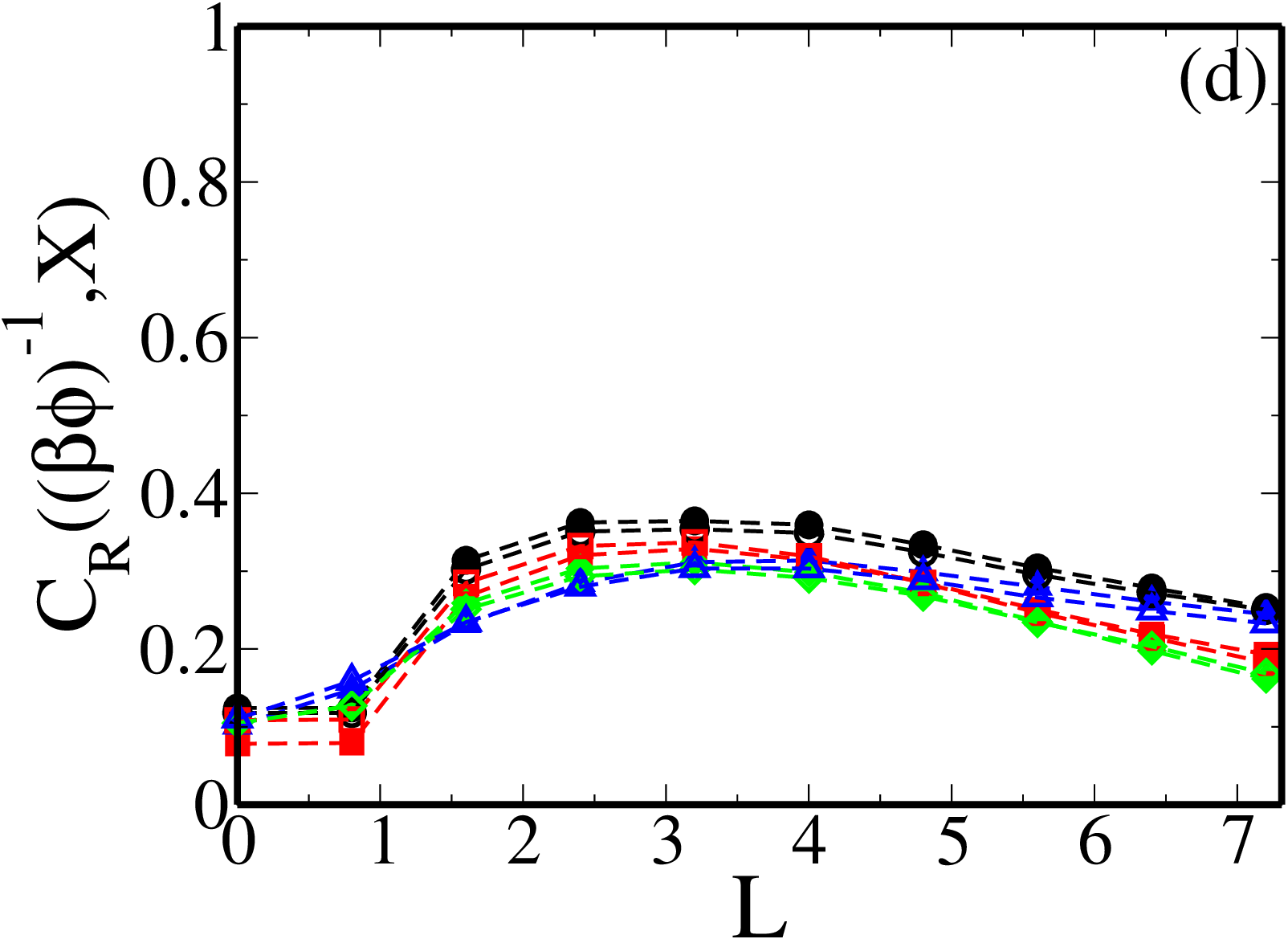}
\end{subfigure}
\vspace{0.2cm}
\begin{subfigure}{0.225\textwidth}
\includegraphics[width=0.98\linewidth]{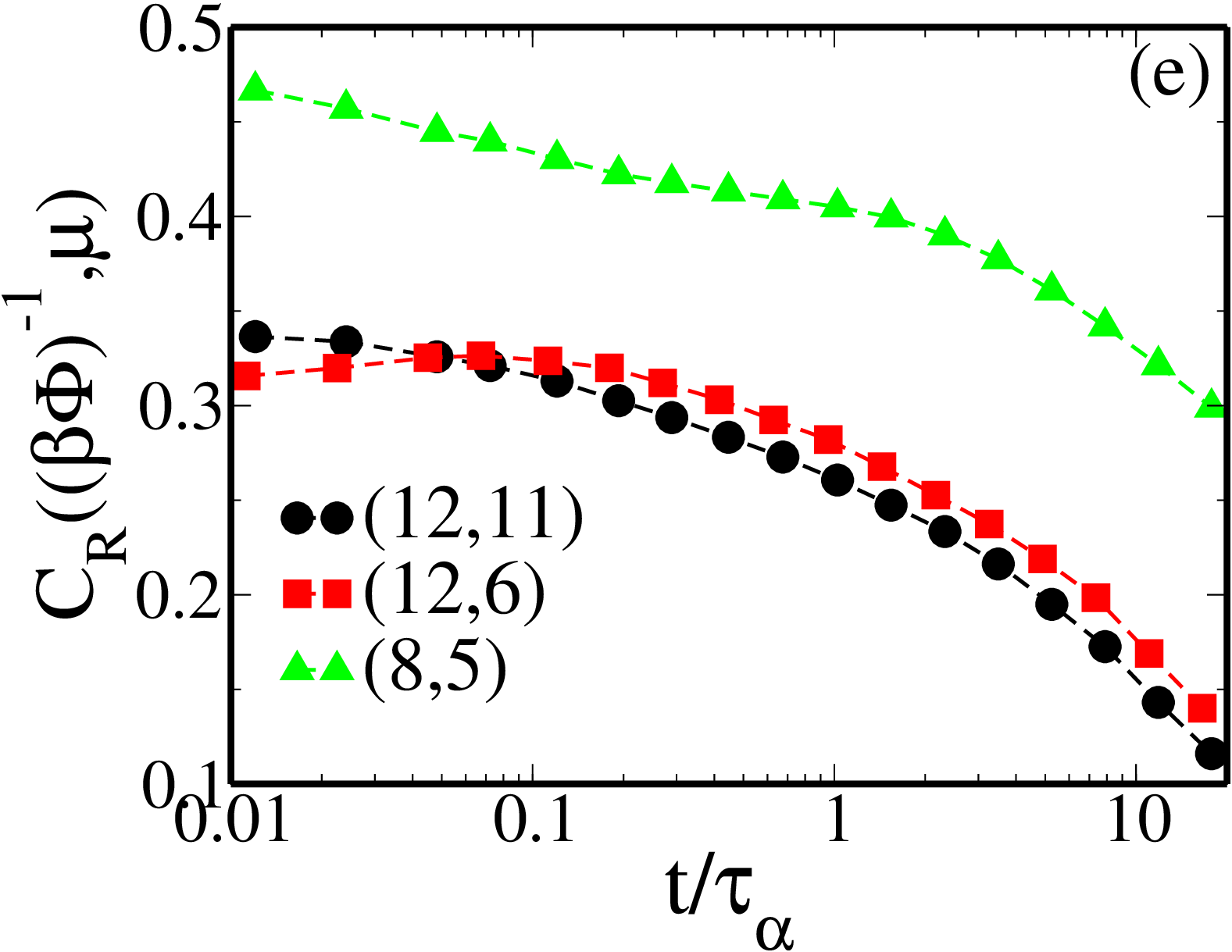}
\end{subfigure}
\hspace{0.2cm}
\begin{subfigure}{0.232\textwidth}
\includegraphics[width=0.98\linewidth]{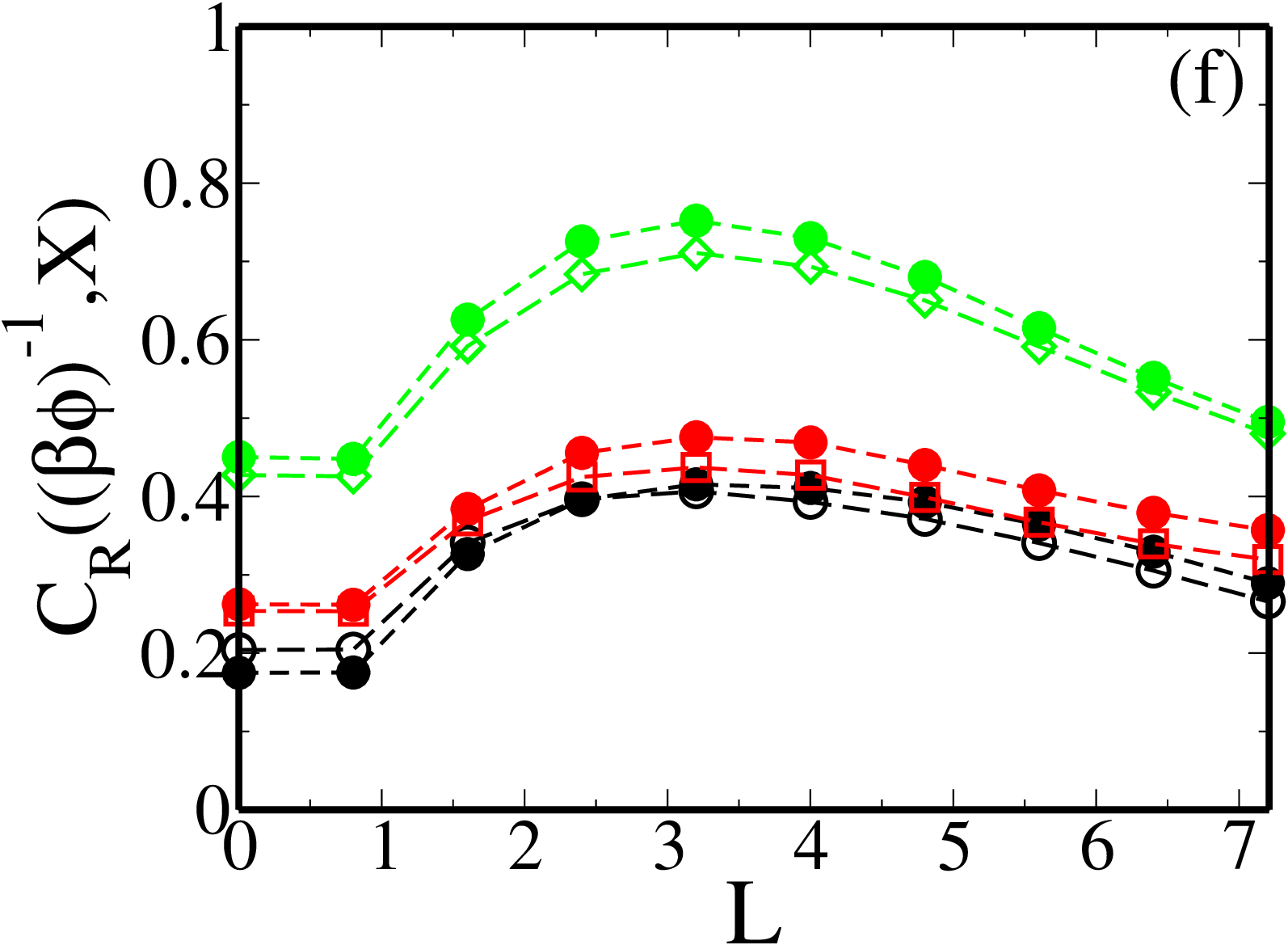}
\end{subfigure}
\vspace{0.2cm}
\caption{\textit{{\textbf{Left panels:} Spearman rank correlation,$C_{R}((\beta\phi)^{-1},\mu)$, between the SOP, $(\beta\phi)^{-1}$, and mobility,$\mu$, computed at $L = 1$ for three different sets of systems:(a)LJ and (c) WCA, each examined at four different densities, and (e) modified LJ $(q,p)$ system. This figure complements the main results shown in Fig.\ref{Fig_correlation}, and demonstrates that the observed density-dependent behavior is not sensitive to the choice of $L$. \textbf{Right panels:} Spearmann rank correlation,$C_{R}((\beta\phi)^{-1},X)$, between the SOP, $(\beta\phi)^{-1}$, and the dynamic quantity $X$, where $X$ represents either the instantaneous mobility $\mu$ evaluated at the $\alpha$-relaxation time (closed symbols) or the particle-level $\alpha$-relaxation times averaged over 200 isoconfigurational ensembles (open symbols) for the same systems shown in left panels. The colour coding in the right panels matches that of the left panels.}
}}
\label{Fig_partau}
\end{figure}
\begin{table}[h]
\centering
\caption{\textit{The working temperature, \( T \), is selected for each system to ensure that the \( \alpha \)-relaxation times are comparable across different densities. This choice facilitates a consistent comparison of structural-dynamic correlations and dynamic heterogeneity analyses among the various systems studied.}}
\vspace{0.2cm}
\label{tab:workingtemp}
\begin{tabular}{|c|c|}
\hline
\textbf{System} & \textbf{T} \\
\hline
LJ, $\rho = 1.1$  & 0.30 \\
\hline
LJ, $\rho = 1.2$  & 0.47 \\
\hline
LJ, $\rho = 1.4$  & 1.00 \\
\hline
LJ, $\rho = 1.6$  & 1.85 \\
\hline
WCA, $\rho = 1.1$ & 0.134 \\
\hline
WCA, $\rho = 1.2$ & 0.31 \\
\hline
WCA, $\rho = 1.4$ & 0.88 \\
\hline
WCA, $\rho = 1.6$ & 1.78 \\
\hline
$(12,11)$             & 0.90 \\
\hline
$(12,6)$             & 0.47 \\
\hline
$(8,5)$             & 0.2475 \\
\hline
\end{tabular}
\end{table}

\textbf{Appendix V: Detection of spinoidal point - Pressure vs. density curve }\\
\\
Fig.\ref{Fig_pressure} shows the variation of inherent structure pressure, $P_{eIS}$, as a function of density, $\rho$, for the Lennard-Jones (LJ) system (black circles) and the (8,5) system (red squares). For each data point, $P_{eIS}$ was obtained by averaging more than 1000 independent configurations of the inherent structure to ensure statistical reliability. The plot reveals a distinct minimum in $P_{eIS}$, located at $\rho = 1.095$ for the LJ system and at $\rho = 1.16$ for the (8,5) system. These values are similar to those reported earlier \cite{Sastry2000_prl_press,Sengupta2011_jcp}, and these minima are indicative of the mechanical instability associated with the spinodal point in the energy landscape. In particular, the densities at which we perform our simulations, $\rho = 1.1$ for the LJ system and $\rho = 1.2$ for the (8,5) system, lie close to these spinodal densities, highlighting that the systems are being studied near the edge of stability.\\
\\
\begin{figure}
\centering
\vspace{0.2cm}
\begin{subfigure}{0.48\textwidth}
\includegraphics[width=1.0\linewidth]{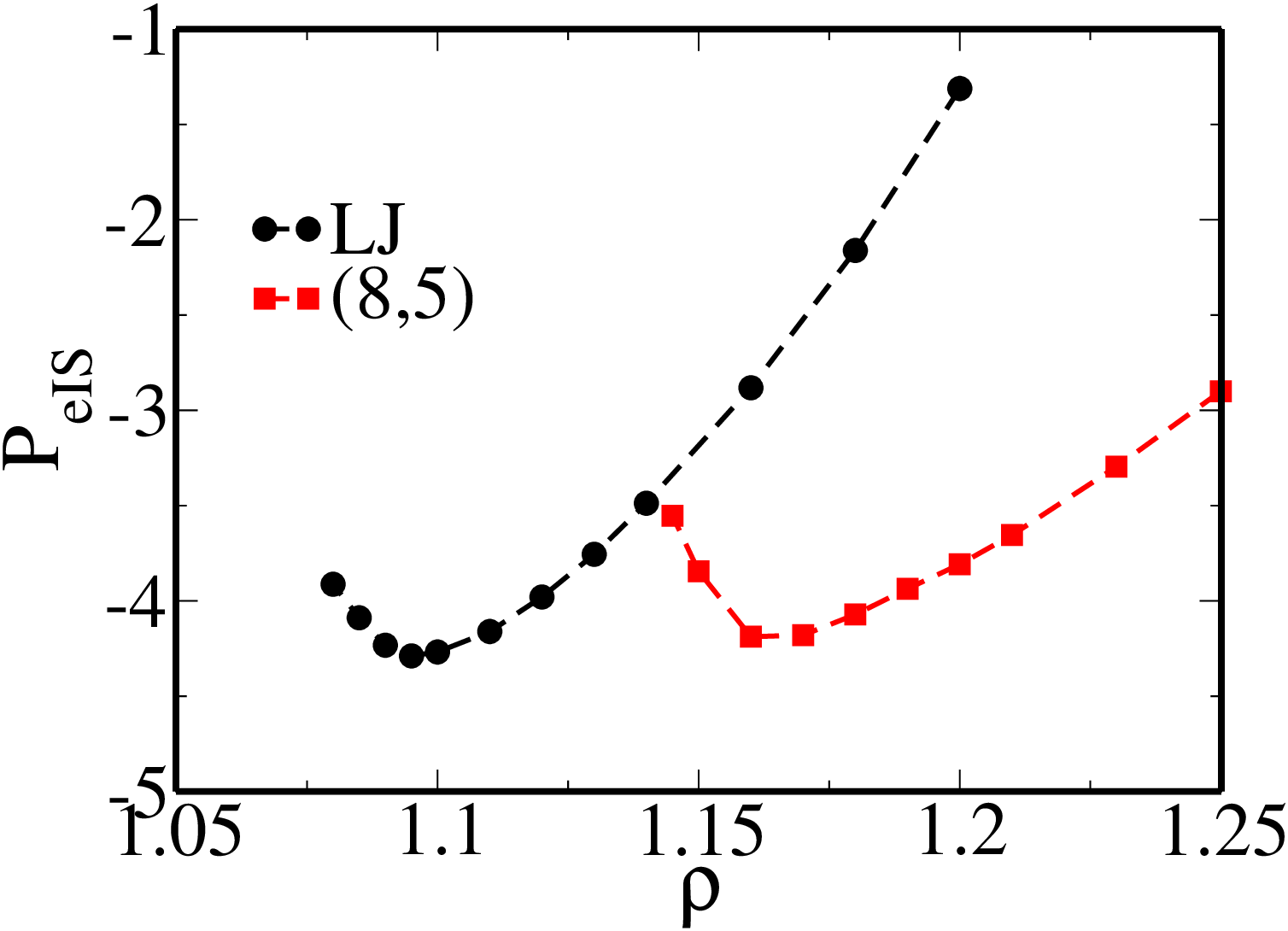}
\end{subfigure}
\vspace{0.2cm}
\caption{\textit{The inherent structure pressure, $P_{\text{eIS}}$, is plotted as a function of density, $\rho$, for the LJ system (black circles) and the (8,5) system (red squares). The minimum of $P_{eIS}$ occurs at $\rho = 1.095$ for the LJ system and at $\rho = 1.16$ for the (8,5) system.
}
}
\label{Fig_pressure}
\end{figure}

{\bf ACKNOWLEDGMENT}\\
\\
S.~M.~B. thanks, Science and Engineering Research Board (SERB, Grant No. SPF/2021/000112 ) for the funding. SS acknowledges support through the JC Bose Fellowship (JBR/2020/000015) from the Science and Engineering Research Board, Department of Science and Technology, India. S.~M.~B. would like to thank Sanat Kumar and Indrajit Tah for discussions. M.~S. acknowledges CSIR for the research fellowships.\\[2mm]
{\bf AVAILABILITY OF DATA}\\
\\
The data that support the findings of this study are available from the corresponding author upon reasonable request.\\[2mm]
\section{REFERENCES}

\end{document}